# Nonlinear social evolution and the emergence of collective action

Benjamin Allen [ID][a,*], Abdur-Rahman Khwaja [ID][a], James L. Donahue [ID][a], Theodore J. Kelly[a], Sasha R. Hyacinthe[a], Jacob Proulx[a], Cassidy Lattanzio [ID][a], Yulia A. Dementieva [ID][a] and Christine Sample [ID][a]

[a]Department of Mathematics, Emmanuel College, Boston, MA 02115, USA
*To whom correspondence should be addressed: Email: allenb@emmanuel.edu
**Edited By:** David Rand

**Abstract**

Organisms from microbes to humans engage in a variety of social behaviors, which affect fitness in complex, often nonlinear ways. The question of how these behaviors evolve has consequences ranging from antibiotic resistance to human origins. However, evolution with nonlinear social interactions is challenging to model mathematically, especially in combination with spatial, group, and/or kin assortment. We derive a mathematical condition for natural selection with synergistic interactions among any number of individuals. This result applies to populations with arbitrary (but fixed) spatial or network structure, group subdivision, and/or mating patterns. In this condition, nonlinear fitness effects are ascribed to collectives, and weighted by a new measure of collective relatedness. For weak selection, this condition can be systematically evaluated by computing branch lengths of ancestral trees. We apply this condition to pairwise games between diploid relatives, and to dilemmas of collective help or harm among siblings and on spatial networks. Our work provides a rigorous basis for extending the notion of "actor", in the study of social evolution, from individuals to collectives.

**Keywords:** evolutionary dynamics, social evolution, collective action, coalescent theory, evolutionary game theory

**Significance Statement**

The way organisms interact affects their ability to survive and reproduce. These interactions can be quite complicated, with each organism' fitness depending on the combined actions of multiple others. We provide a mathematical modeling approach to analyzing natural selection with complex interactions and population structures. Our approach highlights the role of collectives—groups of individuals whose actions combine to affect the fitness of others. We derive a mathematical condition that shows how the behavior of collectives—like that of individuals—is shaped by natural selection. Applying this condition to interactions between family members, and among local communities in a network, illuminates the critical roles played by synergy and genetic relatedness in the evolution of collective behavior.

## Introduction

The evolution of social behavior—behavior that affects the fitness of others—has been a focus of inquiry since the origins of evolutionary biology (1, 2). Examples range from pairwise interactions such as food sharing (3), grooming (4), and contests over mating or food (5), to large-scale collective actions such as pack hunting (6), living bridges (7), biofilm formation (8), and fruiting body aggregation (9). Understanding how these behaviors evolve requires linking their costs and benefits—for those interacting and others in the population—to the long-term fate of genes involved.

Social evolution has been investigated using a variety of theoretical approaches, including kin selection (10–12), multilevel selection (13–16), evolutionary game theory (17–19), and population genetics (20, 21). A particularly influential approach is inclusive fitness theory (10, 22–25), which quantifies selection on social behavior in terms of the fitness effects an actor has on itself and on others, weighted by genetic relatedness. These approaches illuminate how the evolution of social behavior depends on patterns of genetic assortment (10, 26, 27), which in turn emerge from the population's family (11, 27, 28), group (14, 16, 29), spatial (22, 30), and/or network structure (31–35).

The simplest interactions to analyze are those in which each individual has a well-defined, additive effect on the fitness of each other individual. In this case, the aggregate effect on each individual's fitness depends linearly on the phenotypes of those involved. However, real-world social behavior is often nonlinear (6, 36–41). The effects of multiple actors may combine synergistically (more than the sum of separate contributions) (36), antisynergistically (less than the sum) (37, 38), or may switch between synergistic and antisynergistic regimes (6, 39, 40). Nonlinear social evolution has consequences ranging from cancer treatment (42) and antibiotic resistance (43) to the origins of human society (41).

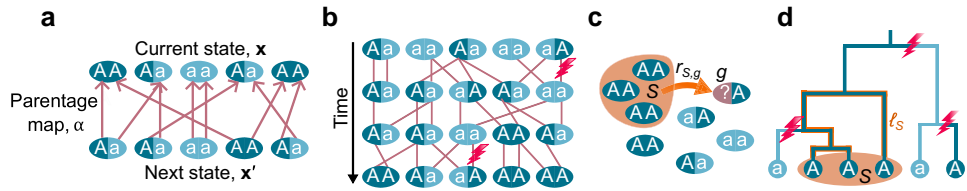

**Fig. 1.** Modeling framework. a) We consider a population of alleles at a specific locus. Alleles can be of type *A* or *a*. Each allele resides in a particular genetic site, within an individual. Each time-step, some alleles are replaced by copies of others, as a result of interaction, reproduction, mating, and/or death. This is recorded in a parentage map, $\alpha$, indicating the parent allele of each site in the new state. b) The process of selection is represented as a Markov chain. State transitions are determined by sampling a parentage map $\alpha$ from a probability distribution, which depends on the current state and captures all effects of social interaction, spatial structure, mating pattern, and so on. With mutation, there is a unique stationary distribution over states. c) Multilateral genetic assortment is quantified by collective relatedness $r_{S,g}$, which characterizes the likelihood that site $g$ contains allele *A* when all sites in set *S* do. d) Under neutral drift, collective relatedness can be computed using the expected branch lengths, $\ell_S$, of the tree representing *S*'s ancestry. The smaller the coalescence length $\ell_S$, the more likely that sites in *S* contain the same allele.

Nonlinear interactions are integral to the evolutionary game theory (17, 18, 44–46), population genetics (20, 21, 47, 48), and multilevel selection (14, 16, 49, 50) approaches to social evolution, and have been incorporated into some inclusive fitness approaches (26, 27, 51). However, nonlinear interactions are challenging to model mathematically, with computational complexity growing as the number of interacting agents increases (27, 46, 52). In particular, nonlinearity makes it difficult to ascribe inclusive fitness quantities to individual actors (12, 25, 53). These challenges are multiplied in populations with complex, heterogeneous structure (54–56).

Here, we derive a condition to determine the outcome of selection involving nonlinear interactions among any number of individuals. This condition—Eq. 10 below—is derived from a general modeling framework (57) that allows for heterogeneous spatial or network structure and arbitrary mating patterns. As in classical inclusive fitness theory (10), our condition involves a sum of fitness effects caused by an actor, weighted by their relatedness to each affected individual. However, since nonlinear effects are collectively produced, the actors here are collectives—arbitrary subsets of the population. This provides a way of understanding nonlinear social evolution in terms of competing individual and collective interests.

## Modeling framework

We build upon a general mathematical framework for natural selection (57), allowing for arbitrary spatial structure, mating patterns, and fitness-affecting interactions (Fig. 1). This framework encompasses classical models of well-mixed populations (58–60), as well as models with heterogeneous spatial structure (55) and nonrandom mating (61), but excludes models with changing population size or structure.

## States and transitions

Taking a gene's-eye view, we imagine a population of alleles, of types *A* or *a*, competing at a single genetic locus. Each allele lives at a particular genetic site, within an individual. Haploids contain one site each, diploids two. Sites may also be labeled with additional information such as sex, spatial location, and/or group membership.

The sites are indexed by a fixed set $G$, of size $n$. The allele occupying each site $g \in G$ is indicated by a binary variable $x_g$, equal to 1 if $g$ contains *A* and 0 if *a*. The overall population state is captured by collecting all variables $x_g$ into a binary vector, **x**.

In each state **x**, individuals may interact, migrate, mate, reproduce, and/or die. On the gene level, some alleles are replaced by copies of others. The new allele in each site $g$ is either survived or copied from an allele previously occupying some site $\alpha(g)$ (Fig. 1a). Here, $\alpha$ is a parentage map (57) from $G$ to itself, indicating the site from which each allele is inherited. Additionally, some (possibly empty) subset $U \subseteq G$ of sites undergo mutation, interchanging *A* and *a*. This results in a new state, **x**′, whereupon the process repeats.

The probability that parentage map $\alpha$ and mutation set $U$ occur in state **x** is denoted by $p_\mathbf{x}(\alpha, U)$. These probabilities capture all consequences of social interaction, competition, mating, and reproduction on the transmission of alleles to the next state. Sampling a pair $(\alpha, U)$ in each state **x**, and constructing the next state **x**′ accordingly, yields a Markov chain representing natural selection.

## Mutation and selection parameters

The mutation rate is quantified by a parameter $0 \leq u \leq 1$. For $u = 0$, mutation is absent, and either *A* or *a* ultimately becomes fixed, with probabilities depending on the initial state. For $u > 0$, the Markov chain converges to a unique stationary probability distribution over states.

Another parameter, $\delta \geq 0$, quantifies the intensity of selection. For neutral drift ($\delta = 0$), the probabilities $p(\alpha, U)$ do not depend on the state **x**. Some of our results apply to arbitrary selection intensity. Others pertain to weak selection ($\delta \ll 1$), meaning that social interactions have small—but still potentially nonlinear (62)—effects on fitness.

## Reproductive value

Even under neutral drift, some sites may contribute more than others to the future gene pool. We quantify this by assigning each site $g$ a *reproductive value*, $v_g$ (63). Under neutral drift, a site's reproductive value must equal the total reproductive value of itself (if it survives) and its offspring. This leads to the recurrence relation

$$v_g = \sum_\alpha p(\alpha) \sum_{h \in \alpha^{-1}(g)} v_h. \tag{1}$$

Equation 1, together with the normalization $\sum_{g \in G} v_g = n$, uniquely determines all reproductive values, $v_g$ (57). A homogeneous population, which looks the same from the perspective of any individual, has all reproductive values equal to one.

## Quantifying selection

We quantify selection on two time-scales. On the scale of a single time-step, we define the *fitness increment* of each site $g$ in state **x** as the expected change in reproductive value from $g$ to $g$'s progeny:

$$w_g(\mathbf{x}) = -v_g + \sum_a p_\mathbf{x}(a) \sum_{h \in a^{-1}(g)} v_h. \quad (2)$$

Overall, selection in a given state $\mathbf{x}$ is quantified by the expected change, $\Delta(\mathbf{x})$, in the total reproductive value of $A$ alleles, $\sum_{g \in G} x_g v_g$. This change can be computed using a variant of the Price equation (64):

$$\Delta(\mathbf{x}) = \frac{1}{n} \sum_{g \in G} w_g(\mathbf{x})(x_g - \bar{x}), \quad (3)$$

where $\bar{x} = \frac{1}{n} \sum_{h \in G} x_h$ is the (unweighted) frequency of $A$.

On the time-scale of the entire selection process, we say that selection favors allele $A$ if, in the low-mutation limit, $A$ has greater stationary frequency than $a$. Equivalently, selection favors $A$ if, for $u = 0$, $A$ is more likely to replace $a$ than vice versa, when starting from a single mutant. We prove (SI Appendix, Theorem 3.3) that selection favors $A$ if and only if

$$\left. \frac{d\mathbb{E}[\Delta(\mathbf{x})]}{du} \right|_{u=0} > 0, \quad (4)$$

where $\mathbb{E}$ denotes expectation over the stationary distribution, echoing similar results in other frameworks (22, 24, 65).

## Results

### Collective relatedness

With nonlinear interactions, selection depends on multilateral patterns of genetic assortment (26, 27, 48, 51, 55). To quantify these patterns, we define the collective relatedness, $r_{S,g}$, of a set $S$ of sites to a single site $g$:

$$r_{S,g} = \lim_{u \to 0} \frac{\mathbb{E}[\underline{x}_S(x_g - \bar{x})]}{\mathbb{E}[\bar{x}(1 - \bar{x})]}. \quad (5)$$

Above, $\underline{x}_S = \prod_{h \in S} x_h$ has value 1 if all sites in $S$ contain allele $A$, and 0 otherwise. The numerator in Eq. 5 quantifies whether $g$ is more or less likely than an average site to contain allele $A$, when all of $S$ does. The denominator is the expected allelic variance in the population. Equation 5 generalizes standard pairwise relatedness measures—based on covariance (11), identity-by-descent (22), and geometry (66)—and builds upon previous efforts to extend relatedness beyond pairs (26, 27, 51, 67).

Collective relatedness is difficult to calculate for arbitrary selection intensity. For neutral drift ($\delta = 0$), however, it can be computed using coalescent theory (68, 69). The key quantity is the *coalescence length*, $\ell_S$, defined as the expected total branch length of a tree representing the ancestry of set a $S$ (Fig. 1d). Tracing one step back in time leads to the recurrence relation

$$\ell_S = |S| + \sum_a p(a) \ell_{a(S)}. \quad (6)$$

Equation 6, together with $\ell_S = 0$ for singleton sets $S$, uniquely determines all coalescence lengths (70). If mutation rates vary over sites, the coalescence lengths are scaled accordingly; see SI Appendix, Eq. 4.8. Employing a well-known relationship between coalescence length and identity-by-descent probability (70, 71), we derive (SI Appendix, Eq. 4.15) a formula for neutral collective relatedness in terms of coalescence lengths:

$$r_{S,g} = \frac{\bar{\ell}_S - \ell_{S \cup \{g\}}}{\bar{\ell}}. \quad (7)$$

Above, $\bar{\ell}_S$ is the average of $\ell_{S \cup \{h\}}$ as $h$ runs over all sites in $G$, and $\bar{\ell}$ is the average of $\ell_{\{h,k\}}$ over all pairs $h, k \in G$.

### Main result

To obtain a condition for selection, we write the fitness increments, $w_g(\mathbf{x})$, uniquely in the polynomial form (72)

$$w_g(\mathbf{x}) = \sum_{S \subseteq G} c_{S,g} \underline{x}_S. \quad (8)$$

Each coefficient, $c_{S,g}$, represents a synergistic effect on $g$'s fitness that arises only if all sites in $S$ contain allele $A$. As $S$ runs over all subsets of $G$, the terms $c_{S,g} \underline{x}_S$ form the building blocks of any nonlinear dependence of $w_g$ on the state $\mathbf{x}$. Substituting this representation into Eq. 3 yields

$$\mathbb{E}[\Delta(\mathbf{x})] = \frac{1}{n} \sum_{g \in G} \sum_{S \subseteq G} c_{S,g} \mathbb{E}[\underline{x}_S(x_g - \bar{x})]. \quad (9)$$

Dividing by $\mathbb{E}[\bar{x}(1 - \bar{x})]$, taking $u \to 0$, and applying Eqs. 4 and 5, we arrive at our main result (SI Appendix, Theorem 5.1): Selection favors allele $A$ if and only if

$$\sum_{g \in G} \sum_{S \subseteq G} c_{S,g} r_{S,g} > 0. \quad (10)$$

This condition has two complementary interpretations. The first is recipient-centered: for a given site $g$, the sum $\sum_{S \subseteq G} c_{S,g} r_{S,g}$ characterizes the expected fitness effect of all social interactions experienced by an $A$ allele in this site. $A$ is favored if the total effect on $A$ alleles, over all sites $g$, is positive. The second is actor-centered: for a given set $S$ of sites, the sum $\sum_{g \in G} c_{S,g} r_{S,g}$ has the form of an inclusive fitness effect (10), in that $S$'s contribution, $c_{S,g}$, to the fitness of each site $g$, is weighted by collective relatedness, $r_{S,g}$. However, in contrast to standard inclusive fitness theory, the actor, $S$, is not an individual but a collective—a set of genetic sites that can synergistically affect the fitness of themselves and others.

Equation 10 is valid for any selection intensity, but difficult to evaluate because the collective relatedness coefficients, $r_{S,g}$, depend on the selection process. For weak selection, however, collective relatedness can be computed at neutrality using Eqs. 6 and 7 (SI Appendix, Theorem 5.2). This provides a method to evaluate weak selection on any nonlinear fitness-affecting behavior, with arbitrary spatial, network, group, and/or mating structure. If the synergistic fitness effects $c_{S,g}$ vanish for sets $S$ above a fixed size, this computation takes polynomial time.

#### Phenotypic condition

For diploids, we derive an equivalent condition at the level of phenotypes. We consider a fixed set $I$ of individuals. $AA$ individuals have phenotype 1, $aa$'s have phenotype 0, and $Aa$'s have phenotype 1 with some probability $h$ (representing the degree of genetic dominance) and 0 otherwise. Overall, an individual $i$, with genetic sites $i_1$ and $i_2$, has phenotype 1 with probability

$$\varphi_i = h(x_{i_1} + x_{i_2}) + (1 - 2h) x_{i_1} x_{i_2}. \quad (11)$$

In analogy with Eq. 5, we define the collective relatedness of a set $J$ of individuals to an individual $i$:

$$r_{J,i} = \lim_{u \to 0} \frac{\mathbb{E}\left[\left(\prod_{j \in J} \varphi_j\right)(\bar{x}_i - \bar{x})\right]}{\mathbb{E}[\bar{x}(1 - \bar{x})]}, \quad (12)$$

where $\bar{x}_i = (x_{i_1} + x_{i_2})/2$ is the frequency of allele $A$ in individual $i$. This leads to a phenotypic analog of Eq. 10 (SI Appendix, Theorem 5.5): selection favors allele $A$ if

$$\sum_{i \in I} \sum_{J \subseteq I} c_{J,i} r_{J,i} > 0, \qquad (13)$$

where $c_{J,i}$ is a synergistic effect on $i$'s fitness that arises when all individuals in set $J$ have phenotype 1.

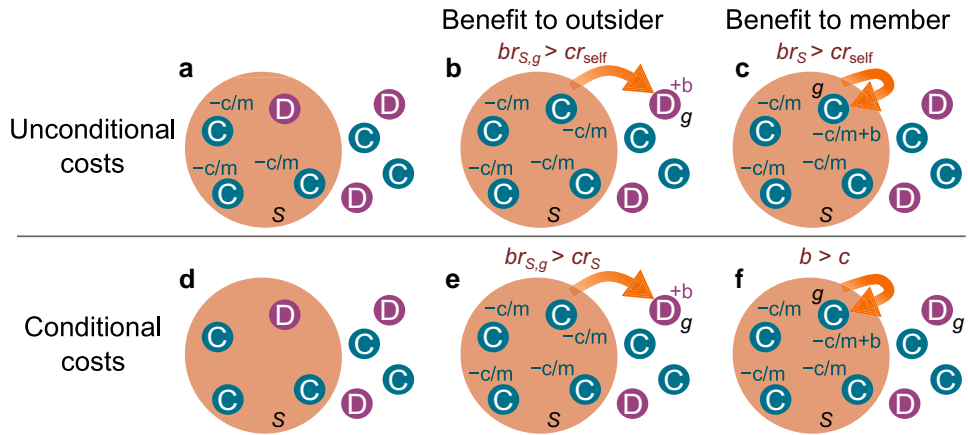

**Fig. 2.** Collective Action dilemma. A collective $S$, of size $m$, may help or harm a target $g$. There are two heritable strategies: (C)ontribute or (D)o not contribute. If all members of $S$ contribute, $g$ receives benefit $b$; otherwise no benefit is received. a) For unconditional costs, each Contributor in $S$ pays cost $c/m$. b) Applying Eq. 10, selection favors collective action if $br_{S,g} > cr_{\text{self}}$. c) If the target $g$ belongs to $S$, then $r_{S,g}$ is replaced by $S$'s intrarelatedness $r_S$. d) Conditional costs are only paid if the benefit would be achieved. e) Action is favored if $br_{S,g} > cr_S$. f) If the target belongs to $S$, the condition becomes $br_S > cr_S$, which reduces to $b > c$.

### Games between diploid relatives

As a first example, let us consider an interaction between diploid relatives, represented by the game matrix

$$\begin{array}{c} \\ C \\ D \end{array} \begin{pmatrix} C & D \\ b-c+d & -c \\ b & d \end{pmatrix}. \qquad (14)$$

In this game, Cooperators (C) pay cost $c$ to provide benefit $b$ to the other player, while Defectors (D) do not. Additionally, both players receive a synergistic effect, $d$, if they play the same strategy. Allowing $b$, $c$, and $d$ to be arbitrary, any 2 × 2 matrix game can be written in the form of Eq. 14, up to an additive constant that does not affect selection.

Given a representative pair of interaction partners, $i, j \in I$, with respective phenotypes $\varphi_i$ and $\varphi_j$ (1 for C, 0 for D), the expected payoff to $i$ is

$$f_i = -c\varphi_i + b\varphi_j + d(\varphi_i\varphi_j + (1-\varphi_i)(1-\varphi_j)). \qquad (15)$$

Under weak selection, the fitness increment of each individual $i$ is proportional to $f_i - \bar{f}$, where $\bar{f}$ is population average payoff. Applying Eq. 13, weak selection favors cooperation if

$$-cr_{\{i\},i} + br_{\{j\},i} + d\left(2r_{\{i,j\},i} - r_{\{i\},i} - r_{\{j\},i}\right) > 0. \qquad (16)$$

We quantify the kin relationship between partners by the probabilities, $p_1$ and $p_2$, that their maternally inherited and paternally inherited alleles, respectively, descend from the same allele copy in a recent common ancestor. For example, maternal half-siblings have $p_1 = 1/2$ and $p_2 = 0$, while full cousins have $p_1 = p_2 = 1/8$. Computing neutral collective relatedness according to Eq. 12 with the aid of Eq. 7 (SI Appendix, Section 8.6), we find that (i) self-relatedness is $r_{\{i\},i} = 1/2$ for diploids; (ii) relatedness to partner is $r_{\{j\},i} = r/2$, where $r = (p_1 + p_2)/2$ is Wright's (73) coefficient of relationship (one-half for full siblings, one-eighth for cousins, etc.); and (iii) the pair's collective relatedness to each partner is

$$r_{\{i,j\},i} = \frac{1+r}{4} + \frac{\left(h - \frac{1}{2}\right)(r - p_1 p_2)}{6}, \qquad (17)$$

where $h$ is the degree of dominance for the Cooperator phenotype. Substituting into Eq. 16, weak selection favors cooperation if and only if

$$-c + br + \frac{2d\left(h - \frac{1}{2}\right)(r - p_1 p_2)}{3} > 0. \qquad (18)$$

This condition augments Hamilton's rule (10)—that cooperation is favored if the benefit, multiplied by relatedness to the target, exceeds the cost—with an additional term capturing the joint effects of synergy and genetic dominance. If $h = 1/2$ (no dominance) or $d = 0$ (no synergy), the third term vanishes and Hamilton's rule, $br > c$, is recovered. The factor $r - p_1 p_2$ in the third term is nonnegative, and strictly positive unless partners are clones ($p_1 = p_2 = 1$) or unrelated ($p_1 = p_2 = 0$). Cooperation is therefore promoted if it is synergistic ($d > 0$) and mostly dominant ($h > 1/2$), or antisynergistic ($d < 0$) and mostly recessive ($h < 1/2$). Although we have described this scenario in terms of cooperation, Eq. 18 applies to any two-player, two-strategy game played by diploid relatives. This result extends previous analyses of evolutionary games between relatives, as we discuss in SI Appendix, Section 8.7.

### Collective Action Dilemma

To illustrate the application of Eq. 10 to collective help or harm, we introduce the "Collective Action Dilemma" (Fig. 2). A collective $S$, of size $m$, may help or harm a target $g$, inside or outside of $S$. Members of $S$ may contribute, or not, to this action. If all contribute, $g$ receives a "benefit" $b$ (positive for help, negative for harm); otherwise, no benefit is received. This action costs $c/m$ to each of $S$'s members, for a total cost of $c > 0$. Costs may be unconditional, meaning each contributor in $S$ pays independently of others; or conditional, meaning contributors assess support for collective action (e.g. via quorum-sensing (74)) and pay only if the benefit would be achieved. This scenario resembles a threshold public goods game (45), except that the benefit goes to a specific target rather than being divided equally among collective members. For targets outside $S$, action represents collective altruism ($b > 0$) or spite ($b < 0$).

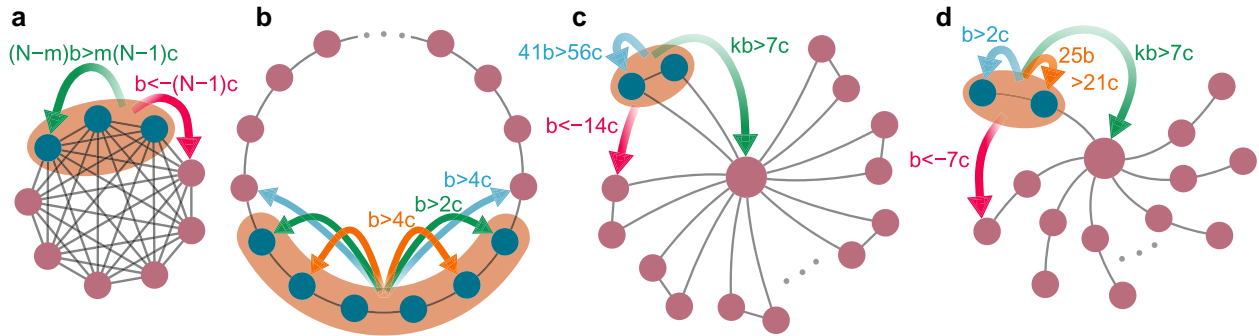

**Fig. 3.** Collective action on networks. We analyze the Collective Action Dilemma, with unconditional costs, on a fixed network of size $N$. a) In a well-mixed population, represented by a complete graph, a collective of size $m$ is favored to help a member if $(N-m)b > m(N-1)c$, and favored to harm an outsider if $b < -(N-1)c$. b) For a large ($N \gg 1$) cycle network, a connected collective of four or more nodes is favored to help its own boundary nodes if $b > 2c$, and the neighbors of these boundary nodes if $b > 4c$; neither help nor harm are favored to other nodes. c) On a windmill network with $k \gg 1$ blades, a blade is favored to help the hub if $kb > 7c$. This can occur even if the benefit is a small fraction of the cost. In contrast, help to a node within the blade is only favored if $41b > 56c$. Harmful behavior is favored toward nodes in other blades if $b < -14c$. d) The "spider" network displays similar behavior, but help is more readily favored to the inner node of a leg ($25b > 21c$) than to the outer node ($b > 2c$). Results in panels b–d refer to large populations ($N \gg 1$); results for finite $N$ are derived in SI Appendix, Sections 9.5–9.7, and shown in Fig. S1.

Applying Eq. 10, we obtain a collective form of Hamilton's rule (10): collective action is favored if

$$br_{S,g} > cr_{\text{self}} \quad \text{(unconditional costs)}$$
$$br_{S,g} > cr_S \quad \text{(conditional costs).} \quad (19)$$

Above, $r_{\text{self}} = (1/m)\sum_{h \in S} r_{\{h\},h}$ is the average self-relatedness among $S$'s members, and

$$r_S = \lim_{u \to 0} \frac{\mathbb{E}[\underline{x}_S - \underline{x}_S \bar{x}]}{\mathbb{E}[\bar{x}(1-\bar{x})]} \quad (20)$$

is the common value of $r_{S,g}$ for all members $g$ of $S$. Analogous conditions apply at the phenotype level, using Eqs. 12 and 13.

For collectives of two or more, $r_S < r_{\text{self}}$. This means, intuitively, that collective action is more easily selected when costs are conditional. If the target belongs to $S$, then $r_{S,g} = r_S$, and action is favored for unconditional costs if $br_S > cr_{\text{self}}$, and for conditional costs if $br_S > cr_S$, or simply $b > c$.

### Well-mixed haploid population

We first consider the Collective Action Dilemma in a haploid, well-mixed population of size $N$ (Fig. 3a). Using Eqs. 6 and 7 (SI Appendix, Section 7.3), we compute $r_{\text{self}} = 1$, $r_S = (N-m)/(m(N-1))$, and $r_{S,g} = -1/(N-1)$ for $g$ outside of $S$. Since $r_{S,g} < 0$, collective help to outsiders is never favored according to Eq. 19. Harm to outsiders is favored if $b < -(N-1)c$ for unconditional costs, or $mb < -(N-m)c$ for conditional costs, but these conditions become infeasible for large $N$. Collective help to a member (with unconditional costs) is favored if $(N-m)b > m(N-1)c$. In particular, the benefit must exceed $m$ times the cost. Interestingly, this condition applies even if intermediate benefits arise when only some in $S$ contribute (SI Appendix, Theorem 7.1).

### Diploid siblings

J.B.S. Haldane famously quipped that he would jump into a river to save two brothers, or eight cousins. But what if Haldane must collaborate with one or more siblings to save another (28, 75)? We represent this as a Collective Action Dilemma, with a collective $J$ of full siblings and a target sibling $i$ outside the collective.

For unconditional costs, applying Eq. 19 at the phenotype level, collective $J$ is favored to help individual $i$ if $br_{J,i} > cr_{\text{self}}$. Evaluating Eq. 12 by way of Eq. 7, we obtain (SI Appendix, Section 8.4) $r_{\text{self}} = 1/2$, and $r_{J,i} = 1/4$ for $i$ outside of $J$. Substituting, weak selection favors $J$ to collectively help $i$, with unconditional costs, if $b > 2c$. Thus, two of Haldane's siblings will sacrifice themselves to save four, three to save six, and so on. This condition applies even if intermediate benefits accrue when only some in $J$ contribute, because the relatedness to the target sibling, $r_{J,i} = 1/4$, is independent of the collective's size.

For conditional costs, help is favored if $br_{J,i} > cr_J$, or equivalently, $b > 4cr_J$. The values of $r_J$ are given in Table 1. For large collectives, $r_J$ approaches $1/4$; in this case, help to a sibling is favored whenever there is a net benefit, $b > c$.

## Collective action on networks

Network structure—representing spatial or social relationships—profoundly affects the evolution of social behavior (31, 32, 35). Exact results have been derived for pairwise interactions (32, 33, 35), but are difficult to obtain for interactions beyond pairs (34, 54, 56, 76). A general finding is that network structure promotes selection for cooperative behavior among neighbors (31, 32, 35) and in localized groups (34, 54, 56, 76). However, little theory exists for how spatial collectives evolve to act toward outsiders, or toward different members within the collective.

Here, we analyze the Collective Action Dilemma with unconditional costs on networks, played by a given collective $S$ and target node $g$. Strategies reproduce via death–Birth updating (31, 32, 35): First, a node is chosen, with uniform probability, to be replaced. Then, a neighbor is chosen with probability proportional to (payoff) × (edge weight) to reproduce into the vacancy.

### Condition for selection

The collective Hamilton's rule, Eq. 19, does not directly apply on networks, because death–Birth updating induces two additional effects on fitness. First, since higher degree nodes have more opportunities to reproduce, each node $h$ has reproductive value proportional to its degree, $d_h$ (77). Second, when a site becomes vacant, the neighbors competing to fill the vacancy are two-step neighbors of each other (31, 32). Consequently, any effect on $h$'s payoff induces a compensating effect on $h$'s two-step neighbors

**Table 1.** Collective intrarelatedness, $r_J$, of $m$ full siblings in a diploid population.

| # of siblings, $m$ | 1 | 2 | 3 | 4 | 5 | $m \to \infty$ |
|---|---|---|---|---|---|---|
| Arbitrary dominance, $h$ | $\frac{1}{2}$ | $\frac{17+2h}{48}$ | $\frac{57+12h-4h^2}{192}$ | $\frac{209+38h+12h^2-24h^3}{768}$ | $\frac{801+104h+88h^2-32h^3-80h^4}{3072}$ | $\frac{1}{4}$ |
| Recessive ($h = 0$) | $\frac{1}{2}$ | $\frac{17}{48}$ | $\frac{19}{64}$ | $\frac{209}{768}$ | $\frac{267}{1024}$ | $\frac{1}{4}$ |
| No dominance ($h = \frac{1}{2}$) | $\frac{1}{2}$ | $\frac{3}{8}$ | $\frac{31}{96}$ | $\frac{19}{64}$ | $\frac{433}{1536}$ | $\frac{1}{4}$ |
| Dominant ($h = 1$) | $\frac{1}{2}$ | $\frac{19}{48}$ | $\frac{65}{192}$ | $\frac{235}{768}$ | $\frac{881}{3072}$ | $\frac{1}{4}$ |

Collective relatedness is computed using Eqs. 7 and 12 in SI Appendix, Section 8.4.

(33). Including these effects in Eq. 10, weak selection favors S to help or harm $g$ if (SI Appendix, Eq. 9.9)

$$bd_g\left(r_{S,g} - r_{S,g}^{(2)}\right) > \frac{c}{m}\sum_{h\in S} d_h\left(r_h - r_h^{(2)}\right). \quad (21)$$

Above, $r_{S,g}^{(2)}$ is the mean collective relatedness of S to two-step random walk neighbors of $g$, $r_h = r_{\{h\},h}$ is the self-relatedness of node $h$, and $r_h^{(2)} = r_{\{h\},h}^{(2)}$ is the mean relatedness of $h$ to its own two-step neighbors. The left-hand side of Eq. 21 compares the effect of collective action on the target $g$ (weighted by collective relatedness from S) to that on $g$'s two-step neighbors. The right-hand side makes the same comparison for costs paid. Equation 21 can be rewritten in Hamilton-like form as

$$b\kappa_{S,g} > c, \qquad \kappa_{S,g} = \frac{md_g\left(r_{S,g} - r_{S,g}^{(2)}\right)}{\sum_{h\in S} d_h\left(r_h - r_h^{(2)}\right)}, \quad (22)$$

where $\kappa_{S,g}$ is the cost–benefit threshold—also known as scaled relatedness (24) or compensated relatedness (33)—for S to act on $g$.

If $r_{S,g} > r_{S,g}^{(2)}$—meaning S is more related to $g$ than to $g$'s two-step neighbors—then $\kappa_{S,g}$ is positive, and S can be favored to help $g$ for sufficiently large benefit. If $r_{S,g} < r_{S,g}^{(2)}$, then $\kappa_{S,g}$ is negative and only harm can be favored.

*Simple networks*

Applying Eq. 21 to theoretical network families reveals key features of spatial collective action (Fig. 3; SI Appendix, Sections 9.5–9.7 and Fig. S1). First, collective help is most strongly favored to targets near the collective's boundary. For example, on a large cycle (Fig. 3b), a collective of four or more connected nodes is favored to help its own boundary node if $b > 2c$, or the immediate neighbors of a boundary node if $b > 4c$. Second, help is never favored (on any network) to interior members two or more steps from the collective's boundary, because then $r_{S,g} = r_{S,g}^{(2)} = r_S$ and the left-hand side of Eq. 21 vanishes. Third, extreme collective altruism—with benefit much less than the cost—can be favored to highly connected neighbors (Fig. 3c and d). Such "hubs" have high reproductive value, making them critical for the spread of alleles. Fourth, collective harm can be favored to distant targets, when the collective is more related to the target's two-step neighbors than to the target itself (Fig. 3c and d).

*Spatial networks*

For a more realistic model of 2D spatial structure, we turn to Delaunay networks (78), formed by placing random points in a square and linking neighbors (Fig. 4). Delaunay networks have been used to model cancer cells in solid tumors, which cooperate by producing growth hormones (40). They can also represent, for example, the spatial layout of chambers in communal nests of sociable weavers (*Philetairus socius*)—wherein males collectively maintain the nest and disperse locally within it (80)—or microbes on solid substrates that produce beneficial diffusible goods (81).

For localized collectives in Delaunay networks, we find that spatial structure promotes help to the majority of collective members (Fig. 4c), in line with previous findings from spatial public goods games (34, 56, 76). However, in some cases, spatial structure can inhibit help to a collective's interior sites. This is because these interior sites are less relevant to whether the contributor allele spreads beyond the collective.

To those outside the collective (Fig. 4d), spatial structure can promote help to neighbors, and in rare cases neighbors-of-neighbors, but usually promotes harm to more distant targets. Costly help and harm both become less favored as collective size increases.

The same patterns emerge in observed spatial networks (80) of chambers in sociable weaver nests (Fig. S2). Our findings accord, for example, with the observation that microbial public goods production is favored only when diffusion is limited (81), or that sociable weavers localize their nest-maintenance efforts to areas near their home chamber (80).

## Complexity of selection

The condition for selection, Eq. 10, involves a sum over all subsets and sites in the population. In full generality, the number of terms grows exponentially with the population size. However, four factors can substantially reduce the number of terms.

First, in realistic scenarios, the vast majority of collectives S will have negligible relatedness, $r_{S,g}$, and/or potential for synergistic effects, $c_{S,g}$, to all or most targets $g$. In spatially structured populations, for example, only nearby individuals are likely to have sufficient collective relatedness and synergistic potential to contribute significantly to selection. In theoretical models, one need only consider collectives S that are involved in the interactions being modeled, which are typically a small fraction of all possible subsets.

Second, for models with symmetry (82), only one term is needed for each *class* of collectives. For example, all sets of three consecutive nodes on the cycle (Fig. 3b) comprise a class, since they are equivalent by rotation. Each such class can be represented by a single term in Eq. 10, dramatically reducing the number of terms.

Third, in empirical contexts, statistical significance criteria can be used to eliminate terms. As a statistical analog of the polynomial representation in Eq. 8, one may form the polynomial regression model (SI Appendix, Section 11)

$$W_g = \sum_{S \in \mathcal{S}_g} c_{S,g}\, \underline{x}_S + \epsilon_{g,\mathbf{x}}. \quad (23)$$

Above, $W_g$ is the realized fitness increment of site $g$, equal to the total reproductive value of $g$'s offspring minus $g$'s own reproductive value, and $\mathcal{S}_g$ is a user-defined set of subsets thought to play a causal role in $g$'s fitness. The $c_{S,g}$ may then be estimated via least-squares regression. Terms that do not meet a significance threshold may be removed from $\mathcal{S}_g$, and hence also from Eq. 10. Symmetry (82) may be used to further reduce the number of terms in Eq. 23.

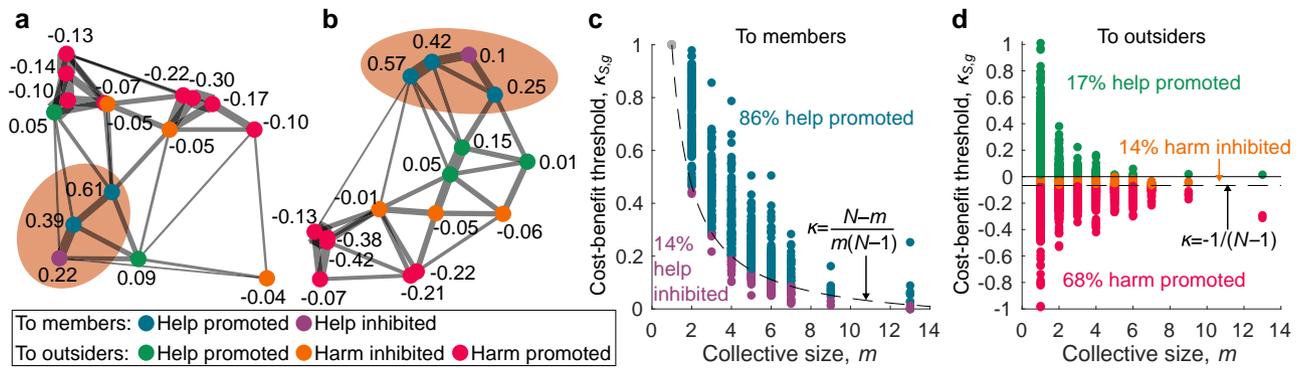

**Fig. 4.** Collective action on spatial networks. a–b) Delaunay networks (78) are a model of 2D spatial structure, formed by randomly placing points in a square and joining neighbors together. We identified network subcommunities using a spatial variant of the Girvan–Newman algorithm (79) (SI Appendix, Section 9.8.1). We then computed the cost–benefit thresholds $\kappa_{S,g}$, in the Collective Action Dilemma, from each subcommunity $S$ to each target node $g$. Collective action is favored if $b\kappa_{S,g} > c$; larger values indicate greater propensity for action (positive for help, negative for harm). These are compared to the corresponding well-mixed thresholds, $\kappa = (N - m)/(m(N - 1))$ for members and $\kappa = -1/(N - 1)$ for outsiders, to determine the effects of network structure. c) We generated 50 Delaunay networks of size 16, comprising 298 subcommunities. Spatial structure promotes help, in the sense $\kappa_{S,g} > (N - m)/(m(N - 1))$, to 86% of internal targets. The remaining 14% tend to be further inside the collective, away from the boundary. (Percentages exclude "collectives" of size one, which necessarily have $\kappa_{S,g} = 1$ to their one member.) d) Spatial structure promotes help ($\kappa_{S,g} > 0$) to 17% of external targets, of which 94% of which are neighbors of the collective and the rest are two-step neighbors. Harm is promoted ($\kappa_{S,g} < -1/(N - 1)$) to the majority of external targets. Selection for costly help or harm to outsiders decreases with collective size.

Fourth, suppose that alleles affect phenotypes by only a small amount, $\epsilon$. This is a stronger assumption than weak selection in the sense used here (62). In this case, the synergistic effects $c_{S,g}$ of a collective $S$ of size $m$ are of order $\epsilon^m$. The phenotypic condition, Eq. 13, can then be approximated to any order $k$ in $\epsilon$ by disregarding collectives larger than size $k$ (SI Appendix, Section 5.6). In particular, to first order in $\epsilon$, only singleton "collectives" contribute, and Eq. 13 reduces to the approximate condition

$$\sum_{j \in I} \sum_{i \in I} c_{ji} r_{ji} > 0. \tag{24}$$

In this approximation, the inner sum, $\sum_{i \in I} c_{ji} r_{ji} > 0$, can be understood as the inclusive fitness effect of individual $j$ (10, 22). Thus, if phenotypic differences are very small, Eq. 10 reduces to the condition that the total inclusive fitness effect of all individuals is positive. The larger the phenotypic differences, however, the more significantly larger collectives contribute to selection in Eq. 10.

## Discussion

Our work provides mathematical and conceptual tools for understanding natural selection with nonlinear social interactions. In particular, Eq. 10 shows how selection for social behavior depends on genetic assortment (quantified by $r_{S,g}$) and synergy (quantified by $c_{S,g}$). Evaluating this condition—by means of Eqs. 6 and 7—allows for systematic analysis of weak selection in models of nonlinear social behavior.

Our results apply to a broad class of models, with haploid or diploid genetics, and any fixed spatial structure, mating pattern, and/or fitness-affecting interactions. In contrast to approaches that analyze change from a given population state (26, 27, 48, 64), Eq. 10 applies to the overall selection process, from mutant appearance to fixation.

Many social behaviors of interest—including collective actions such as biofilm formation (8)—alter the population structure itself. It will therefore be important to extend Eq. 10 to populations with dynamic structure. A promising first step is provided by Su et al. (83), who generalize the formalism of sites and parentage maps to dynamic structures.

## Synergy and collective actors

The evolution of social behavior is most often studied at the level of individual actors. Inclusive fitness theory, in particular, adopts an actor-centric perspective by quantifying selection in terms of an individual's effects on the fitness of itself and others (10, 25, 84). This actor-centric perspective can be useful in guiding intuition (25, 84, 85). However, in nonlinear interactions, fitness effects do not cleanly separate into distinct contributions from individual actors (25, 84–87). While it is possible to assign fitness effects to individuals using linear regression (88, 89), the resulting regression coefficients have limited interpretive value (85–87).

Our approach resolves this difficulty by extending the notion of "actor" to collectives. Any synergistic effect, $c_{S,g}$, arising whenever an allele is shared among set $S$, is ascribed to $S$ collectively rather than to its members.

Nonlinear social interactions generally involve overlapping collectives at multiple scales (Fig. 5). The dependence of a site $g$'s fitness on the state **x**, in Eq. 8, may involve linear terms, $c_{\{h\},g} x_h$, ascribed to individual sites, quadratic terms, $c_{\{h,j\},g} x_h x_j$, ascribed to pairs, cubic terms ascribed to triples, and so on. Similarly, a synergistic or conditional interaction between two collectives, $S$ and $T$, gives rise to fitness effects, $c_{S \cup T,g}$, ascribed to the joint collective, $S \cup T$.

## Collective inclusive fitness

The overall contribution of a collective $S$ to selection can be quantified by its "collective inclusive fitness effect" $w_S^{\text{IF}} = \sum_{g \in G} c_{S,g} r_{S,g}$. As in standard inclusive fitness theory (10), this effect can be decomposed into a term, $r_S \sum_{g \in S} c_{S,g}$, accounting for $S$'s interest in its own members' fitness, plus terms $c_{S,g} r_{S,g}$ accounting for $S$'s interest in the fitness each external target $g$. Selection favors an allele if the total inclusive fitness effect of all collectives is positive, $\sum_{S \subseteq G} w_S^{IF} > 0$.

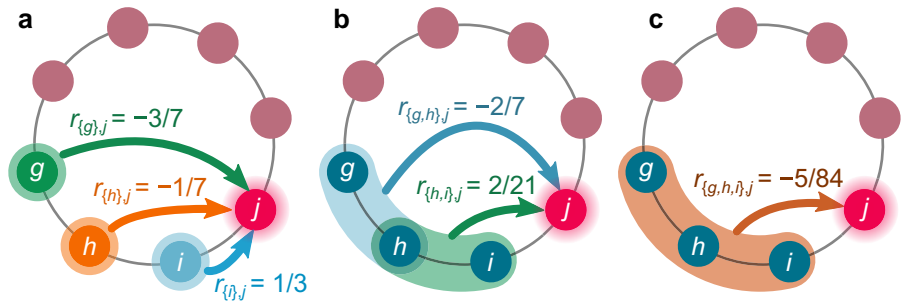

**Fig. 5.** Conflicting inclusive fitness interests within a collective. Suppose the fitness of a site $j$ depends on the alleles in three others, $g$, $h$, and $i$. Eq. 10 then has three terms arising from single sites, $c_{\{g\},j} r_{\{g\},j} + c_{\{h\},j} r_{\{h\},j} + c_{\{i\},j} r_{\{i\},j}$, three from pairwise synergistic effects, $c_{\{g,h\},j} r_{\{g,h\},j} + c_{\{g,i\},j} r_{\{g,i\},j} + c_{\{h,i\},j} r_{\{h,i\},j}$, and one, $c_{\{g,h,i\},j} r_{\{g,h,i\},j}$, from all three together. a) If $g$, $h$, $i$, and $j$ are consecutive nodes on a cycle of size 8, then $i$ is positively related to $j$, indicating potential for helpful behavior, while $g$ and $h$ are negatively related to $j$, indicating potential for harm. b) The collective $\{h, i\}$ is positively related to $j$, while $\{g, h\}$ is negatively related to $j$. c) Together, $\{g, h, i\}$ are negatively related to $j$. The outcome of selection reflects an aggregation over these individual and collective interests.

In some circumstances, it has been shown that selection leads individuals to act as if maximizing inclusive fitness (90, 91). Does this principle apply to collective actors? The answer is no in general, for two reasons. First, with multiple alleles, there is no guarantee that selection will favor one over all others. Rather, nontransitive competition (92, 93) may lead to evolutionary cycles or chaos (94), with no quantity maximized. Second, even if an allele is favored over all others, it does not necessarily follow that this allele causes any collective—let alone all of them—to act as if maximizing inclusive fitness. This is because Eq. 10, in the form $\sum_{S \subseteq G} w_S^{IF} > 0$, aggregates over the inclusive fitness interests of all collectives. When these interests diverge—as in conflict over worker reproduction in ant colonies (95)—selection averages over these interests, without leading to maximizing behavior for any of them (Fig. 5).

There is one case where selection leads to maximizing behavior. If a mutant allele affects the actions of only a single class of collective (e.g. three consecutive nodes on a cycle), this allele is favored if $w_S^{IF} > 0$, where $S$ can be any representative of this class (SI Appendix, Theorem 10.1). Over many such mutations, with the actions of all other collectives held fixed, selection would lead collectives in this class toward maximizing behavior. However, this result requires the unrealistic assumption that mutations affect only one class of collectives while leaving fixed the behavior of all others—including these collectives' members and subsets. Without this assumption, Eq. 10 implies that selection leads not to maximizing behavior, but to conflict and compromise over competing individual and collective prerogatives.

### Collective adaptation

Although collective maximizing behavior is selected only in unrealistic scenarios, our results highlight a route to collective adaptation (96, 97) in a more flexible sense. The greater the collective intrarelatedness, $r_S$, and capacity for synergistic fitness effects, $c_{S,g}$, the more a set $S$ is predicted to evolve collective behaviors that align with its inclusive fitness interests. We see this principle at work in spatial networks (Fig. 4), wherein local subcommunities —especially smaller ones—can evolve collective behaviors that benefit the group and its immediate neighbors, even at cost to individual members.

An even stronger capacity for collective adaptation is predicted in reproductively isolated groups, underscoring a key finding of multilevel selection theory (13–16, 96–98). Such groups have high intrarelatedness, $r_S$, and many possibilities for synergistic behavior (37, 41, 96). In our framework, collective adaptation does not require competition between groups; on the contrary, cooperation can be selected between groups that are closely related (high $r_{S,g}$ for neighbors $g$ of $S$; see Fig. 4d). What is required instead is synergistic fitness effects, i.e. nonzero $c_{S,g}$ for the group $S$ in question. If, in contrast, all fitness effects are linear ($c_{T,g} = 0$ for all non-singleton sets $T$), then inclusive fitness effects vanish for all nonsingleton collectives, and any intragroup cooperation is explainable in terms of individual-level adaptations.

### Collective individuality

By conceptualizing collectives as actors on par with individuals, our framework may be useful in understanding the "paradox of the organism" (99, 100)—that organisms persist as integrated adaptive units despite the potential for intraorganismal conflict. Considering multicellular organisms as collectives of cells, Eq. 10 enables simultaneous analysis of selection pressures at organismal and suborganismal levels. Viewed in this light, the origin of multicellularity (101, 102) and other transitions in individuality (103, 104) may be understood as the emergence of radical new forms of collective action. In this sense, all action is collective action.

### Acknowledgments

The authors are grateful to J. Arvid Ågren, Alex McAvoy, Jorge Peña, Joshua Plotkin, and Qi Su for feedback and discussions, to Christoph Hauert and anonymous referees for insightful comments, and to Julia Shapiro for help with figure design.

### Supplementary Material

Supplementary material is available at *PNAS Nexus* online.

### Funding

This project was supported by Grant 62220 from the John Templeton Foundation. S.R.H. was supported by the Clare Boothe Luce Program of the Henry Luce Foundation.

### Author Contributions

B.A. conceived the project. B.A., A.-R.K., J.L.D., T.J.K., S.R.H., J.P., Y.A.D., and C.S. analyzed the model. B.A. and C.L. designed the figures. B.A., Y.A.D., and C.S. supervised student researchers. B.A. wrote the manuscript.

## Preprints

A preprint of this manuscript was posted at https://arxiv.org/abs/2302.14700.

## Data Availability

Full results for all Delaunay and sociable weaver networks we analyzed are available on Zenodo at https://zenodo.org/doi/10.5281/zenodo.10866984. We made use of publicly available datasets from van Dijk et al. (105) (https://datadryad.org/stash/dataset/doi:10.5061/dryad.c0r18). Coalescence lengths and cost–benefit thresholds on networks were computed using MATLAB (version R2022a). MATLAB Code to compute coalescence lengths and cost–benefit thresholds on networks is available at https://github.com/Emmanuel-Math-Bio-Research-Group/Collective-Action and on Zenodo at https://zenodo.org/doi/10.5281/zenodo.10866984.

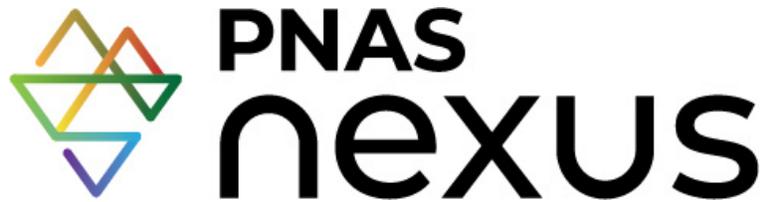

# Supporting Information for

## Nonlinear social evolution and the emergence of collective action

**Benjamin Allen, Abdur-Rahman Khwaja, James L. Donahue, Theodore J. Kelly, Sasha R. Hyacinthe, Jacob Proulx, Cassidy Lattanzio, Yulia A. Dementieva, Christine Sample**

**Benjamin Allen**
**E-mail: allenb@emmanuel.edu**

**This PDF file includes:**

Supporting text
Figs. S1 to S2
Table S1
SI References

## Supporting Information Text

## Contents

This Supplementary Information contains a mathematical description of our modeling framework, proofs of our main results, computation of examples, and comparisons to previous work. The derivations are mostly self-contained, with a few reliances on results proven in previous works [1–6]. We assume familiarity with discrete mathematics [7] (sets, relations, functions, etc.), multivariable calculus, and probability, especially finite Markov chains [8]. Readers may also refer to Refs. [2, 3] for a more pedagogical exposition of the kind of mathematical framework used here.

Section 1 presents our mathematical modeling framework. Sections 2 and 3 prove key mathematical results, culminating in a proof of equivalence between various criteria for selection (Theorems 3.3 and 3.4). Section 4 presents our definition of collective relatedness, and its relationship to other assortment measures such as coalescence length and identity-by-descent. Our main result—Eq. (10) of the main text—is proven in Section 5, Theorems 5.1 and 5.2. Section 6 introduces the Collective Action Dilemma, which is then applied to well-mixed haploid populations (Section 7), diploid siblings (Section 8), and network-structured populations (Section 9). Finally, we turn to the topics of inclusive fitness maximization (Section 10) and empirical estimation of fitness effects (Section 11).

Our notation is catalogued in Table S1. We also use standard set-theoretic notation, e.g. $\in$ for element, $\subseteq$ for subset, and $\varnothing$ for empty set. The cardinality (size) of a finite set $S$ is denoted $|S|$. For two sets $S$ and $T$, we let $S^T$ denote the set of all $T$-indexed tuples of the form $(s_t)_{t \in T}$, with each $s_t \in S$. For a set mapping $f : S \to T$, we denote the image of a subset $S' \subseteq S$ by $f(S') \subseteq T$, and the preimage of a subset $T' \subseteq T$ by $f^{-1}(T') \subseteq S$. We use the shorthand $f^{-1}(t) = f^{-1}(\{t\})$ for the preimage of a singleton subset $\{t\} \subseteq T$.

## 1. Modeling framework

We start by introducing the modeling framework from which our results are derived. The framework used here is a variant of others developed in earlier works [2–5]. It defines a class of models of natural selection, in populations of fixed size and spatial structure, with haploid, diploid, or other genetics. The relationship to other general frameworks for natural selection is discussed in Section 1.9.

**Table S1. Glossary of notation**

| Symbol | Description | Section |
|---|---|---|
| $\alpha$ | Parentage mapping | 1.3 |
| $c_{S,g}$ | Synergistic effect of set $S$ on site $g$'s fitness | 5.1 |
| $\delta$ | Selection intensity parameter | 1.6 |
| $\Delta(\mathbf{x})$ | Selection increment in state $\mathbf{x}$ | 3.3 |
| $G$ | Set of genetic sites | 1.1 |
| $\imath_S,\ \imath_S^A,\ \imath_S^a$ | Identity-by-state indicator functions | 4.1 |
| $I$ | Set of individuals | 1.7 |
| $\ell_S$ | Genetic dissimilarity/coalescence length of set $S$ | 4.2 |
| $\mathcal{M}$ | Selection Markov chain | 1.5 |
| $\mathcal{M}_0$ | Selection Markov chain with $u=0$ | 2.1 |
| $\mu$ | Mutant appearance distribution | 2.2.2 |
| $n$ | Number of genetic sites | 1.1 |
| $N$ | Number of individuals | 1.1 |
| $\nu_g,\ \nu_S$ | Mutation rate at site $g$ or set $S$, respectively | 2.2.2 |
| $p_{\mathbf{x}}(\alpha, U)$ | Probability of transition $(\alpha, U)$ in state $\mathbf{x}$ | 1.3 |
| $\pi_{\mathcal{M}}$ | Stationary probability distribution of $\mathcal{M}$ | 2.1 |
| $\varphi_i$ | Probability that individual $i$ has phenotype 1 | 1.7 |
| $\Phi_i$ | Phenotype of individual $i$ | 1.7 |
| $\mathbf{\Phi}$ | Phenotypic population state | 1.7 |
| $r_{S,g}$ | Collective relatedness of set $S$ to site $g$ | 4.3 |
| $\rho_A,\ \rho_a$ | Fixation probability of allele $A$ or $a$, respectively | 2.3 |
| $u$ | Mutation parameter | 1.6 |
| $U$ | Set of mutated sites | 1.3 |
| $v_g$ | Reproductive value (RV) of site $g$ | 3.1 |
| $w_g(\mathbf{x})$ | Fitness increment of site $g$ in state $\mathbf{x}$ | 3.2 |
| $x_g$ | Allele in site $g$ (1 for $A$, 0 for $a$) | 1.2 |
| $\mathbf{x}$ | Population state | 1.2 |
| $\langle\ \rangle$ | Low-mutation operator | 2.2.3 |
| $\circ$ | Indicates neutral drift ($\delta = 0$) | 3.1 |
| $'$ | Indicates $\delta$-derivative at $\delta = 0$ | 3.5 |
| $^-$ | unweighted average over sites | 3.1 |
| $\hat{\ }$ | RV-weighted average over sites | 3.1 |
| $\smile$ | Indicates dependence on phenotypic state $\mathbf{\Phi}$ | 1.7 |

**1.1. Genetic sites and individuals.** Adopting a gene's-eye view, we consider a population of alleles at a particular genetic locus. There is a set $G$ of $n$ genetic sites, at which alleles can live. Each site houses a single allele.

Individuals are represented by a fixed set $I$ of size $N$. Each individual $i \in I$ contains a set $G_i \subseteq G$ of sites. The number of sites, $n_i = |G_i|$, corresponds to $i$'s ploidy (e.g., $n_i = 1$ if $i$ is haploid, $n_i = 2$ if diploid). Although we focus on haploids and diploids, our formalism allows for any ploidy. The sets $G_i$ form a partition of the overall set $G$ of sites, so that each genetic site resides in exactly one individual.

**1.2. Alleles and states.** There are two competing allele types, $A$ and $a$. The allele type occupying each site $g \in G$ is indicated by a binary variable $x_g \in \{0, 1\}$, with 0 corresponding to $a$ and 1 corresponding to $A$. The overall population state is the binary vector $\mathbf{x} = (x_g)_{g \in G}$. The set of all possible states is $\{0, 1\}^G$.

**1.3. Transitions.** A transition from one state to another involves two components. The first is a set mapping $\alpha : G \to G$, indicating the site from which each allele in the new state is inherited (in the case of new offspring) or retained (if the allele survives). Thus, $\alpha(g) = h$ indicates that the new occupant of site $g$ is the same as, or a copy of, the allele that occupied site $h$ in the previous time-step. Our framework does not formally distinguish between these two possibilities, as they both result in allele transmission from $g$ to $h$. We call $\alpha$ the *parentage map*, understanding a surviving allele to be its own "parent".

The second component is the subset $U \subseteq G$ of sites that undergo mutation during the transition. In general, $U$ can be any subset of $G$, including the empty set $\varnothing$ (indicating that no mutations occur).

The probability that parentage map $\alpha$ and mutation set $U$ occur in state $\mathbf{x}$ is denoted $p_\mathbf{x}(\alpha, U)$. For each state $\mathbf{x}$, $\{p_\mathbf{x}(\alpha, U)\}_{(\alpha, U)}$ comprises a joint probability distribution on the set of possible combinations of $\alpha$ and $U$.

We allow the probabilities $p_\mathbf{x}(\alpha, U)$ to depend on the state $\mathbf{x}$ in arbitrary fashion, subject only to a Fixation Axiom introduced in the next subsection. In this way, our framework encompasses a wide variety of models of selection. Spatial, network, or group structure, behavioral interactions, migration, mate choice, and allele transmission (Mendelian or not) are represented implicitly in the probability distributions $p_\mathbf{x}$.

We denote the marginal probabilities of a parentage map $\alpha$ and a mutation set $U$, respectively, by

$$p_\mathbf{x}(\alpha) = \sum_{U \subseteq G} p_\mathbf{x}(\alpha, U) \quad \text{and} \quad p_\mathbf{x}(U) = \sum_{\alpha : G \to G} p_\mathbf{x}(\alpha, U). \tag{1.1}$$

We also use the notation $\mathbb{P}_\mathbf{x}$ and $\mathbb{E}_\mathbf{x}$ for probabilities and expectations, respectively, under the distribution $\{p_\mathbf{x}(\alpha, U)\}_{(\alpha, U)}$ for a given state $\mathbf{x}$. For example, the expected number of (self + offspring) of site $g \in G$, in a single transition from state $\mathbf{x}$, can be written as $\mathbb{E}_\mathbf{x}[|\alpha^{-1}(g)|]$.

**1.4. Fixation Axiom.** For the population to evolve as a unit, it should be possible (with positive probability) for at least one genetic site $g$ to spread its progeny throughout the population. We formalize this as an axiom:

**Fixation Axiom.** There exists a site $g \in G$ such that, for each $h \in G$, there is a finite sequence of parentage maps $\alpha_1, \ldots, \alpha_m$, with $p_\mathbf{x}(\alpha_k) > 0$ for each $k = 1, \ldots, m$ and $\mathbf{x} \in \{0, 1\}^G$, and $\alpha_1 \circ \cdots \circ \alpha_m(h) = g$.

**1.5. Selection Markov chain.** This framework defines a class of models of natural selection. A particular model within this framework is defined by specifying a set $G$ of sites and a collection of probability distributions $\{p_\mathbf{x}(\alpha, U)\}_{(\alpha, U)}$, for each state $\mathbf{x} \in \{0, 1\}^G$, satisfying the Fixation Axiom.

Given these ingredients, the selection process is represented as a Markov chain $\mathcal{M} = (\mathbf{X}^t)_{t=0}^\infty$ on the set of states $\{0, 1\}^G$. We call this the *selection Markov chain*. Given an initial state $\mathbf{X}^0$, each subsequent state $\mathbf{X}^t$ is constructed from the previous state, $\mathbf{X}^{t-1}$, as follows: First, a parentage map $\alpha^t$ and mutation set $U^t$

**6 of 72**                                                                   **Allen et al.**

are sampled from the probability distribution $p_{\mathbf{X}^{t-1}}$. Then, each non-mutated site $g \notin U^t$ inherits the allele type of the parent, $X_g^t = X_{\alpha^t(g)}^{t-1}$, and each mutated site $g \in U^t$ inherits the *opposite* allele from the parent, $X_g^t = 1 - X_{\alpha^t(g)}^{t-1}$. Putting these cases together, we have

$$X_g^t = \begin{cases} X_{\alpha^t(g)}^{t-1} & \text{if } g \notin U^t \\ 1 - X_{\alpha^t(g)}^{t-1} & \text{if } g \in U^t. \end{cases} \quad (1.2)$$

In particular, if there is no mutation ($U^t = \varnothing$), then each site inherits its parent's allele, and we can write Eq. (1.2) compactly as

$$\mathbf{X}^t = \mathbf{X}_{\alpha^t}^{t-1}. \quad (1.3)$$

Above, we have introduced a new notation: for any state $\mathbf{x} \in \{0,1\}^G$ and set mapping $\tau : G \to G$, $\mathbf{x}_\tau$ denotes the state with allele $x_{\tau(g)}$ in each site $g \in G$.

We denote the transition probability from state $\mathbf{x}$ to state $\mathbf{y}$ in the selection Markov chain $\mathcal{M}$ by $P_{\mathbf{x} \to \mathbf{y}}$.

**1.6. Mutation and selection parameters.** To vary the effects of mutation and selection, we allow the probabilities $p_{\mathbf{x}}(\alpha, U)$ to additionally depend on two parameters: a mutation parameter $u$ with $0 \leq u \leq 1$, and a selection intensity parameter $\delta$ with $0 \leq \delta \leq \epsilon$ for some $\epsilon > 0$. The case $\delta = 0$ represents neutral drift, while $u = 0$ means mutation does not occur.

In addition to requiring that the Fixation Axiom be satisfied for all combinations of $u$ and $\delta$, we impose the following additional assumptions on the behavior of the probability distributions $p_{\mathbf{x}}$ with respect to these parameters. First, the probabilities of replacement and mutation should vary smoothly, up to second order, with respect to the mutation and selection parameters:

(D1) Each probability $p_{\mathbf{x}}(\alpha, U)$ is jointly twice-differentiable, with continuous second partial derivatives, in both $u$ and $\delta$.

Second, under neutral drift ($\delta = 0$), alleles $A$ and $a$ should be interchangeable, and should not affect probabilities of replacement or mutation:

(N1) For $\delta = 0$, the probabilities $p_{\mathbf{x}}(\alpha, U)$ are independent of the population state $\mathbf{x}$.

Third, no mutation should occur in the case $u = 0$:

(M1) For $u = 0$, $p_{\mathbf{x}}(U) = 0$ for all $U \neq \varnothing$.

Fourth, for $u > 0$, it should be possible for new mutations to arise and sweep to fixation:

(M2) For $u > 0$, there exists some $g \in G$, satisfying the conditions of the Fixation axiom, such that $\mathbb{P}_{\mathbf{x}}[g \in U] > 0$.

Fifth, when the mutation parameter is small, the probability of two or more mutations should be much less than the probability of one:

(M3) For each $\mathbf{x} \in \{0,1\}^G$ and each fixed $\delta \geq 0$, $\lim_{u \to 0} \frac{\mathbb{P}_{\mathbf{x}}[|U| \geq 2]}{u} = 0$.

Sixth, since $u$ parameterizes mutation only, the marginal probability of each parentage map in a given state should be independent of $u$:

(M4) For each $\alpha : G \to G$ and each $\mathbf{x} \in \{0,1\}^G$, $p_{\mathbf{x}}(\alpha)$ is independent of $u$.

Seventh, for $u < 1$, it should be possible for no mutation to occur:

(M5) For $u < 1$ and all $\mathbf{x} \in \{0, 1\}$, $\mathbb{P}_\mathbf{x}[U = \varnothing] > 0$.

Finally, in order to isolate the effects of selection, we assume that probabilities of mutation are the same in the two monoallelic states:

(M6) For each $U \subseteq G$ and $u \geq 0$, $p_\mathbf{a}(U) = p_\mathbf{A}(U)$.

Assumption (M6) removes the possibility that mutation can favor one trait over another; thus any differences in the frequency of $A$ versus $a$ must be due to selection alone. Without (M6), the effects of mutation and selection are difficult to disentangle [9].

Many of our results pertain to weak selection, i.e., small positive values of the selection parameter $\delta$. By assumptions (N1) and (D1), replacement and mutation probabilities under weak selection are close their corresponding neutral drift values. This formulation of weak selection does not imply that the phenotypic effects of alleles $A$ and $a$ are small, but that these phenotypic effects only weakly affect replacement and mutation. This subtle distinction has important consequences for modeling social behavior; see Wild and Traulsen [10] for discussion. Small phenotypic effects represent a special case of weak selection, and we analyze this case in Section 5.6.

**1.7. Individual phenotypes.** Although our analysis is mostly conducted at the gene level, we will also employ a notion of individual phenotypes. We suppose there are two possible phenotypes, numbered 1 and 0. The phenotype $\Phi_i \in \{0, 1\}$ of each individual $i \in I$ depends stochastically on the binary vector $\mathbf{x}_{|G_i} = (x_g)_{g \in G_i}$ of $i$'s alleles. The probability that individual $i \in I$ has phenotype 1 is denoted $\varphi_i(\mathbf{x}_{|G_i}) = \mathbb{P}_\mathbf{x}[\Phi_i = 1]$. We require that individuals with all $A$ alleles always have phenotype 1, $\varphi_i(1, \ldots, 1) = 1$, and those with all $a$ alleles always have phenotype 0, $\varphi_i(0, \ldots, 0) = 0$. Each individual's phenotype is probabilistically independent of all others', conditional on the (genotypic) population state $\mathbf{x}$.

For haploid individuals, the phenotypic formalism is equivalent to the genotypic one: Those with allele $A$ always have phenotype 1, $\varphi_i(1) = 1$, and those with allele $a$ always have phenotype 0, $\varphi_i(0) = 0$.

For a diploids, the probability that individual $i \in I$, with genetic sites $G_i = \{i_1, i_2\}$, has phenotype 1 can be written

$$\varphi_i(x_{i_1}, x_{i_2}) = h(x_{i_1} + x_{i_2}) + (1 - 2h)x_{i_1}x_{i_2}. \tag{1.4}$$

Above, $h$ is a genetic dominance parameter, $0 \leq h \leq 1$, giving the probability that an $Aa$ heterozygote has phenotype 1. In particular, $h = 0$ if allele $A$ is recessive, $h = 1$ if $A$ is dominant, and $h = 1/2$ in the case of no dominance. (We exclude the possibility of under- or over-dominance.) As required, $AA$ individuals have phenotype 1, $\varphi_i(1, 1) = 1$; $aa$'s have phenotype 0, $\varphi_i(0, 0) = 0$; and $Aa$'s have phenotype 1 with probability $h$, $\varphi_i(1, 0) = \varphi_i(0, 1) = h$.

We collect the phenotypes of all individuals into a binary vector $\mathbf{\Phi} = (\Phi_i)_{i \in I} \in \{0, 1\}^I$, representing the phenotypic state of the population. The phenotypic state $\mathbf{\Phi}$ depends stochastically on the genotypic state $\mathbf{x}$, according to the probabilities $\varphi_i(\mathbf{x}_{|G_i})$ for each $i \in I$.

The probability that transition event $(\alpha, U)$ occurs in phenotypic state $\mathbf{\Phi}$ is denoted $\breve{p}_\mathbf{\Phi}(\alpha, U)$. Here and throughout, the accent $\breve{\phantom{x}}$ is used to indicate quantities that depend on the phenotypic state $\mathbf{\Phi} \in \{0, 1\}^I$ rather than on the genotypic state $\mathbf{x} \in \{0, 1\}^G$.

This phenotype-level formalism is a special case of the gene-level framework introduced in Sections 1.1–1.5. The gene-level framework can be recovered from the phenotypic one by setting $p_\mathbf{x}(\alpha, U) = \mathbb{E}_\mathbf{x}[\breve{p}_\mathbf{\Phi}(\alpha, U)]$, with expectation taken over the probability distribution on $\mathbf{\Phi}$ in state $\mathbf{x}$.

**1.8. Symmetry and class structure.** Many models of selection possess some form of symmetry. For example, in classical models of well-mixed populations, individuals are considered interchangeable. Then, instead of the full state $\mathbf{x}$, one need only record the frequency of each genotype. Other models may classify individuals by age, sex, or other factors, but consider any two individuals in the same class to be interchangeable.

Following Allen [6], we formally define a symmetry as a permutation of genetic sites that preserves probabilities of parentage and mutation:

**Definition.** A *symmetry* is a permutation $\sigma$ of $G$ (a one-to-one mapping from $G$ to itself) that satisfies

$$p_{\mathbf{x}}(\alpha, U) = p_{\mathbf{x}_\sigma}(\sigma^{-1} \circ \alpha \circ \sigma, \sigma^{-1}(U)), \tag{1.5}$$

for each state $\mathbf{x} \in \{0,1\}^G$, parentage mapping $\alpha : G \to G$, and mutation subset $U \subseteq G$.

This definition asserts that $\sigma$ is a symmetry if and only if the probabilities of parentage and mutation are preserved when each site $g$ is relabeled as $\sigma(g)$. (Recall from Section 1.5 that $\mathbf{x}_\sigma$ is the state that has allele $x_{\sigma(g)}$ in each site $g$.) To understand Eq. (1.5), suppose that for some sites $g, h, k \in G$, events $\alpha(g) = h$ and $k \in U$ occur in state $\mathbf{x}$. The equivalent transition in state $\mathbf{x}_\sigma$ has $\alpha(\sigma(g)) = \sigma(h)$ and $\sigma(k) \in U$, or equivalently, $\sigma^{-1} \circ \alpha \circ \sigma(g) = h$ and $k \in \sigma^{-1}(U)$. Therefore, the probability that parentage map $\alpha$ and mutation set $U$ occur in state $\mathbf{x}$ must equal the probability that parentage map $\sigma^{-1} \circ \alpha \circ \sigma$ and mutation set $\sigma^{-1}(U)$ occur in state $\mathbf{x}_\sigma$.

We illustrate this definition of symmetry with some examples:

- In a diploid individual, the two alleles at a given locus are usually assumed to be interchangeable (barring sex-linkage and related phenomena). Accordingly, if $G_i = \{i_1, i_2\}$ is the set of genetic sites in individual $i \in I$, then a permutation $\sigma$ that interchanges these sites (that is, $\sigma(i_1) = i_2$ and $\sigma(i_2) = i_1$) while leaving other sites fixed ($\sigma(g) = g$ for $g \notin G_i$) is a symmetry.

- Consider a two-sex population model, in which individuals of the same sex are considered to be interchangeable. Then a permutation that interchanges the sites corresponding to two male individuals, or two female individuals, is a symmetry.

- For the cycle-structured population in Figure 2b of the main text, any rotation or reflection of the cycle is a symmetry.

- For the Windmill-structured in Figure 2c of the main text, a permutation that interchanges two "blades", or the two nodes on a single blade, is a symmetry.

The set of all symmetries, denoted $\mathrm{Sym}(G, p)$, forms a group [6] in the mathematical sense of group theory [11]. This means that for any symmetry $\sigma$, the inverse permutation $\sigma^{-1}$ is also a symmetry, and for any two symmetries $\sigma$ and $\tau$, the composition $\sigma \circ \tau$ is also a symmetry.

If two sites $g, h \in G$ are related by $\sigma(g) = h$ for some symmetry $\sigma \in \mathrm{Sym}(G, p)$, then $g$ and $h$ are equivalent with respect to the selection process. Using this notion of equivalence, the set of sites $G$ can be partitioned into *classes*, such that any two sites $g, h$ in the same class are equivalent. These are classes in the mathematical sense of equivalence class [7], as well as in the sense of a class-structured population [12, 13].

In a homogeneous population, all sites belong to a single class; an example is the cycle-structured population in Fig. 2b of the main text. Heterogeneous populations have multiple classes; for example, the two-sex model described above has one class for each sex, and the windmill-structured population (Fig. 2c of the main text) has one class for the hub and one for the blade nodes.

Symmetries preserve transition probabilities in the selection Markov chain $\mathcal{M}$, in the sense that

$$P_{\mathbf{x} \to \mathbf{y}} = P_{\mathbf{x}_\sigma \to \mathbf{y}_\sigma}, \tag{1.6}$$

for any states $\mathbf{x}, \mathbf{y} \in \{0,1\}^G$ and symmetry $\sigma \in \mathrm{Sym}(G, p)$, according to Theorem 3.2 of Allen [6].

**1.9. Relationship to other frameworks for natural selection.** The framework described above encompasses a general class of models of natural selection. It allows for arbitrary forms of population structure (islands, lattices, networks), with no *a priori* assumption of homogeneity or symmetry. Haploid, diploid, haplodiploid, or even polyploid genetics are allowed, with arbitrary mating pattern. Generations may be discrete or overlapping—indeed, any number of individuals may be replaced in a given time-step. The probabilities of transition events may depend in arbitrary fashion on the population state—subject only to the Fixation Axiom—allowing for complex nonlinear social behaviors.

However, as in any modeling framework, there are limiting assumptions. Selection is restricted to a single biallelic locus, and the population size and structure are fixed. We also assume a constant environment by supposing that the probabilities of transition events depend only on the population state **x**, and not on any environmental variables.

Below, we summarize the relationship of our framework other general frameworks for modeling natural selection, including previous formulations of the present framework.

*1.9.1. Allen & McAvoy (2019).* The framework introduced here is a variant of ones developed in previous works [2–5], in particular that of Allen and McAvoy [3]. However, there are two key differences to highlight.

First, we do not explicitly record births and deaths. That is, we make no formal distinction between an allele surviving into the next generation versus being replaced by a copy of itself. This simplifies notation, and also allows for surviving alleles to move between sites (e.g. via migration of adult individuals), which was not allowed in previous formulations.

Second, we do not specify any particular relationship between reproduction and mutation. Previous formulations assumed that offspring alleles acquire mutations independently with some fixed probability. By instead allowing for an arbitrary joint probability distribution on the set $U$ of mutated sites and the parentage map $\alpha$, we allow for mutation of surviving alleles, mutation rates that vary across sites, and non-independence of mutation events in a particular state.

Despite these differences, some results proven in previous works [2–6] carry over to the present framework with minimal modification.

*1.9.2. Kirkpatrick, Johnson, and Barton (2002).* Kirkpatrick et al. [14], building on earlier work of Barton and Turelli [15], developed a framework for representing expected genetic change due to selection. Although Kirkpatrick et al. [14] focus on multilocus genetic interactions within individuals, their methodology also applies to social interactions among individuals [16]. They employ a general representation of population structure, in which the probabilities of mating, replacement, migration, and mutation may depend in arbitrary fashion on the current population state. The trait(s) under selection may involve any number of genetic loci, with arbitrary epistasis.

The main result of Kirkpatrick et al. [14] is a system of recurrence equations for how the expected statistical moments of the next population state ($\mathbf{X}^{t+1}$, in our notation), depend on those of the current state, $\mathbf{X}^t$. For weak selection, they derive a quasi-linkage equilibrium approximation, which can be used to solve for these statistical moments in terms of the overall mutant allele frequency. These moments can then be used to compute the expected change in this mutant frequency, under weak selection, from a given population state.

Our approach differs from Kirkpatrick et al.'s [14] in the timescale of change considered, and in the methods used to compute this change. Whereas Kirkpatrick et al.'s [14] framework models expected change from a given population state, ours focuses on the overall selection process, as represented by the selection Markov chain $\mathcal{M} = (\mathbf{X}^t)_{t=0}^\infty$. Instead of focusing on mutant increase or decrease from a given level, we characterize success in terms of stationary allele frequencies, or equivalently, fixation probabilities (see Theorems 3.3 and 3.4 below). Additionally, instead of recurrence equations for statistical moments, we use coalescent theory [17, 18] to compute statistics of genetic assortment (see Section 4.2). However, we do not consider multilocus traits here, as Kirkpatrick et al.'s framework does.

*1.9.3. Wakano et al. (2013).* Wakano et al. [19] developed a framework to formalize inclusive fitness theory for a class of finite population models. Their framework comprises a special case of ours. As we do, they represent natural selection as a Markov chain on population states. Similarly as in our framework, states are represented by binary vectors indicating the allele in each individual, and transitions are represented by a parentage mapping. Relative to our framework, Wakano et al. require the following additional assumptions: (i) individuals are haploid, (ii) the population is homogeneous (in the sense described in our Section 1.8), (iii) generations are discrete (each individuals is replaced every time step), (iv) the parent of each individual is determined independently from others, and (v) the two alleles are very close in phenotype, in the sense we describe in Section 5.6. We recover their main result—Theorem 2 of Ref. [19]—as Eq. (5.38) below. Our framework generalizes that of Wakano et al. [19] to include arbitrary nonlinear interactions in heterogeneously structured populations, haploid or diploid.

**1.9.4. Ohtsuki (2014).** Ohtsuki [20] also introduced a framework for the evolution of multilateral social interactions, modeled as symmetric multiplayer games. This framework represents an infinitely large, homogeneous, haploid population with two competing alleles (corresponding to strategies in the game). In each state of the process, groups of $m$ players are formed, possibly with some assortment among kin. These groups interact according to a symmetric $m$-player game, and then each produces a number of offspring depending on their payoff. The next adult population is then uniformly sampled from among these offspring.

Ohtsuki's [20] framework assumes a symmetric population structure, such that any subset of $k \leq m$ players within a group have the same statistics of genetic assortment. This assortment is quantified by probabilities, $\theta_{k_1 \to k_2}$, that $k_1$ individuals in a group have $k_2$ distinct ancestors in the recent past, which are given *a priori*. The game is also symmetric, in that one's payoff depends only on one's own strategy and the number of each strategy in the rest of the group. In similar (but less general) fashion to the Kirkpatrick et al.'s [14] framework, Ohtsuki derives an expression for the expected change in allele frequency from a given state, under weak selection.

Compared to Ohtsuki's [20] framework, ours is significantly more flexible with regard to spatial structure, genetics, and mating system. Additionally, while Ohtsuki's framework models deterministic change in an infinite population, ours uses a stochastic (Markov chain) representation of the overall selection process in a finite population.

**1.9.5. van Cleve (2015).** Building on earlier work by Rousset and others [21–25], van Cleve [26] developed a framework to model natural selection between two alleles in a finite, homogeneous, haploid population. As in our framework, the selection process is represented as a Markov chain over population states, and success is quantified using fixation probabilities (or equivalently, stationary frequencies in the low-mutation limit). Also as in our framework, genetic assortment is computed via coalescent theory. While some of van Cleve's results—in particular, his Eq. (24)—pertain to arbitrary nonlinear interactions, his applications focus on two-player games and on alleles with small phenotypic effect (see Section 5.6). Our framework generalizes that of van Cleve [26] to allow for spatial heterogeneity, diploid genetics, and arbitrary mating patterns.

**1.9.6. Lehmann et al (2016).** Lehmann et al. [27] also introduced a framework for modeling natural selection, which has since been expanded in a number of directions [28–30]. This framework models selection between two alleles (mutant and resident) at a single genetic locus. The population is spread over infinitely many identical islands, with uniform dispersal between them. Within each island, the population size and environmental state may vary, and individuals may belong to different classes (sex, caste, etc.). Stochastic environmental and evolutionary dynamics on each island are represented by an arbitrary Markov chain. Selection is quantified in terms of the invasion fitness (asymptotic geometric growth rate when rare) of a mutant allele [27].

Compared to our framework, the Lehmann et al. [27] framework is less general with respect to overall population structure (requiring infinite identical islands), but more flexible in allowing for population and environmental fluctuations within each island.

## 2. Stationarity and fixation

In this section we establish the long-term behavior of the selection Markov chain $\mathcal{M}$, with particular emphasis on the low-mutation limit.

**2.1. Asymptotic behavior of the selection Markov chain.** The long-term behavior of the selection process depends on whether mutation is present or absent. For this reason, we denote the no-mutation ($u = 0$), case of the selection Markov chain by $\mathcal{M}_0$.

If mutation is present ($u > 0$), the selection Markov chain $\mathcal{M}$ converges to a unique stationary distribution. If not, the population is eventually taken over by one of the competing alleles. We formalize these observations in the following theorem:

**Theorem 2.1.** *$\mathcal{M}_0$ has absorbing states $\mathbf{a}$ and $\mathbf{A}$, and all other states are transient in $\mathcal{M}_0$. For $0 < u < 1$, $\mathcal{M}$ has a unique stationary distribution, $\pi_\mathcal{M}$, that satisfies*

$$\lim_{t \to \infty} \mathbb{P}[\mathbf{X}^t = \mathbf{x} | \mathbf{X}^0 = \mathbf{y}] = \pi_\mathcal{M}(\mathbf{x}), \tag{2.1}$$

*for any pair of states $\mathbf{x}, \mathbf{y} \in \{0,1\}^G$.*

*Proof.* The claim regarding $\mathcal{M}_0$ is Theorem 1 of Allen and Tarnita [2]. For the $0 < u < 1$ case, it suffices to show that $\mathcal{M}$ has a unique closed communicating class, and is aperiodic on this class. Consider an arbitrary initial state $\mathbf{X}^0 = \mathbf{y}$. We will show that states $\mathbf{A}$ and $\mathbf{a}$ are both accessible from $\mathbf{y}$, which will prove that $\mathcal{M}$ has a unique closed communicating class containing $\mathbf{A}$ and $\mathbf{a}$. Consider a site $g \in G$ and sequence $\alpha_1, \ldots, \alpha_k$ satisfying the properties specified in the Fixation Axiom and Assumption (M2). Suppose the transition events $(\alpha_1, \varnothing), \ldots, (\alpha_k, \varnothing)$ all occur in sequence from initial state $\mathbf{y}$; this sequence has positive probability by (M5) and the assumed properties of $g$. The resulting state is $\mathbf{a}$ or $\mathbf{A}$, respectively, if $y_g = 0$ or $y_g = 1$. Now suppose the next transition event has $g \in U$ (this has positive probability by Assumption (M2)), and that the sequence $(\alpha_1, \varnothing), \ldots, (\alpha_k, \varnothing)$ again follows afterwards. The resulting state is now either $\mathbf{A}$ or $\mathbf{a}$, respectively, if $y_g = 0$ or $y_g = 1$. In either case, both $\mathbf{A}$ and $\mathbf{a}$ are accessible from $\mathbf{y}$. Therefore, $\mathcal{M}$ has a unique closed communicating class.

To prove aperiodicity, we note that if no mutation ($U = \varnothing$) occurs in a transition from state $\mathbf{a}$ or $\mathbf{A}$, the state is unchanged. Since $\mathbb{P}_\mathbf{a}[U = \varnothing] > 0$ and $\mathbb{P}_\mathbf{A}[U = \varnothing] > 0$ by Assumption (M5), the transition probabilities $P_{\mathbf{a} \to \mathbf{a}}$ and $P_{\mathbf{A} \to \mathbf{A}}$ are positive. This proves that $\mathcal{M}$ is aperiodic. □

Stationary probabilities are preserved under symmetry (as defined in Section 1.8), in the sense that for any symmetry $\sigma \in \mathrm{Sym}(G, p)$, state $\mathbf{x} \in \{0,1\}^G$ and $u > 0$, [6]

$$\pi_\mathcal{M}(\mathbf{x}) = \pi_\mathcal{M}(\mathbf{x}_\sigma). \tag{2.2}$$

**2.2. The low-mutation limit.** The low-mutation limit, $u \to 0$, corresponds to biological settings in which the waiting time to mutation, in the relevant alleles, exceeds the time to fixation.

*2.2.1. Limiting stationary distribution.* The stationary distribution, $\pi_\mathcal{M}$, has a well-defined limit as $u \to 0$ [1, 2], which we denote by $\pi_{\mathcal{M}_0}$:

$$\pi_{\mathcal{M}_0}(\mathbf{x}) = \lim_{u \to 0} \pi_\mathcal{M}(\mathbf{x}). \tag{2.3}$$

This limiting distribution, $\pi_{\mathcal{M}_0}$, is nonzero only for the monoallelic states $\mathbf{A}$ and $\mathbf{a}$ [1, 2] (intuituively, for rare mutation, only one allele will typically be present). While $\pi_{\mathcal{M}_0}$ is a stationary distribution for the mutation-free Markov chain $\mathcal{M}_0$, it is not unique in this regard; indeed, any probability distribution supported only on the absorbing states $\mathbf{A}$ and $\mathbf{a}$ is stationary for $\mathcal{M}_0$.

*2.2.2. Appearance of mutations.* In light of the previous subsection, the mutations of relevance in the $u \to 0$ limit are those that arise in the monoallelic states $\mathbf{A}$ and $\mathbf{a}$. The rate at which mutations arise in these states may vary from site to site:

**Definition.** For each site $g \in G$, we define the *mutation rate $\nu_g$ at $g$* as

$$\nu_g = \left.\frac{d\,\mathbb{P}_\mathbf{A}[g \in U]}{du}\right|_{u=0} = \left.\frac{d\,\mathbb{P}_\mathbf{a}[g \in U]}{du}\right|_{u=0}. \tag{2.4a}$$

More generally, for any subset $S \subseteq G$, we define

$$\nu_S = \frac{d\,\mathbb{P}_{\mathbf{a}}[S \cap U \neq \varnothing]}{du}\bigg|_{u=0} = \frac{d\,\mathbb{P}_{\mathbf{A}}[S \cap U \neq \varnothing]}{du}\bigg|_{u=0}. \tag{2.4b}$$

The second equalities in Eqs. (2.4a)–(2.4b) are due to Assumption (M6). It follows from Assumption (M3) that $\nu_S = \sum_{g \in S} \nu_g$ for each $S \subseteq G$. Assumption (M3) also implies that the transition probability from $\mathbf{a}$ to any state $\mathbf{x}$, which we denote by $P_{\mathbf{a} \to \mathbf{x}}$, can be expanded under low mutation as

$$P_{\mathbf{a} \to \mathbf{x}} = \begin{cases} 1 - u\nu_G + \mathcal{O}(u^2) & \text{if } \mathbf{x} = \mathbf{a} \\ u\nu_g + \mathcal{O}(u^2) & \text{if } x_g = 1 \text{ and } x_h = 0 \text{ for all } h \neq g \\ \mathcal{O}(u^2) & \text{otherwise.} \end{cases} \tag{2.5a}$$

Similarly, for transitions from $\mathbf{A}$, we have

$$P_{\mathbf{A} \to \mathbf{x}} = \begin{cases} 1 - u\nu_G + \mathcal{O}(u^2) & \text{if } \mathbf{x} = \mathbf{A} \\ u\nu_g + \mathcal{O}(u^2) & \text{if } x_g = 0 \text{ and } x_h = 1 \text{ for all } h \neq g \\ \mathcal{O}(u^2) & \text{otherwise.} \end{cases} \tag{2.5b}$$

The states that arise after a new mutation can be characterized using the following probability distributions:

**Definition.** The *mutant appearance distribution from state* $\mathbf{a}$, $\mu_{A;a}$, is a probability distribution on $\{0,1\}^G - \{\mathbf{a}\}$, characterizing the limit as $u \to 0$ of the probability of reaching a given state via a transition away from $\mathbf{a}$:

$$\mu_{A;a}(\mathbf{x}) = \lim_{u \to 0} \frac{P_{\mathbf{a} \to \mathbf{x}}}{1 - P_{\mathbf{a} \to \mathbf{a}}}. \tag{2.6a}$$

The *mutant appearance distribution from* $\mathbf{A}$, $\mu_{a;A}$, is defined similarly as a probability distribution on $\{0,1\}^G - \{\mathbf{A}\}$:

$$\mu_{a;A}(\mathbf{x}) = \lim_{u \to 0} \frac{P_{\mathbf{A} \to \mathbf{x}}}{1 - P_{\mathbf{A} \to \mathbf{A}}}. \tag{2.6b}$$

The *two-sided mutant appearance distribution*, $\mu$, on $\{0,1\}^G$, by first sampling from either $\mu_{A;a}$ or $\mu_{a;A}$ with probabilities $\pi_{\mathcal{M}_0}(\mathbf{a})$ and $\pi_{\mathcal{M}_0}(\mathbf{A})$, respectively,

$$\mu(\mathbf{x}) = \pi_{\mathcal{M}_0}(\mathbf{a})\mu_{A;a}(\mathbf{x}) + \pi_{\mathcal{M}_0}(\mathbf{A})\mu_{a;A}(\mathbf{x}). \tag{2.6c}$$

It follows from Eq. (2.5) that the mutant appearance distributions are concentrated on states with one mutant allele, whose location $g$ is chosen proportionally to the mutation rate $\nu_g$:

$$\mu_{A;a}(\mathbf{x}) = \begin{cases} \frac{\nu_g}{\nu_G} & \text{if } x_g = 1 \text{ and } x_h = 0 \text{ for all } h \neq g \\ 0 & \text{otherwise} \end{cases} \tag{2.7a}$$

$$\mu_{a;A}(\mathbf{x}) = \begin{cases} \frac{\nu_g}{\nu_G} & \text{if } x_g = 0 \text{ and } x_h = 1 \text{ for all } h \neq g \\ 0 & \text{otherwise.} \end{cases} \tag{2.7b}$$

**2.2.3. Rare-mutation lemma.** A lemma proven in Ref. [5], applied to the framework described here, provides a useful relationship between the low-mutation ($u \to 0$) and no-mutation ($u = 0$) cases of $\mathcal{M}$:

**Lemma 2.2.** *Let $f : \{0,1\}^G \to \mathbb{R}$ be any function with $f(\mathbf{A}) = f(\mathbf{a}) = 0$. Then*

$$\frac{d\,\mathbb{E}_{\pi_{\mathcal{M}}}[f]}{du}\bigg|_{u=0} = \nu_G \sum_{t=0}^{\infty} \mathbb{E}_{\mathcal{M}_0}[f(\mathbf{X}^t)|\mathbf{X}^0 \sim \mu], \tag{2.8}$$

*and the sum on the right converges absolutely.*

This lemma allows us to pass back and forth between low-mutation limits under the stationary distribution, $\pi_{\mathcal{M}}$, and expected sums over the mutation-free process, $\mathcal{M}_0$. Motivated by this result, we define an operator $\langle \ \rangle$, on functions $f : \{0,1\}^G \to \mathbb{R}$ satisfying $f(\mathbf{A}) = f(\mathbf{a}) = 0$, so that $\langle f \rangle$ is equal to both sides of Eq. (2.8):

$$\langle f \rangle = \left. \frac{d\,\mathbb{E}_{\pi_{\mathcal{M}}}[f]}{du} \right|_{u=0} = \nu_G \sum_{t=0}^{\infty} \mathbb{E}_{\mathcal{M}_0}[f(\mathbf{X}^t)|\mathbf{X}^0 \sim \mu]. \quad (2.9)$$

The $\langle \ \rangle$ operator is preserved under symmetry, in the following sense: For any symmetry $\sigma \in \mathrm{Sym}(G, p)$ (see Section 1.8), define $f_\sigma(\mathbf{x}) = f(\mathbf{x}_\sigma)$. Then Eq. (2.2) implies

$$\langle f \rangle = \langle f_\sigma \rangle. \quad (2.10)$$

**2.3. Fixation probability.** Fixation probability refers to the probability that a mutant lineage, starting from a single allele copy, will eventually take over the population. We define this formally as follows:

**Definition.** The *fixation probabilities* $\rho_A$ and $\rho_a$ are defined as the probabilities of becoming absorbed in states $\mathbf{A}$ and $\mathbf{a}$, respectively, in $\mathcal{M}_0$ with initial state sampled from the appropriate mutant appearance distribution:

$$\rho_A = \lim_{t \to \infty} \mathbb{P}_{\mathcal{M}_0}[\mathbf{X}^t = \mathbf{A} | \mathbf{X}^0 \sim \mu_{A;a}]$$
$$\rho_a = \lim_{t \to \infty} \mathbb{P}_{\mathcal{M}_0}[\mathbf{X}^t = \mathbf{a} | \mathbf{X}^0 \sim \mu_{a;A}].$$

Theorem 2 of Fudenberg and Imhof [1] allows us to express the limiting stationary distribution $\pi_{\mathcal{M}_0}$ in terms of fixation probabilities:

$$\pi_{\mathcal{M}_0}(\mathbf{x}) = \begin{cases} \dfrac{\rho_A}{\rho_A + \rho_a} & \mathbf{x} = \mathbf{A} \\ \dfrac{\rho_a}{\rho_A + \rho_a} & \mathbf{x} = \mathbf{a} \\ 0 & \text{otherwise.} \end{cases} \quad (2.11)$$

## 3. Selection

We now begin to quantify selection, in particular states as well as in the overall selection process. All results here pertain either to no mutation ($u = 0$) or to the low-mutation limit ($u \to 0$) limit. The main results are Theorems 3.3 and 3.4, which show the equivalence between different measures of success.

**3.1. Neutral drift and reproductive value.** Neutral drift ($\delta = 0$) serves as a baseline against which to measure selection. According to Assumption (D2), the probabilities of transition events—and hence all quantities derived from them—are independent of the state $\mathbf{x}$ under neutral drift. We indicate neutral drift using a superscript $\circ$; for example, $p^\circ(\alpha, U)$ denotes the neutral probability of transition event $(\alpha, U)$.

The neutral stationary distribution, $\pi_{\mathcal{M}}^\circ$, is symmetric with respect to interchange of alleles $A$ and $a$; see Propositions 2 and 3 of Ref. [3] for a formal statement and proof. It follows that for each $g \in G$,

$$\mathbb{E}_{\pi_{\mathcal{M}}}^\circ[x_g] = \frac{1}{2}. \quad (3.1)$$

That is, in the neutral stationary distribution, each site is equally likely to contain either allele [31].

Reproductive value (RV) [12, 32, 33] plays an important role in quantifynig selection. We assign each site $g \in G$ a reproductive value $v_g$, which quantifies its expected contribution to the future gene pool under neutral drift. These reproductive values can be obtained by solving the recurrence equations [3, 34]:

$$v_g = \sum_\alpha p^\circ(\alpha) \sum_{h \in \alpha^{-1}(g)} v_h \qquad \text{for all } g \in G \quad (3.2a)$$

$$\sum_{g \in G} v_g = n. \quad (3.2b)$$

Under this definition, $v_g$ is proportional (by a factor of $n$) to the probability that a mutation arising at $g$ will become fixed under neutral drift [3, 34]. Accordingly, $v_g$ can be interpreted as the expected number of descendants of site $g$ as $t \to \infty$, under neutral drift.

Symmetry preserves reproductive value, in the sense that for any symmetry $\sigma \in \mathrm{Sym}(G, p)$ (as defined in Section 1.8),

$$v_g = v_{\sigma(g)}. \tag{3.3}$$

This follows from a result on symmetry of fixation probabilities, proven in Section 4 of Ref. [6]. Consequently, sites in the same class have the same reproductive value. This reduces the system of equations Eq. (3.2) to one equation per class. In a homogeneous population (for which there is only one class), each site has reproductive value 1.

Reproductive value provides a natural weighting on genetic sites. Specifically, the *RV-weighted frequency* of allele $A$, defined as

$$\hat{x} = \frac{1}{n} \sum_{g \in G} x_g v_g, \tag{3.4}$$

provides a useful summary statistic of the population state $\mathbf{x}$. We observe that $\hat{x} = 0$ in state $\mathbf{x} = \mathbf{a}$ and $\hat{x} = 1$ in state $\mathbf{x} = \mathbf{A}$. We denote the RV-weighted frequency at time $t$ in $\mathcal{M}$ by the random variable $\hat{X}^t$. $\hat{X}^t$ is a Martingale for neutral drift without mutation, meaning that $\mathbb{E}^\circ_{\mathcal{M}_0}[\hat{X}^{t+1} | \mathbf{X}^t = \mathbf{x}] = \hat{x}$.

**3.2. Fitness increments.** The reproductive success of a given site in a given state is quantified using fitness increments:

**Definition.** The *fitness increment* of site $g \in G$ in state $\mathbf{x}$ is

$$w_g(\mathbf{x}) = \mathbb{E}_{\mathbf{x}} \left[ \sum_{h \in \alpha^{-1}(g)} v_h \right] - v_g. \tag{3.5}$$

In words, $w_g(\mathbf{x})$ is the expected difference in reproductive value between the allele occupying $g$ and this allele's offspring (counting itself, if it survives) in the next time-step.

The total fitness increment of all sites is zero, for each state $\mathbf{x}$, since the total reproductive value of all sites is constant:

$$\sum_{g \in G} w_g(\mathbf{x}) = \mathbb{E}_{\mathbf{x}} \left[ \sum_{g \in G} \sum_{h \in \alpha^{-1}(g)} v_h \right] - \sum_{g \in G} v_g = \sum_{h \in G} v_h - \sum_{g \in G} v_g = 0. \tag{3.6}$$

The second equality above uses the fact that for any parentage map $\alpha : G \to G$, each $h \in G$ is an element of $\alpha^{-1}(g)$ for exactly one $g \in G$, namely, $g = \alpha(h)$.

For neutral drift, the fitness increment of each site is zero, $w_g^\circ = 0$ for all $g \in G$, as a consequence of Eq. (3.2a).

Fitness increments are preserved by symmetry in the following sense:

**Proposition 3.1.** *For any symmetry $\sigma \in \mathrm{Sym}(G, p)$, site $g \in G$, and state $\mathbf{x} \in \{0,1\}^G$,*

$$w_{\sigma(g)}(\mathbf{x}) = w_g(\mathbf{x}_\sigma). \tag{3.7}$$

*Proof.* We begin with the right-hand side:

$$w_g(\mathbf{x}_\sigma) = \sum_{\alpha:G\to G} p_{\mathbf{x}_\sigma}(\alpha) \sum_{h\in\alpha^{-1}(g)} v_h - v_g \qquad \text{by Eq. (3.5)}$$

$$= \sum_{\alpha':G\to G} p_{\mathbf{x}_\sigma}(\sigma^{-1}\circ\alpha'\circ\sigma) \sum_{h\in(\sigma^{-1}\circ\alpha'\circ\sigma)^{-1}(g)} v_h - v_g$$

$$\text{with } \alpha' = \sigma\circ\alpha\circ\sigma^{-1}$$

$$= \sum_{\alpha':G\to G} p_{\mathbf{x}}(\alpha') \sum_{h\in(\sigma^{-1}\circ\alpha'\circ\sigma)^{-1}(g)} v_h - v_g \qquad \text{by Eq. (1.5)}$$

$$= \sum_{\alpha':G\to G} p_{\mathbf{x}}(\alpha') \sum_{h\in\sigma^{-1}\circ(\alpha')^{-1}\circ\sigma(g)} v_h - v_g \qquad \text{inverting the composition}$$

$$= \sum_{\alpha':G\to G} p_{\mathbf{x}}(\alpha') \sum_{h'\in(\alpha')^{-1}(\sigma(g))} v_{\sigma^{-1}(h')} - v_g \qquad \text{with } h' = \sigma(h)$$

$$= \sum_{\alpha':G\to G} p_{\mathbf{x}}(\alpha') \sum_{h'\in(\alpha')^{-1}(\sigma(g))} v_{h'} - v_{\sigma(g)} \qquad \text{by Eq. (3.3)}$$

$$= w_{\sigma(g)}(\mathbf{x}).$$

Above, the steps introducing $\alpha'$ and $h'$ use the fact that $\sigma$ is a bijection to re-index the respective summations. □

**3.3. Selection increments.** Selection between alleles $A$ and $a$, in a given state $\mathbf{x}$, is quantified in terms of the expected change in RV-weighted frequency:

**Definition.** We define the *selection increment* in state $\mathbf{x}$, denoted $\Delta(\mathbf{x})$, as the expected change (absent mutation) in $\hat{x}$ from state $\mathbf{x}$:

$$\Delta(\mathbf{x}) = \mathbb{E}_{\mathcal{M}_0}[\hat{X}^{t+1} - \hat{X}^t | \mathbf{X}^t = \mathbf{x}]. \tag{3.8}$$

Fitness and selection increments are related to each other by the following lemma, an instance of the Price equation [35]:

**Lemma 3.2.** *For $u = 0$, in each state $\mathbf{x}$,*

$$\Delta(\mathbf{x}) = \frac{1}{n}\sum_{g\in G} x_g w_g(\mathbf{x}) = \frac{1}{n}\sum_{g\in G}(x_g - \bar{x})\, w_g(\mathbf{x}). \tag{3.9}$$

*Proof.* We first observe that, from Eq. (1.3) and Eq. (3.4),

$$\mathbb{E}[\hat{X}^{t+1}|\mathbf{X}^t = \mathbf{x}] = \frac{1}{n}\mathbb{E}_{\mathbf{x}}\left[\sum_{g\in G}\sum_{h\in\alpha^{-1}(g)} v_h x_{\alpha(h)}\right] = \frac{1}{n}\mathbb{E}_{\mathbf{x}}\left[\sum_{g\in G}\sum_{h\in\alpha^{-1}(g)} v_h x_g\right]. \tag{3.10}$$

Substituting this and Eq. (3.4) into Eq. (3.8), we have

$$\Delta(\mathbf{x}) = \frac{1}{n}\mathbb{E}_{\mathbf{x}}\left[\sum_{g\in G}\sum_{h\in\alpha^{-1}(g)} v_h x_g - \sum_{g\in G} v_g x_g\right]$$

$$= \frac{1}{n}\sum_{g\in G} x_g\left(\mathbb{E}_{\mathbf{x}}\left[\sum_{h\in\alpha^{-1}(g)} v_h\right] - v_g\right)$$

$$= \frac{1}{n}\sum_{g\in G} x_g w_g(\mathbf{x}).$$

This proves the first equality in Eq. (3.9). The second follows from Eq. (3.6). □

We observe that $\Delta(\mathbf{A}) = \Delta(\mathbf{a}) = 0$, reflecting the fact that, absent mutation, the state cannot change from $\mathbf{A}$ or $\mathbf{a}$. For neutral drift ($\delta = 0$), the selection increment is zero in each state, $\Delta^\circ = 0$, since $w_g^\circ = 0$ for each $g \in G$.

**3.4. Equivalence of success criteria.** The success of an allele can be quantified in a number of ways. Commonly used measures are based on expected frequency, on fixation probability, and on fitness quantities. Here we show that these measures are equivalent in our framework, in the $u \to 0$ limit. Similar equivalencies have been proven for other models and frameworks [2, 3, 9, 19, 21, 22, 26, 27, 36].

**Theorem 3.3.** *The following success criteria are equivalent:*

(i) $\pi_{\mathcal{M}_0}(\mathbf{A}) > \frac{1}{2} > \pi_{\mathcal{M}_0}(\mathbf{a})$,

(ii) $\rho_A > \rho_a$,

(iii) $\langle \Delta \rangle > 0$.

The brackets $\langle \ \rangle$ in criterion (iii) refer to the operator defined in Eq. (2.9).

*Proof.* The equivalence (i) $\Leftrightarrow$ (ii) follows directly from Eq. (2.11). For (ii) $\Leftrightarrow$ (iii), we first develop expressions for $\rho_A$ and $\rho_a$ in terms of $\Delta(\mathbf{x})$. Since the weighted frequency $\hat{x}$ is 0 and 1, respectively, in states $\mathbf{a}$ and $\mathbf{A}$, we can express fixation probability in terms of the limiting expectation of $\hat{X}^t$:

$$\rho_A = \lim_{t \to \infty} \mathbb{E}_{\mathcal{M}_0}[\hat{X}^t \mid \mathbf{X}^0 \sim \mu_{A;a}] \tag{3.11a}$$

$$\rho_a = 1 - \lim_{t \to \infty} \mathbb{E}_{\mathcal{M}_0}[\hat{X}^t \mid \mathbf{X}^0 \sim \mu_{a;A}]. \tag{3.11b}$$

We let $\hat{\mu}$ denote the expected RV-weighted frequency of a new mutant type:

$$\hat{\mu} = \mathbb{E}_{\mu_{A;a}}[\hat{X}] = \mathbb{E}_{\mu_{a;A}}[1 - \hat{X}] = \frac{1}{n\nu_G} \sum_{g \in G} \nu_g v_g. \tag{3.12}$$

Then, using Eq. (3.11a) and Eq. (3.8), we can express the fixation probability of $A$ as

$$\rho_A = \mathbb{E}_{\mu_{A;a}}[\hat{X}] + \sum_{t=0}^{\infty} \mathbb{E}_{\mathcal{M}_0}[\hat{X}^{t+1} - \hat{X}^t \mid \mathbf{X}^0 \sim \mu_{A;a}]$$

$$= \hat{\mu} + \sum_{t=0}^{\infty} \mathbb{E}_{\mathcal{M}_0}[\Delta(\mathbf{X}^t) \mid \mathbf{X}^0 \sim \mu_{A;a}]. \tag{3.13}$$

Similarly, using Eq. (3.11b), the fixation probability of $a$ can be written as

$$\rho_a = 1 - \left( \mathbb{E}_{\mu_{a;A}}[\hat{X}] + \sum_{t=0}^{\infty} \mathbb{E}_{\mathcal{M}_0}[\hat{X}^{t+1} - \hat{X}^t \mid \mathbf{X}^0 \sim \mu_{a;A}] \right)$$

$$= \mathbb{E}_{\mu_{a;A}}[1 - \hat{X}] - \sum_{t=0}^{\infty} \mathbb{E}_{\mathcal{M}_0}[\Delta(\mathbf{X}^t) \mid \mathbf{X}^0 \sim \mu_{a;A}]$$

$$= \hat{\mu} - \sum_{t=0}^{\infty} \mathbb{E}_{\mathcal{M}_0}[\Delta(\mathbf{X}^t) \mid \mathbf{X}^0 \sim \mu_{a;A}]. \tag{3.14}$$

Combining these expressions with Eq. (2.6c), Eq. (2.9), and Eq. (2.11) provides the relationship between $\langle \Delta \rangle$ and the fixation probabilities:

$$\begin{aligned}
\langle \Delta \rangle &= \nu_G \sum_{t=0}^{\infty} \mathbb{E}_{\mathcal{M}_0}[\Delta(\mathbf{X}^t) | \mathbf{X}^0 \sim \mu] \\
&= \nu_G \left( \pi_{\mathcal{M}_0}(\mathbf{a}) \sum_{t=0}^{\infty} \mathbb{E}_{\mathcal{M}_0}[\Delta(\mathbf{X}^t) | \mathbf{X}^0 \sim \mu_{A;a}] \right. \\
&\qquad \left. + \pi_{\mathcal{M}_0}(\mathbf{A}) \sum_{t=0}^{\infty} \mathbb{E}_{\mathcal{M}_0}[\Delta(\mathbf{X}^t) | \mathbf{X}^0 \sim \mu_{a;A}] \right) \\
&= \frac{\nu_G}{\rho_A + \rho_a} \\
&\quad \times \left( \rho_a \sum_{t=0}^{\infty} \mathbb{E}_{\mathcal{M}_0}[\Delta(\mathbf{X}^t) | \mathbf{X}^0 \sim \mu_{A;a}] + \rho_A \sum_{t=0}^{\infty} \mathbb{E}_{\mathcal{M}_0}[\Delta(\mathbf{X}^t) | \mathbf{X}^0 \sim \mu_{a;A}] \right) \\
&= \frac{\nu_G}{\rho_A + \rho_a} \left( \rho_a(\rho_A - \hat{\mu}) - \rho_A(\rho_a - \hat{\mu}) \right) \\
&= \frac{\nu_G \hat{\mu}}{\rho_A + \rho_a} (\rho_A - \rho_a).
\end{aligned}$$

The right-hand side (and hence the left-hand side) is positive if and only if Condition (ii) holds, proving (ii) $\Leftrightarrow$ (iii). $\square$

The above result makes use of Assumption (M6), that mutation rates are the same in the two monoallelic states. Without Assumption (M6), mutation can bias allele frequencies, leading to a more complex relationship between fixation probability and expected frequency [3, 4, 9].

**3.5. Weak selection.** We turn now to weak selection, which we represent as a first-order perturbation, in $\delta$, around neutral drift ($\delta = 0$). We use primes (′) to indicate $\delta$-derivatives at $\delta = 0$; for example, $\Delta'(\mathbf{x}) = \frac{d\Delta(\mathbf{x})}{d\delta}|_{\delta=0}$. By Assumption (D1), the probabilities of transition events (and all quantities derived from them) admit a weak-selection expansion of the form

$$p_{\mathbf{x}}(\alpha, U) = p^{\circ}(\alpha, U) + \delta p'_{\mathbf{x}}(\alpha, U) + \mathcal{O}(\delta^2) \qquad (\delta \to 0). \tag{3.15}$$

Using this perturbation idea, we obtain a weak-selection analogue of Theorem 3.3:

**Theorem 3.4.** *The following weak-selection success criteria are equivalent:*

*(i)* $\pi'_{\mathcal{M}_0}(\mathbf{A}) > 0 > \pi'_{\mathcal{M}_0}(\mathbf{a})$,

*(ii)* $\rho'_A > \rho'_a$,

*(iii)* $\langle \Delta' \rangle^{\circ} > 0$.

*Proof.* The equivalence $(i) \Leftrightarrow (ii)$ follows directly from taking $\delta$-derivatives, at $\delta = 0$, of the quantities in the corresponding conditions of Theorem 3.3. To prove $(ii) \Leftrightarrow (iii)$ we first observe that, for every fixed $0 < u < 1$,

$$\frac{d\,\mathbb{E}_{\pi_{\mathcal{M}}}[\Delta]}{d\delta}\bigg|_{\delta=0} = \sum_{\mathbf{x}\in\{0,1\}^G} \pi^{\circ}_{\mathcal{M}}(\mathbf{x})\,\Delta'(\mathbf{x}) = \mathbb{E}^{\circ}_{\pi_{\mathcal{M}}}[\Delta'], \tag{3.16}$$

where the first equality uses the fact that $\Delta^{\circ}(\mathbf{x}) = 0$ for each state $\mathbf{x}$. Taking $u$-derivatives at $u = 0$ gives

$$\frac{d\langle\Delta\rangle}{d\delta}\bigg|_{\delta=0} = \frac{\partial^2\,\mathbb{E}_{\pi_{\mathcal{M}}}[\Delta]}{\partial u\,\partial\delta}\bigg|_{\substack{u=0\\\delta=0}} = \frac{d\,\mathbb{E}^{\circ}_{\pi_{\mathcal{M}}}[\Delta']}{du}\bigg|_{u=0} = \langle\Delta'\rangle^{\circ}. \tag{3.17}$$

Now, taking $\delta$-derivatives of Conditions (ii) and (iii) in Theorem 3.3, we have

$$\rho'_A > \rho'_a \quad \Longleftrightarrow \quad \left.\frac{d\langle\Delta\rangle}{d\delta}\right|_{\delta=0} > 0 \quad \Longleftrightarrow \quad \langle\Delta'\rangle^\circ > 0, \tag{3.18}$$

as desired. $\square$

If the criteria of Theorem 3.4 are met, we say that *weak selection favors A over a*. Theorem 3.4 is a variation of Theorem 8 of Allen & McAvoy [3], and generalizes Proposition 4.1 of Taylor [36] and Eq. (16) of van Cleve [26].

## 4. Genetic assortment and relatedness

We now turn to measures of genetic assortment, building up to our definition of collective relatedness.

### 4.1. Identity-by-state.

A set of alleles are *identical by state (IBS)* if they are copies of each other, whether by co-ancestry or another reason. We quantify IBS using indicator functions $\imath_S$, such that $\imath_S(\mathbf{x}) = 1$ if all sites in $S$ contain the same allele in state $\mathbf{x}$ (that is, if $x_g = x_h$ for all $g, h \in S$), and $\imath_S(\mathbf{x}) = 0$ otherwise. We also define allele-specific IBS functions $\imath_S^A$ and $\imath_S^a$, so that $\imath_S^A(\mathbf{x})$ (respectively, $\imath_S^a(\mathbf{x})$), equals one if all sites in $S$ contain $A$ (respectively, $a$) in state $\mathbf{x}$, and zero otherwise.

We can express these IBS functions algebraically: for nonempty $S \subseteq G$,

$$\imath_S^A(\mathbf{x}) = \prod_{g \in S} x_g \tag{4.1a}$$

$$\imath_S^a(\mathbf{x}) = \prod_{g \in S} (1 - x_g) \tag{4.1b}$$

$$\imath_S(\mathbf{x}) = \imath_S^A(\mathbf{x}) + \imath_S^a(\mathbf{x}). \tag{4.1c}$$

If $S$ is a singleton set, then $\imath_S(\mathbf{x}) = 1$ for all states $\mathbf{x}$ (a single site contains the same allele as itself). In the vacuous case $S = \varnothing$, we have $\imath_\varnothing^A(\mathbf{x}) = \imath_\varnothing^a(\mathbf{x}) = \imath_\varnothing(\mathbf{x}) = 1$ for each $\mathbf{x}$.

From Eq. (4.1) we obtain a relationship between $\imath_S^A(\mathbf{x})$ and $\imath_S^a(\mathbf{x})$,

$$\imath_S^a(\mathbf{x}) = \sum_{T \subseteq S} (-1)^{|T|} \imath_T^A(\mathbf{x}). \tag{4.2}$$

We also obtain the identities

$$x_g \imath_S^A(\mathbf{x}) = \imath_{S \cup \{g\}}^A(\mathbf{x}) \quad \text{and} \quad x_g \imath_S^a(\mathbf{x}) = \imath_S^a(\mathbf{x}) - \imath_{S \cup \{g\}}^a(\mathbf{x}). \tag{4.3}$$

For any set mapping $\tau : G \to G$, the identity-by-state functions satisfy

$$\imath_{\tau(S)}^A(\mathbf{x}) = \imath_S^A(\mathbf{x}_\tau) \quad \text{and} \quad \imath_{\tau(S)}^a(\mathbf{x}) = \imath_S^a(\mathbf{x}_\tau). \tag{4.4}$$

### 4.2. Genetic dissimilarity and coalescence length.

The identity-by-state functions $\imath_S^A(\mathbf{x})$ and $\imath_S^a(\mathbf{x})$ quantify genetic assortment in a given state $\mathbf{x}$. Here, we introduce a measure that applies to the overall selection process:

**Definition.** The *genetic dissimilarity* of a nonempty subset $S \subseteq G$ is defined as

$$\ell_S = \langle 1 - \imath_S \rangle. \tag{4.5}$$

We can interpret this genetic dissimilarity measure in two ways. First, interpreting the brackets $\langle \; \rangle$ according to the left-hand side of Eq. (2.8), $\ell_S$ quantifies the effect of mutation (when rare) on the stationary probability that the sites in $S$ are *not* all identical-by-state:

$$\ell_S = \frac{d}{du} \mathbb{E}_{\pi_\mathcal{M}}[1 - \imath_S]|_{u=0}. \tag{4.6}$$

Second, using the left-hand side of Eq. (2.8), $\ell_S$ is proportional to the expected duration of time for which the sites in $S$ do not all contain the same allele:

$$\ell_S = \nu_G \sum_{t=0}^{\infty} \mathbb{E}_{\mathcal{M}_0}[1 - \imath_S(\mathbf{X}^t)|\mathbf{X}^0 \sim \mu]. \tag{4.7}$$

We observe from Eq. (4.5) that $\ell_S$ is always nonnegative, and is zero if $S$ is a singleton set.

In the neutral ($\delta = 0$) case, there is a third interpretation of $\ell_S$ in terms of the coalescent [17, 18]— a stochastic process representing the ancestry of a present-day population traced backwards in time. Specifically, $\ell_S^\circ$ is the expected total branch length of the coalescent tree of $S$, where each branch is scaled by the site-specific mutation rates $\nu_g^\circ$. These coalescence lengths are uniquely determined by the recurrence relation

$$\ell_S^\circ = \begin{cases} \nu_S^\circ + \sum_\alpha p^\circ(\alpha)\, \ell_{\alpha(S)}^\circ & |S| \geq 2 \\ 0 & |S| = 1. \end{cases} \tag{4.8}$$

The equivalence of this third interpretation of $\ell_S$ to the first two, under neutral drift, is proven in Theorem 5.1 of Allen and McAvoy [5]. In that work, $\ell_S^\circ$ is denoted as $m_S'$, and $\nu_S^\circ$ as $\nu_S'$. Eq. (4.8) appears as Eq. (5.9) in Ref. [5], while uniqueness of the solution to Eq. (4.8) is proven in Theorem C.1 of that work.

**4.3. Collective relatedness.** Having introduced short-term and long-term measures of genetic assortment, we are prepared to define collective relatedness.

*4.3.1. Definition and alternative expressions.* We introduce the following notation for the average identity-by-state of a given set $S$ to all sites $g$:

$$\bar{\imath}_S(\mathbf{x}) = \frac{1}{n} \sum_{g \in G} \imath_{S \cup \{g\}}(\mathbf{x}), \qquad \bar{\imath}_S^A(\mathbf{x}) = \frac{1}{n} \sum_{g \in G} \imath_{S \cup \{g\}}^A(\mathbf{x}), \qquad \bar{\imath}_S^a(\mathbf{x}) = \frac{1}{n} \sum_{g \in G} \imath_{S \cup \{g\}}^a(\mathbf{x}). \tag{4.9}$$

With that notation, we define collective relatedness as follows:

**Definition.** For a nonempty set of sites $S \subseteq G$ and a site $g \in G$, the *collective relatedness of $S$ to $g$ with respect to allele $A$* is defined as

$$r_{S,g}^A = \frac{\langle \imath_S^A(\mathbf{x})\,(x_g - \bar{x}) \rangle}{\langle \bar{x}(1 - \bar{x}) \rangle} = \frac{\langle \imath_{S \cup \{g\}}^A(\mathbf{x}) - \bar{\imath}_S^A(\mathbf{x}) \rangle}{\langle \bar{x}(1 - \bar{x}) \rangle}. \tag{4.10a}$$

The *collective relatedness of $S$ to $g$ with respect to allele $a$* is obtained by interchanging the roles of $A$ and $a$ in Eq. (4.10a):

$$r_{S,g}^a = \frac{-\langle \imath_S^a(\mathbf{x})\,(x_g - \bar{x}) \rangle}{\langle \bar{x}(1 - \bar{x}) \rangle} = \frac{\langle \imath_{S \cup \{g\}}^a(\mathbf{x}) - \bar{\imath}_S^a(\mathbf{x}) \rangle}{\langle \bar{x}(1 - \bar{x}) \rangle}. \tag{4.10b}$$

Collective relatedness quantifies the likelihood that, when all of $S$ contains the same allele, $g$ does as well. This likelihood is normalized in two ways. First, $\bar{x}$ is subtracted in the numerator, so that $r_{S,g}^A$ is positive (resp., negative) if $g$ is more (resp., less) likely to contain $A$ when all of $S$ does, and similarly for $r_{S,g}^a$. Second, the result is divided by the variance, $\bar{x}(1 - \bar{x})$, in $x_g$ over $G$ (i.e. the allelic variance in the population). Alternative normalizations, such as subtracting the RV-weighted mean, $\hat{x}$, instead of the frequency, $\bar{x}$, could be used instead without substantially affecting our main results.

Collective relatedness can be expressed in a number of equivalent ways. First, by applying Eq. (2.9) and L'Hôpital's rule to Eq. (4.10a), we obtain Eq. (5) of the main text (which uses $r_{S,g}$ as shorthand for $r^A_{S,g}$).

Second, the variance, $\bar{x}(1-\bar{x})$, that appears in the denominator can be rewritten as:

$$\bar{x}(1-\bar{x}) = \frac{1}{2}\left(1 - (\bar{x})^2 - (1-\bar{x})^2\right)$$

$$= \frac{1}{2}\left(1 - \frac{1}{n^2}\sum_{h,k \in G}(x_h x_k + (1-x_h)(1-x_k))\right)$$

$$= \frac{1}{2}\left(1 - \frac{1}{n^2}\sum_{h,k \in G} \imath_{\{h,k\}}(\mathbf{x})\right).$$

Letting $\bar{\ell} = \frac{1}{n^2}\sum_{h,k \in G}\ell_{\{h,k\}}$ denote the average coalescence length between all pairs, Eq. (4.5) gives

$$\langle \bar{x}(1-\bar{x})\rangle = \frac{\bar{\ell}}{2}. \tag{4.11}$$

We can then rewrite Eq. (4.10) as

$$r^A_{S,g} = \frac{2}{\bar{\ell}}\left\langle \imath^A_{S \cup \{g\}} - \bar{\imath}^A_S\right\rangle \tag{4.12a}$$

$$r^a_{S,g} = \frac{2}{\bar{\ell}}\left\langle \imath^a_{S \cup \{g\}} - \bar{\imath}^A_S\right\rangle. \tag{4.12b}$$

Combining Eq. (4.1c), Eq. (4.5), and Eq. (4.12), we obtain an elegant expression for the average of $r^A_{S,g}$ and $r^a_{S,g}$:

$$\frac{r^A_{S,g} + r^a_{S,g}}{2} = \frac{\bar{\ell}_S - \ell_{S \cup \{g\}}}{\bar{\ell}}, \tag{4.13}$$

where $\bar{\ell}_S = \frac{1}{n}\sum_{h \in G}\ell_{S \cup \{h\}}$. For neutral drift, the symmetry of $\pi_\mathcal{M}$ under interchange of $A$ and $a$ implies that $\left(r^A_{S,g}\right)^\circ = \left(r^a_{S,g}\right)^\circ$. Denoting this neutral collective relatedness by $r^\circ_{S,g}$, Eq. (4.13) gives

$$r^\circ_{S,g} = \frac{\bar{\ell}^\circ_S - \ell^\circ_{S \cup \{g\}}}{\bar{\ell}^\circ}, \tag{4.14}$$

which is Eq. (7) of the main text. The neutral collective relatednesses $r^\circ_{S,g}$ can be computed by solving Eq. (4.8) for the coalescence lengths $\ell^\circ_S$, and then applying Eq. (4.14).

### 4.3.2. Properties of collective relatedness. Collective relatedness has the following properties:

- The average collective relatedness of a given set $S$ to all sites is zero:

$$\frac{1}{n}\sum_{g \in G} r^A_{S,g} = \frac{1}{n}\sum_{g \in G} r^a_{S,g} = 0. \tag{4.15}$$

- A collective has the same relatedness to each of its members. Specifically, for any nonempty $S \subseteq G$, we have $r^A_{S,g} = r^A_S$ and $r^a_{S,g} = r^a_S$ for each $g \in S$, where the intra-relatedness quantities $r^A_S$ and $r^a_S$ are given by

$$r^A_S = \frac{\left\langle \imath^A_S(\mathbf{x}) - \bar{\imath}^A_S(\mathbf{x})\right\rangle}{\langle \bar{x}(1-\bar{x})\rangle}, \qquad r^a_S = \frac{\left\langle \imath^a_S(\mathbf{x}) - \bar{\imath}^a_S(\mathbf{x})\right\rangle}{\langle \bar{x}(1-\bar{x})\rangle}.$$

- The average relatedness of each site to itself is 1:
$$\frac{1}{n}\sum_{g\in G} r^A_{\{g\},g} = \frac{1}{n}\sum_{g\in G} r^a_{\{g\},g} = 1.$$

However, the relatedness of an individual site to itself is not necessarily 1, even under neutral drift. Instead, we have
$$r^A_{\{g\},g} = r^a_{\{g\},g} = \frac{\langle x_g - x_g \bar{x}\rangle}{\frac{1}{n}\sum_{h\in G} \langle x_h - x_h\bar{x}\rangle} = \frac{\bar{\ell}_{\{g\}}}{\bar{\ell}}.$$

Thus, a site $g$ has self-relatedness greater than 1 if and only if $\bar{\ell}_{\{g\}} > \bar{\ell}$, meaning that the average coalescence length from $g$ to all sites exceeds the average coalescence length of all pairs.

- The "collective" relatedness of the empty set to any site is zero under neutral drift: $r^\circ_{\varnothing,g} = 0$. This follows from Eq. (4.14), noting that $\ell_{\varnothing\cup\{h\}}(\mathbf{x}) = \ell_{\{h\}}(\mathbf{x}) = 0$ for all $h \in G$. Away from neutral drift, $r^A_{\varnothing,g}$ and $r^a_{\varnothing,g}$ are not necessarily zero; instead, we have
$$r^A_{\varnothing,g} = -r^a_{\varnothing,g} = \frac{\langle x_g - \bar{x}\rangle}{\langle \bar{x}(1-\bar{x})\rangle}. \tag{4.16}$$

Thus $r^A_{\varnothing,g}$ is positive if site $g$ is more likely than the average site to hold an $A$ allele in transient states of the selection process, and likewise for $r^a_{\varnothing,g}$.

- Symmetry preserves relatedness, in the sense that if $\sigma : G \to G$ is a symmetry (see Section 1.8), then
$$r^A_{S,g} = r^A_{\sigma(S),\sigma(g)} \quad \text{and} \quad r^a_{S,g} = r^a_{\sigma(S),\sigma(g)}, \tag{4.17}$$

for any $S \subseteq G$ and $g \in G$. This is demonstrated, for $r^A_{S,g}$, as follows:
$$\begin{aligned}
r^A_{S,g} &= \frac{\langle \imath^A_{S\cup\{g\}}(\mathbf{x}) - \bar{\imath}^A_S(\mathbf{x})\rangle}{\langle \bar{x}(1-\bar{x})\rangle} && \text{by Eq. (4.10a)} \\
&= \frac{\langle \imath^A_{S\cup\{g\}}(\mathbf{x}_\sigma) - \bar{\imath}^A_S(\mathbf{x}_\sigma)\rangle}{\langle \bar{x}(1-\bar{x})\rangle} && \text{by Eq. (2.10)} \\
&= \frac{\langle \imath^A_{\sigma(S)\cup\{\sigma(g)\}}(\mathbf{x}) - \bar{\imath}^A_{\sigma(S)}(\mathbf{x})\rangle}{\langle \bar{x}(1-\bar{x})\rangle} && \text{by Eq. (4.4)} \\
&= r^A_{\sigma(S),\sigma(g)}.
\end{aligned}$$

The result for $r^a_{S,g}$ is proved similarly.

***4.3.3. Collective phenotypic relatedness.*** It is also useful to define collective relatedness at the phenotype level, using the formalism introduced in Section 1.7. We denote the frequency of allele $A$ in individual $i \in I$ as $\bar{x}_i$:
$$\bar{x}_i = \frac{1}{n_i}\sum_{g\in G_i} x_g. \tag{4.18}$$

**Definition.** The *collective phenotypic relatedness* of a set of individuals $J \subseteq I$ to an individual $i \in I$, with respect to phenotype 1, is defined as
$$r^1_{J,i} = \frac{\left\langle \mathbb{E}\left[\prod_{j\in J}\Phi_j\right](\bar{x}_i - \bar{x})\right\rangle}{\langle \bar{x}(1-\bar{x})\rangle} = \frac{\left\langle \prod_{j\in J}\varphi_j(\mathbf{x}_{|G_j})(\bar{x}_i - \bar{x})\right\rangle}{\langle \bar{x}(1-\bar{x})\rangle}. \tag{4.19a}$$

Analogously, the collective phenotypic relatedness of $J$ to $i$ with respect to phenotype 0 is defined as
$$r^0_{J,i} = \frac{\left\langle \mathbb{E}\left[\prod_{j\in J}(1-\Phi_j)\right](\bar{x}_i - \bar{x})\right\rangle}{\langle \bar{x}(1-\bar{x})\rangle} = \frac{\left\langle \prod_{j\in J}(1-\varphi_j(\mathbf{x}_{|G_j}))(\bar{x}_i - \bar{x})\right\rangle}{\langle \bar{x}(1-\bar{x})\rangle}. \tag{4.19b}$$

**4.4. Relationship to other relatedness measures.** Collective relatedness, as introduced here, is closely related to established definitions of pairwise relatedness based on covariance [37–40], identity-by-descent [21, 23, 36, 41], and geometric considerations [42]. A number of these pairwise relatedness measures can be recovered from collective relatedness, in the case that the "collective" is a single site or individual.

***4.4.1. Identity-by-descent.*** Relatedness is often quantified using identity-by-descent (IBD) [20, 23, 43–45]. Two alleles are identical by descent if no mutation separates them from their common ancestor.

For the neutral case of this framework, Allen and McAvoy [5] obtained recurrence relations for the stationary probability, $q_S^\circ$, that all alleles in $S$ are identical by descent:

$$q_S^\circ = \begin{cases} \sum_{\substack{(\alpha, U) \\ U \cap S = \varnothing}} p^\circ(\alpha)\, q_{\alpha(S)}^\circ & |S| \geq 2 \\ 1 & |S| = 1. \end{cases} \tag{4.20}$$

Allen and McAvoy [5] also proved (in their Theorem 5.1) that IBD probabilities and coalescence lengths are related by

$$\ell_S^\circ = \lim_{u \to 0} \frac{1 - q_S^\circ}{u}. \tag{4.21}$$

(In that work, $\ell_S^\circ$ is denoted $m_S'$.) Combining this result with Eq. (4.14) yields an expression for neutral collective relatedness in terms of IBD probabilities:

$$r_{S,g}^\circ = \lim_{u \to 0} \frac{q_{S \cup \{g\}}^\circ - \bar{q}_S^\circ}{1 - \bar{q}^\circ}, \tag{4.22}$$

where $\bar{q}_S^\circ = \frac{1}{n} \sum_{h \in G} q_{S \cup \{h\}}^\circ$ and $\bar{q}^\circ = \frac{1}{n^2} \sum_{h,k \in G} q_{\{h,k\}}^\circ$. In the case of a singleton set $S = \{h\}$, we obtain

$$r_{\{h\},g}^\circ = \lim_{u \to 0} \frac{q_{\{h,g\}}^\circ - \bar{q}_{\{h\}}^\circ}{1 - \bar{q}^\circ}, \tag{4.23}$$

which is a standard measure of pairwise relatedness between haploid individuals [19, 21, 23, 36, 41]. Eq. (4.22) generalizes this IBD-based definition of relatedness to collectives.

***4.4.2. Geometric relatedness.*** Grafen [42] introduced a definition of relatedness with a geometric interpretation. Translated into our notation, Grafen's Eq. (7) for the relatedness of individual $j$ to $i$ is

$$R_{ji} = \frac{\langle \varphi_j(\mathbf{x}_{|G_j})(\bar{x}_i - \bar{x}) \rangle}{\langle \varphi_j(\mathbf{x}_{|G_j})(\bar{x}_j - \bar{x}) \rangle}. \tag{4.24}$$

Above, we have replaced Grafen's sums over a "list of occasions" with expected sums over transient states of the process, using the bracket operator defined in Eq. (2.9).

The numerator of Grafen's definition, Eq. (4.24), agrees with that of our Eq. (4.19a) for $r_{J,i}^1$ in the case of a singleton set, $J = \{j\}$. The denominators are differerent, relecting a different choice of normalization. While Grafen's normalization has the property that relatedness to oneself is always one ($R_{ii} = 1$ for all $i \in I$), ours has the advantage of allowing for relatednesses from different actors—including collective actors with different numbers of individuals—to be compared according to the same scale.

***4.4.3. Genetic covariance.*** A number of established definitions of relatedness involve a ratio of covariances, which can in some cases be interpreted as a correlation coefficient [46, 47] or a regression coefficient [37, 39, 40]. We show here that standard regression definitions of relatedness [39, 40] can be recovered as an expectation of our $r_{S,g}^A$, with $g$ sampled uniformly from the population and $S$ sampled from the "social environment" of site $g$.

To obtain this connection, we suppose that each site $g \in G$ has an associated "social environment", characterized by a fixed probability distribution $\{p_{S|g}\}_{\varnothing \subset S \subseteq G}$ over nonempty subsets $S$. These social environments are given *a priori*. They are understood to quantify how frequently each collective interacts with a given site.

For each site $g \in G$, we define the state variable $y_g$ as the probability that a set sampled from $g$'s social environment contains only allele $A$:

$$y_g = \mathbb{E}_{S|g}\left[\imath_S^A(\mathbf{x})\right] = \sum_{\varnothing \subset S \subseteq G} p_{S|g}\, \imath_S^A(\mathbf{x}).$$

We let $\bar{y} = \frac{1}{n}\sum_{g \in G} y_g$ denote the population average of $y_g$.

Now suppose that a site $g$ is sampled uniformly from $G$, and then a nonempty set $S \subseteq G$ is sampled from $\{p_{S|g}\}$. We compute the expectation of $r_{S,g}^A$ under this scheme:

$$\begin{aligned}
\mathbb{E}_{g,S}\left[r_{S,g}^A\right] &= \frac{1}{n}\sum_{g \in G}\sum_{\varnothing \subset S \subseteq G} p_{S|g}\, r_{S,g}^A \\
&= \frac{\left\langle \frac{1}{n}\sum_{g \in G}\sum_{\varnothing \subset S \subseteq G} p_{S|g}\, \imath_S^A(\mathbf{x})\,(x_g - \bar{x})\right\rangle}{\langle \bar{x}(1-\bar{x})\rangle} \\
&= \frac{\left\langle \frac{1}{n}\sum_{g \in G} y_g\,(x_g - \bar{x})\right\rangle}{\langle \bar{x}(1-\bar{x})\rangle} \\
&= \frac{\left\langle \frac{1}{n}\sum_{g \in G} y_g x_g - \bar{y}\bar{x}\right\rangle}{\langle \bar{x}(1-\bar{x})\rangle} \\
&= \lim_{u \to 0}\frac{\mathbb{E}_{\pi_{\mathcal{M}}}\left[\sum_{g \in G} y_g x_g - \bar{y}\bar{x}\right]}{\mathbb{E}_{\pi_{\mathcal{M}}}\left[\bar{x}(1-\bar{x})\right]} \\
&= \lim_{u \to 0}\frac{\mathrm{Cov}_{\pi_{\mathcal{M}},g}[y_g, x_g]}{\mathrm{Var}_{\pi_{\mathcal{M}},g}[x_g]}.
\end{aligned} \qquad (4.25)$$

In the last line, the numerator and denominator are, respectively, the covariance of $x_g$ with $y_g$ and the variance of $x_g$, with state $\mathbf{x}$ is sampled from $\pi_{\mathcal{M}}$ and $g$ sampled uniformly from $G$.

The final expression in Eq. (4.25) has the same form as a standard definition of relatedness based on linear regression [39, 40], which can be written in our framework as

$$r(\mathbf{x}) = \frac{\mathrm{Cov}_g[y_g, x_g]}{\mathrm{Var}_g[x_g]}. \qquad (4.26)$$

Although Eq. (4.25) and Eq. (4.26) have essentially the same form, they differ in that Eq. (4.26) applies to a specific state $\mathbf{x}$, whereas Eq. (4.25) averages over all states, weighted according to the low-mutation limit of the stationary distribution. We also emphasize that while Eq. (4.26) is typically applied in the case of individual actors, Eq. (4.25) allows for collective actors.

Therefore, the expectation of $r_{S,g}^A$, with $g$ sampled uniformly and $S$ sampled from the social environment of $g$, recovers a generalization of a standard regression definition of relatedness [39, 40]. A similar result can be obtained for $r_{S,g}^a$, by replacing $\imath_S^A(\mathbf{x})$ with $\imath_S^a(\mathbf{x})$ and $x_g$ with $1 - x_g$.

***4.4.4. Phenotypic covariance.*** Other common definitions of relatedness [38, 48] are based on covariance between phenotype and genotype of interacting individuals. To relate these definitions to ours, we apply the concept of social environment, from the previous subsection, at the level of phenotypes rather than alleles.

In this context, we represent the social environment of an individual $i \in I$ by a given, fixed probability distribution $\{p_{J|i}\}_{\varnothing \subset J \subseteq I}$ over nonempty sets of individuals $J \subseteq I$. We define $z_i$ as the probability, in a given state $\mathbf{x}$, that a set in $i$'s social environment contains only phenotype 1:

$$z_i = \sum_{J \subseteq I} p_{J|i} \, \mathbb{P}_{\mathbf{x}}[\Phi_j = 1, \; \forall j \in J] = \sum_{J \subseteq I} p_{J|i} \prod_{j \in J} \varphi_j(\mathbf{x}_{|G_j}). \tag{4.27}$$

Now we compute the expectation of $r_{J,i}^1$ as first $i$ is sampled uniformly from $I$, and then $J$ is sampled from $\{p_{J|i}\}$:

$$\begin{aligned}
\mathbb{E}_{i,J}\left[r_{J,i}^1\right] &= \frac{1}{N} \sum_{i \in I} \sum_{\emptyset \subseteq J \subseteq I} p_{J|i} r_{I,j}^1 \\
&= \frac{\left\langle \frac{1}{N} \sum_{i \in I} \sum_{\emptyset \subseteq J \subseteq I} p_{J|i} \prod_{j \in J} \varphi_j(\mathbf{x}_{|G_j}) \left(\bar{x}_i - \bar{x}\right) \right\rangle}{\langle \bar{x}(1 - \bar{x}) \rangle} \\
&= \frac{\left\langle \frac{1}{N} \sum_{i \in I} z_i \left(\bar{x}_i - \bar{x}\right) \right\rangle}{\langle \bar{x}(1 - \bar{x}) \rangle} \\
&= \frac{\left\langle \frac{1}{N} \sum_{i \in I} z_i \bar{x}_i - \bar{z}\bar{x} \right\rangle}{\langle \bar{x}(1 - \bar{x}) \rangle} \\
&= \lim_{u \to 0} \frac{\mathrm{Cov}_{\pi_{\mathcal{M}}, i}[z_i, \bar{x}_i]}{\mathrm{Var}_{\pi_{\mathcal{M}}, g}[x_g]}.
\end{aligned} \tag{4.28}$$

This resulting expression for $\mathbb{E}_{i,J}\left[r_{J,i}^1\right]$ is closely related to relatedness measure developed by Michod and Hamilton [38], which can be expressed in our framework as

$$r(\mathbf{x}) = \frac{\mathrm{Cov}_i[z_i, \bar{x}_i]}{\mathrm{Cov}_i[\varphi_i, \bar{x}_i]}. \tag{4.29}$$

However, our result in Eq. (4.28) differs from Michod and Hamilton's [38] definition, Eq. (4.29), in three ways: (i) Michod and Hamilton's definition applies to individual actors only, whereas ours allows for collective actors; (ii) Eq. (4.29) applies in a particular state, whereas Eq. (4.28) averages over all states in the $u \to 0$ limit of the stationary distribution; (iii) the denominators differ, which amounts to a different choice of normalization.

### 4.4.5. Queller's (1985) coefficient of synergism.
We now turn to measures of relatedness between a pair of individuals and a third individual, which are precursors of the collective relatedness measure introduced here.

Queller [49] introduced a "coefficient of synergism", defined in the final term of his Eq. (3), to quantify how frequently an individual $i$ shares a phenotype with its social partners. This coefficient of synergism, denoted $s(\mathbf{x})$, is defined the same way as $r(\mathbf{x})$ in Eq. (4.29), but with $i$'s "social environment" consisting of pairs of the form $J = \{i, j\}$, so that $z_i$ is given by

$$z_i = \sum_{j \in I} p_{\{i,j\}|i} \, \varphi_i(\mathbf{x}_{|G_i}) \, \varphi_j(\mathbf{x}_{|G_j}). \tag{4.30}$$

In light of the discussion in Section 4.4.4, Queller's $s(\mathbf{x})$ is closely related to our $r_{\{i,j\},i}^1$. The differences are that (i) $s(\mathbf{x})$ averages over pairs in a given state $\mathbf{x}$, whereas $r_{\{i,j\},i}^1$ pertains to a single pair $\{i, j\}$, averaged over the $u \to 0$ limit of the stationary distribution, and (ii) the denominators differ, amounting to a difference in normalization.

***4.4.6. Taylor's (2013) joint relatedness.*** Taylor [50] introduced a measure of joint relatedness between a two haploid individuals and a third. In our notation, for three sites $g, h, k \in G$, Taylor's joint relatedness—as defined in Eq. (9) of Ref. [50]—can be written as

$$R_{gh-k} = \frac{\mathrm{Cov}_{\pi_{\mathcal{M}}}[x_g x_h, x_k]}{\mathrm{Var}_{\pi_{\mathcal{M}}}[x_k]}. \tag{4.31}$$

The idea is that sites $g$ and $h$ jointly produce a synergistic effect on the fitness of site $k$, and this synergistic effect is weighted by the joint relatedness $R_{gh-k}$ in determining the consequences for selection.

Taylor's $R_{gh-k}$ serves a similar role as our collective relatedness $r_{S,k}$ (with $S = \{g, h\}$), and the formulas are similar as well. However, the two definitions are not equivalent, as can be seen by taking the low-mutation limit of Eq. (4.31):

$$\begin{aligned}
\lim_{u \to 0} R_{gh-k} &= \lim_{u \to 0} \frac{\mathrm{Cov}_{\pi_{\mathcal{M}}}[x_g x_h, x_k]}{\mathrm{Var}_{\pi_{\mathcal{M}}}[x_k]} \\
&= \lim_{u \to 0} \frac{\mathbb{E}_{\pi_{\mathcal{M}}}[x_g x_h x_k] - \mathbb{E}_{\pi_{\mathcal{M}}}[x_g x_h] \mathbb{E}_{\pi_{\mathcal{M}}}[x_k]}{\mathbb{E}_{\pi_{\mathcal{M}}}[x_k] - (\mathbb{E}_{\pi_{\mathcal{M}}}[x_k])^2} \\
&= \frac{\pi_{\mathcal{M}_0}(\mathbf{A}) - (\pi_{\mathcal{M}_0}(\mathbf{A}))^2}{\pi_{\mathcal{M}_0}(\mathbf{A}) - (\pi_{\mathcal{M}_0}(\mathbf{A}))^2} \\
&= 1.
\end{aligned}$$

Thus, for each triple $g, h, k \in G$, Taylor's $R_{gh-k}$ converges to 1 in the low-mutation limit. Collective relatedness $r_{S,g}$, which is already defined as a $u \to 0$ limit, provides a more informative quantification of genetic assortment under low mutation.

## 5. Main results

Having quantified selection in Section 3, and genetic assortment in Section 4, we are now in a position to state and prove our main results, Theorems 5.1 and 5.2, and extensions thereof.

**5.1. Synergistic fitness effects.** Our first task is to quantify nonlinear interactions in terms of synergistic effects. For each site $g \in G$, the fitness increment $w_g(\mathbf{x})$ can be uniquely represented [51] in the form

$$w_g(\mathbf{x}) = \sum_{S \subseteq G} c_{S,g}\, \iota_S^A(\mathbf{x}). \tag{5.1}$$

We interpret $c_{S,g}$ as the synergistic effect of set $S$ having allele $A$, relative to $a$, on the fitness of site $g$. An explicit formula is given by

$$c_{S,g} = \sum_{T \subseteq S} (-1)^{|S|-|T|} w_g\left(\mathbf{1}_T\right), \tag{5.2}$$

where $\mathbf{1}_T$, for $T \subseteq G$, is the state with $x_g = 1$ for $g \in T$ and $x_g = 0$ for $g \notin T$.

By Eq. (3.6), the total synergistic fitness effect of any set $S$, over all target sites $g$, is zero:

$$\sum_{g \in G} c_{S,g} = \sum_{T \subseteq S} (-1)^{|S|-|T|} \sum_{g \in G} w_g\left(\mathbf{1}_T\right) = 0. \tag{5.3}$$

Eq. (5.1) represents synergistic fitness effects from the perspective of allele $A$. Alternatively, we may take the perspective of allele $a$, and uniquely write

$$w_g(\mathbf{x}) = \sum_{S \subseteq G} \tilde{c}_{S,g}\, \iota_S^a(\mathbf{x}), \tag{5.4}$$

where the $\tilde{c}_{S,g}$ are given explicitly by

$$\tilde{c}_{S,g} = \sum_{T \subseteq S} (-1)^{|S|-|T|} \, w_g \left( \mathbf{1}_{G-T} \right), \tag{5.5}$$

and satisfy $\sum_{S \subseteq G} \tilde{c}_{S,g} = 0$. Using Eq. (4.2), we obtain that $c_{S,g}$ and $\tilde{c}_{S,g}$ are related by

$$c_{S,g} = (-1)^{|S|} \sum_{T \supseteq S} \tilde{c}_{T,g}, \qquad \tilde{c}_{S,g} = (-1)^{|S|} \sum_{T \supseteq S} c_{T,g}. \tag{5.6}$$

It follows from Eq. (5.6) that the maximal degree of synergy does not depend on whether the representation in Eq. (5.1) or in Eq. (5.4) is used. By this we mean that if there is some $d \geq 0$ such that $c_{S,g} = 0$ whenever $|S| > d$, then it is also true that $\tilde{c}_{S,g} = 0$ whenever $|S| > d$.

Most flexibly, we can represent fitness as

$$w_g(\mathbf{x}) = \sum_{S \subseteq G} \left( c_{S,g}^A \, \iota_S^A(\mathbf{x}) + c_{S,g}^a \, \iota_S^a(\mathbf{x}) \right), \tag{5.7}$$

where the $c_{S,g}^A$ and $c_{S,g}^a$ are subject to

$$\sum_{g \in G} c_{S,g}^A = \sum_{g \in G} c_{S,g}^a = 0 \qquad \text{for each } S \subseteq G, \tag{5.8a}$$

and

$$c_{S,g}^A = c_{S,g}^a = 0, \text{ for all } S \subseteq G \text{ and } g \in G, \text{ when } \delta = 0. \tag{5.8b}$$

Although the representation in Eq. (5.7) is not unique, it will prove useful in later analysis.

With fitness represented this way, the change due to selection can be written using Eq. (3.9), Eq. (4.3), and Eq. (5.8a) as

$$\begin{aligned}
\Delta(\mathbf{x}) &= \frac{1}{n} \sum_{g \in G} x_g \sum_{S \subseteq G} \left( c_{S,g}^A \, \iota_S^A(\mathbf{x}) + c_{S,g}^a \, \iota_S^a(\mathbf{x}) \right) \\
&= \frac{1}{n} \sum_{g \in G} \sum_{S \subseteq G} \left( c_{S,g}^A \, \iota_{S \cup \{g\}}^A(\mathbf{x}) + c_{S,g}^a \left( \iota_S^a(\mathbf{x}) - \iota_{S \cup \{g\}}^a(\mathbf{x}) \right) \right) \\
&= \frac{1}{n} \sum_{g \in G} \sum_{S \subseteq G} \left( c_{S,g}^A \, \iota_{S \cup \{g\}}^A(\mathbf{x}) - c_{S,g}^a \, \iota_{S \cup \{g\}}^a(\mathbf{x}) \right).
\end{aligned} \tag{5.9}$$

In particular, for the representations in Eq. (5.1) and Eq. (5.4), we have

$$\Delta(\mathbf{x}) = \frac{1}{n} \sum_{g \in G} \sum_{S \subseteq G} c_{S,g} \, \iota_{S \cup \{g\}}^A(\mathbf{x}) = -\frac{1}{n} \sum_{g \in G} \sum_{S \subseteq G} \tilde{c}_{S,g} \, \iota_{S \cup \{g\}}^a(\mathbf{x}). \tag{5.10}$$

The second equality can also be obtained directly using Eq. (4.2), Eq. (5.6), and Eq. (5.8a). We caution that the sums over both $S$ and $g$ are required for the second equality to hold; it is not true in general that $\sum_{g \in G} c_{S,g} \, \iota_{S \cup \{g\}}^A(\mathbf{x}) = -\sum_{g \in G} \tilde{c}_{S,g} \, \iota_{S \cup \{g\}}^a(\mathbf{x})$, nor that $\sum_{S \subseteq G} c_{S,g} \, \iota_{S \cup \{g\}}^A(\mathbf{x}) = -\sum_{S \subseteq G} \tilde{c}_{S,g} \, \iota_{S \cup \{g\}}^a(\mathbf{x})$.

**5.2. Condition for success under arbitrary selection strength.** Our main result is the following:

**Theorem 5.1.** *Suppose fitness is represented as in* Eq. (5.7), *subject to* Eq. (5.8a). *Then A is favored over a, in the sense of Theorem* 3.3, *if and only if*

$$\sum_{g \in G} \sum_{S \subseteq G} c_{S,g}^A \, r_{S,g}^A > \sum_{g \in G} \sum_{S \subseteq G} c_{S,g}^a \, r_{S,g}^a. \tag{5.11}$$

In particular, Condition (5.11) becomes $\sum_{g \in G} \sum_{S \subseteq G} c_{S,g} r_{S,g}^A > 0$ for the representation in Eq. (5.1), giving Eq. (10) of the main text. If we instead take allele $a$'s perspective, using the representation in Eq. (5.4), we obtain $\sum_{g \in G} \sum_{S \subseteq G} \tilde{c}_{S,g} r_{S,g}^a < 0$.

*Proof.* Using Eq. (5.8a), we can rewrite Eq. (5.9) as

$$\Delta(\mathbf{x}) = \frac{1}{n} \sum_{g \in G} \sum_{S \subseteq G} \left( c_{S,g}^A \left( \imath_{S \cup \{g\}}^A(\mathbf{x}) - \bar{\imath}_S^A(\mathbf{x}) \right) - c_{S,g}^a \left( \imath_{S \cup \{g\}}^a(\mathbf{x}) - \bar{\imath}_S^a(\mathbf{x}) \right) \right). \tag{5.12}$$

Applying the operator $\langle \, \rangle$ to both sides, we have

$$\langle \Delta \rangle = \frac{1}{n} \sum_{g \in G} \sum_{S \subseteq G} \left( c_{S,g}^A \left\langle \imath_{S \cup \{g\}}^A - \bar{\imath}_S^A \right\rangle - c_{S,g}^a \left\langle \imath_{S \cup \{g\}}^a - \bar{\imath}_S^a \right\rangle \right).$$

By Theorem 3.3, $\rho_A > \rho_a$ if and only if

$$\sum_{g \in G} \sum_{S \subseteq G} c_{S,g}^A \left\langle \imath_{S \cup \{g\}}^A - \bar{\imath}_S^A \right\rangle > \sum_{g \in G} \sum_{S \subseteq G} c_{S,g}^a \left\langle \imath_{S \cup \{g\}}^a - \bar{\imath}_S^a \right\rangle.$$

The result then follows from multiplying both sides by $2/\bar{\ell}$ and applying Eq. (4.12). $\square$

**5.3. Condition for success under weak selection.** Theorem 5.1 holds for arbitrary strength of selection $\delta > 0$. However, the collective relatedness coefficients in Condition (5.11) are difficult to evaluate, because they depend on the stationary distribution $\pi_\mathcal{M}$, which itself depends on the process of selection. For a more tractable condition, we prove a weak-selection version of Theorem 5.1:

**Theorem 5.2.** *Suppose fitness is represented as in* Eq. (5.7), *subject to* Eq. (5.8). *Then weak selection favors A over a in the sense of Theorem 3.4 if and only if*

$$\sum_{g \in G} \sum_{S \subseteq G} (c_{S,g}^A)' r_{S,g}^\circ > \sum_{g \in G} \sum_{S \subseteq G} (c_{S,g}^a)' r_{S,g}^\circ. \tag{5.13}$$

*Proof.* Taking the $\delta$-derivative of Eq. (5.12) at $\delta = 0$, and invoking Eq. (5.8b), we obtain

$$\Delta'(\mathbf{x}) = \frac{1}{n} \sum_{g \in G} \sum_{S \subseteq G} \left( (c_{S,g}^A)' \left( \imath_{S \cup \{g\}}^A(\mathbf{x}) - \bar{\imath}_S^A(\mathbf{x}) \right) - (c_{S,g}^a)' \left( \imath_{S \cup \{g\}}^a(\mathbf{x}) - \bar{\imath}_S^a(\mathbf{x}) \right) \right). \tag{5.14}$$

By Theorem 3.4, weak selection favors $A$ over $a$ if and only if

$$\sum_{g \in G} \sum_{S \subseteq G} \left( (c_{S,g}^A)' \left\langle \imath_{S \cup \{g\}}^A - \bar{\imath}_S^A \right\rangle^\circ - (c_{S,g}^a)' \left\langle \imath_{S \cup \{g\}}^a - \bar{\imath}_S^a \right\rangle^\circ \right) > 0. \tag{5.15}$$

The result follows from multiplying by $2/\bar{\ell}^\circ$ and applying Eq. (4.12) and Eq. (4.14). $\square$

**5.4. Class-based conditions for success.** The conditions in Theorems 5.1 and 5.2 involve a term on each side for every pair of subset $S \subseteq G$ and site $g \in G$. However, symmetry can greatly reduce the number of terms, so that only one term is needed per *interaction class*. An interaction class is a possible distinct variety of collective-site pair. For example, on a cycle (main text, Fig 3b) a collective $S$ of three adjacent nodes, and a target node $g$ adjacent to $S$, comprises an interaction class. One one term is needed for all collective-site pairs of this form.

To define interaction classes formally, we introduce an equivalence relation on pairs $(S, g)$ (with $S \subseteq G$ and $g \in G$). We say $(S, g)$ is equivalent to $(T, h)$ if there is a symmetry $\sigma$ such that $\sigma(S) = T$ and $\sigma(g) = h$. We define interaction classes to be the equivalence classes induced by this relation. That is, the interaction class of a pair $(S, g)$ is the set of pairs

$$[S, g] = \{(T, h) \mid T = \sigma(S) \text{ and } h = \sigma(g) \text{ for some symmetry } \sigma\}. \tag{5.16}$$

In order to formulate the condition for success under symmetry, we must first show that symmetry preserves the collective fitness effects $c_{S,g}$:

**Lemma 5.3.** *The representation of selection increments in* Eq. (5.1) *satisfies*

$$c_{S,g} = c_{\sigma(S),\sigma(g)} \qquad (5.17)$$

*for any site $g \in G$, subset $S \subseteq G$, and symmetry $\sigma \in \mathrm{Sym}(G,p)$. The analogous statement also holds for the representation in* Eq. (5.4).

*Proof.* We will obtain two equivalent expressions for $w_{\sigma(g)}(\mathbf{x})$, and use the uniqueness of the representation in Eq. (5.1) to prove the desired symmetry property. First, using Eq. (3.7),

$$w_{\sigma(g)}(\mathbf{x}) = w_g(\mathbf{x}_\sigma) = \sum_{S \subseteq G} c_{S,g}\, \iota_S^A(\mathbf{x}_\sigma).$$

For the second expression, we write

$$\begin{aligned}
w_{\sigma(g)}(\mathbf{x}) &= \sum_{S \subseteq G} c_{S,\sigma(g)}\, \iota_S^A(\mathbf{x}) && \text{by Eq. (5.7)} \\
&= \sum_{S \subseteq G} c_{\sigma(S),\sigma(g)}\, \iota_{\sigma(S)}^A(\mathbf{x}) && \text{since } \sigma \text{ is a bijection} \\
&= \sum_{S \subseteq G} c_{\sigma(S),\sigma(g)}\, \iota_S^A(\mathbf{x}_\sigma) && \text{by Eq. (4.4).}
\end{aligned}$$

These two expressions for $w_{\sigma(g)}(\mathbf{x})$ differ only in the replacement of $c_{S,g}$ by $c_{\sigma(S),\sigma(g)}$ for each $S \subseteq G$ and $g \in G$. Since the representation in Eq. (5.1) is unique, and since $\sigma$ is a bijection and $g$ and $\mathbf{x}$ were arbitrarily chosen, we conclude that $c_{S,g} = c_{\sigma(S),\sigma(g)}$ for all $S \subseteq G$ and $g \in G$. The analogous statement for Eq. (5.4) is proven similarly. □

In light of Lemma 5.3, we may, without losing generality, choose collective fitness effects in Eq. (5.7) that obey the symmetry property

$$c_{S,g}^A = c_{\sigma(S),\sigma(g)}^A \qquad \text{and} \qquad c_{S,g}^a = c_{\sigma(S),\sigma(g)}^a. \qquad (5.18)$$

This means that, we may assign collective fitness effects $c_{[S,g]}^A$ and $c_{[S,g]}^a$ to each interaction class $[S,g]$, knowing that these effects are the same for all pairs in a given class. Similarly, by Eq. (4.17), we may assign collective relatednesses $r_{[S,g]}^A$ and $r_{[S,g]}^a$ to each interaction class $[S,g]$. Now, letting $n_{[S,g]}$ denote the number of pairs in interaction class $[S,g]$, Condition (5.11) for allele $A$ to be favored can be rewritten as

$$\sum_{[S,g]} n_{[S,g]}\, c_{[S,g]}^A r_{[S,g]}^A > \sum_{[S,g]} n_{[S,g]}\, c_{[S,g]}^a r_{[S,g]}^a. \qquad (5.19)$$

Each side in Condition (5.19) contains a single term for each interaction class. The corresponding weak-selection result, Condition (5.13), can be rewritten in similar fashion.

A special case arises for homogeneous populations, in which any two sites are equivalent by symmetry (see Section 1.8). In this case, one may arbitrarily choose a "focal" site $g \in G$. Any interaction class will have a unique representative $(S,g)$, where $g$ is this focal site. The condition for success can then be restated in terms of the focal site:

$$\sum_{S \subseteq G} c_{S,g}^A r_{S,g}^A > \sum_{S \subseteq G} c_{S,g}^a r_{S,g}^a. \qquad (5.20)$$

**5.5. Conditions for success at the phenotype level.** The conditions in Theorem 5.1 and 5.2 are expressed at the level of genetic sites. Here we derive analogous conditions at the level of phenotypes, using the phenotypic formalism introduced in Section 1.7 and the definition of relatedness from Section 4.3.3.

As in Section 1.7, we use the accent ˘ to indicate quantities that depend on the phenotypic state $\boldsymbol{\Phi} \in \{0,1\}^I$, rather than the (allelic) population state $\mathbf{x} \in \{0,1\}^G$. The fitness increment of each site $g \in G$ in phenotypic state $\boldsymbol{\Phi}$ is defined as

$$\breve{w}_g(\boldsymbol{\Phi}) = \breve{\mathbb{E}}_{\boldsymbol{\Phi}} \left[ \sum_{h \in \alpha^{-1}(g)} v_h \right] - v_g = \sum_{\alpha: G \to G} \breve{p}_{\boldsymbol{\Phi}}(\alpha) \sum_{h \in \alpha^{-1}(g)} v_h - v_g. \tag{5.21}$$

The fitness increments in population sate $\mathbf{x}$ are then recovered by

$$w_g(\mathbf{x}) = \mathbb{E}_{\mathbf{x}}[\breve{w}_g(\boldsymbol{\Phi})]. \tag{5.22}$$

To proceed, we introduce an assumption that sites within a given individual have the same fitness increment:

(I) For each individual $i \in I$ and each phenotypic state $\boldsymbol{\Phi}$, the fitness increment of each site in $i$ is the same: $\breve{w}_g(\boldsymbol{\Phi}) = \breve{w}_h(\boldsymbol{\Phi})$ for each $g, h \in G_i$.

Assumption (I) formalizes the principle of "fair meiosis" in Mendelian inheritance. It excludes the possibility of gene drive, in which certain alleles are more likely than others in the same individual to be transmitted during meiosis [52].

With Assumption (I) in force, we let $\breve{w}_i(\boldsymbol{\Phi})$ denote the fitness increment of each site in individual $i \in I$, so that $\breve{w}_g(\boldsymbol{\Phi}) = \breve{w}_i(\boldsymbol{\Phi})$ for each $g \in G_i$. By Eq. (3.6) we have

$$\sum_{i \in I} n_i \breve{w}_i(\boldsymbol{\Phi}) = 0. \tag{5.23}$$

We next obtain a phenotype-level analogue of Eq. (3.9), which can also be understood as an instance of the Price equation [35].

**Lemma 5.4.** *If Assumption (I) holds, then the selection increment in state $\mathbf{x}$ is given by*

$$\Delta(\mathbf{x}) = \frac{1}{n} \sum_{i \in I} n_i \, \mathbb{E}_{\mathbf{x}} \left[ \breve{w}_i(\boldsymbol{\Phi}) \right] \bar{x}_i = \frac{1}{n} \sum_{i \in I} n_i \, \mathbb{E}_{\mathbf{x}} \left[ \breve{w}_i(\boldsymbol{\Phi}) \right] (\bar{x}_i - \bar{x}). \tag{5.24}$$

*In particular, if each individual has the same ploidy (=number of sites) then*

$$\Delta(\mathbf{x}) = \frac{1}{N} \sum_{i \in I} \mathbb{E}_{\mathbf{x}}[\breve{w}_i(\boldsymbol{\Phi})] \bar{x}_i = \frac{1}{N} \sum_{i \in I} \mathbb{E}_{\mathbf{x}} \left[ \breve{w}_i(\boldsymbol{\Phi}) \right] (\bar{x}_i - \bar{x}), \tag{5.25}$$

*where $N = |I|$ is the number of individuals.*

*Proof.* We begin with Eq. (3.9):

$$\Delta(\mathbf{x}) = \frac{1}{n} \sum_{g \in G} x_g w_g(\mathbf{x})$$

$$= \frac{1}{n} \sum_{i \in I} \sum_{g \in G_i} x_g w_g(\mathbf{x})$$

$$= \frac{1}{n} \sum_{i \in I} \sum_{g \in G_i} x_g \, \mathbb{E}_{\mathbf{x}}[\breve{w}_g(\boldsymbol{\Phi})].$$

Now invoking Assumption (I), we have

$$\Delta(\mathbf{x}) = \frac{1}{n} \sum_{i \in I} \mathbb{E}_\mathbf{x}[\breve{w}_i(\boldsymbol{\Phi})] \sum_{g \in G_i} x_g$$

$$= \frac{1}{n} \sum_{i \in I} n_i \, \mathbb{E}_\mathbf{x}\left[\breve{w}_i(\boldsymbol{\Phi})\right] \bar{x}_i$$

$$= \frac{1}{n} \sum_{i \in I} n_i \, \mathbb{E}_\mathbf{x}\left[\breve{w}_i(\boldsymbol{\Phi})\right] (\bar{x}_i - \bar{x}).$$

The last equality follows from Eq. (5.23). This proves Eq. (5.24). Eq. (5.25) then follows from observing that if $n_i$ is constant over all individuals $i \in I$, then $\frac{n_i}{n} = \frac{1}{N}$. □

As in Section 5.1, we write each fitness increment $\breve{w}_i(\boldsymbol{\Phi})$ uniquely in the form

$$\breve{w}_i(\boldsymbol{\Phi}) = \sum_{J \subseteq I} \breve{c}_{J,i} \left( \prod_{j \in J} \Phi_j \right), \tag{5.26}$$

for some coefficients $\breve{c}_{J,i}$ with $\sum_{i \in I} \breve{c}_{J,i} = 0$ for each $J \subseteq I$. We then obtain a phenotype-level version of Theorem 5.1:

**Theorem 5.5.** *If Assumption (I) holds, then selection favors allele A if and only if*

$$\sum_{i \in I} n_i \sum_{J \subseteq I} \breve{c}_{J,i} \, r^1_{J,i} > 0. \tag{5.27a}$$

*In particular, if all individuals have the same ploidy, selection favors A if and only if*

$$\sum_{i \in I} \sum_{J \subseteq I} \breve{c}_{J,i} \, r^1_{J,i} > 0. \tag{5.27b}$$

*Proof.* Taking the expectation of Eq. (5.26) in state $\mathbf{x}$ gives

$$\mathbb{E}_\mathbf{x}\left[\breve{w}_i(\boldsymbol{\Phi})\right] = \sum_{J \subseteq I} \breve{c}_{J,i} \, \mathbb{E}_\mathbf{x}\left[\prod_{j \in J} \Phi_j\right] = \sum_{J \subseteq I} \breve{c}_{J,i} \left( \prod_{j \in J} \varphi_j(\mathbf{x}_{|G_j}) \right). \tag{5.28}$$

Combining with Eq. (5.24), the selection increment can be written

$$\Delta(\mathbf{x}) = \frac{1}{n} \sum_{i \in I} n_i \sum_{J \subseteq I} \breve{c}_{J,i} \left( \prod_{j \in J} \varphi_j(\mathbf{x}_{|G_j}) \right) (\bar{x}_i - \bar{x}). \tag{5.29}$$

Applying the operator $\langle \, \rangle$ to both sides and invoking Eq. (4.19a), we obtain

$$\langle \Delta \rangle = \frac{1}{n} \langle \bar{x}(1 - \bar{x}) \rangle \sum_{i \in I} n_i \sum_{J \subseteq I} \breve{c}_{J,i} \, r^1_{J,i}. \tag{5.30}$$

The result now follows from Theorem 3.3. □

**5.6. Quantitative phenotypic traits.** So far, we have considered only two discrete phenotypes, 0 and 1. We now consider competition between quantitative (real-valued) phenotypic traits. Specifically, we suppose that alleles $a$ and $A$ are associated with trait values $\Phi_0 \in \mathbb{R}$, and $\Phi_0 + \epsilon$, respetively. Here, $\epsilon$ quantifies the phenotypic effect of allele $A$ relative to $a$. Our aim in this section is to obtain approximate conditions for selection when the relative phenotypic effect $\epsilon$ is small. Small phenotypic effects ($\epsilon \ll 1$) are a special case of weak selection ($\delta \ll 1$), as elucidated by Wild and Traulsen [10].

Similarly as in Section 1.7, we suppose that individual $i \in I$ has phenotype $\Phi_i = \Phi_0 + \epsilon$ for some probability $\varphi_i(\mathbf{x}_{|G_i})$ and otherwise has phenotype $\Phi_i = \Phi_0$. We require that $\varphi_i(0, \ldots, 0) = 0$ and $\varphi_i(1, \ldots, 1) = 1$, so that all-$a$ individuals always have phenotype $\Phi_0$ and all-$A$ individuals always have phenotype $\Phi_0 + \epsilon$.

Now we suppose that the fitness of each individual depends smoothly on the phenotype of each individual. That is, for each $i \in I$, $\breve{w}_i(\boldsymbol{\Phi})$ is a smooth (infinitely differentiable) function of the overall phenotypic state $\boldsymbol{\Phi} \in \mathbb{R}^I$. This allows $\breve{w}_i(\boldsymbol{\Phi})$ to be expressed as a multivariable Taylor series. Specifically, let $\boldsymbol{\Phi}_0 = (\Phi_0, \ldots, \Phi_0)$ denote the phenotypic state that corresponds to the all-$a$ population state, $\mathbf{a}$. Then for any phenotypic state $\boldsymbol{\Phi}$ sufficiently close to $\boldsymbol{\Phi}_0$, and any $m \geq 1$, we have the Taylor approximation

$$\breve{w}_i(\boldsymbol{\Phi}) = \sum_{\substack{J \subseteq I \\ |J| \leq m}} \sum_{\substack{\mathbf{k} \in \mathbb{N}^J \\ |\mathbf{k}| \leq m}} \frac{\partial^{\mathbf{k}} \breve{w}_i(\boldsymbol{\Phi}_0)}{\mathbf{k}!} (\boldsymbol{\Phi} - \boldsymbol{\Phi}_0)^{\mathbf{k}} + \mathcal{O}\left(|\boldsymbol{\Phi} - \boldsymbol{\Phi}_0|^{m+1}\right). \tag{5.31}$$

Each term in this Taylor expansion is characterized by a subset $J \subseteq I$, and a multi-index $\mathbf{k} = (k_j)_{j \in J} \in \mathbb{N}^J$ of positive integers, indicating in which variables $\Phi_j$, to which orders $k_j$, the derivatives of $\breve{w}_i$ are taken. Specifically, $\partial^{\mathbf{k}} \breve{w}_i$ denotes the partial derivative of $\breve{w}_i$, taken to order $k_j$ in $\Phi_j$, for each $j \in J$. The overall order of this derivative is $|\mathbf{k}| = \sum_{j \in J} k_j$. In accordance with the multivariable Taylor's theorem, each derivative is evaluated at $\boldsymbol{\Phi}_0$, multiplied by $(\boldsymbol{\Phi} - \boldsymbol{\Phi}_0)^{\mathbf{k}} = \prod_{j \in J} (\Phi_j - \Phi_0)^{k_j}$, and divided by the factorial $\mathbf{k}! = \prod_{j \in I} k_j!$.

For a state containing only the phenotypes $\Phi_0$ and $\Phi_0 + \epsilon$, we have

$$(\boldsymbol{\Phi} - \boldsymbol{\Phi}_0)^{\mathbf{k}} = \prod_{j \in J} (\Phi_j - \Phi_0)^{k_j} = \begin{cases} \epsilon^{|\mathbf{k}|} & \text{if } \Phi_j = \Phi_0 + \epsilon \text{ for all } j \in J \\ 0 & \text{otherwise.} \end{cases} \tag{5.32}$$

Therefore, recalling that $\Phi_j$ takes value $\Phi_0 + \epsilon$ with probability $\varphi_j(\mathbf{x}_{|G_j})$, we have, on expectation,

$$\mathbb{E}_{\mathbf{x}}\left[(\boldsymbol{\Phi} - \boldsymbol{\Phi}_0)^{\mathbf{k}}\right] = \epsilon^{|\mathbf{k}|} \prod_{j \in J} \varphi_j(\mathbf{x}_{|G_j}). \tag{5.33}$$

Taking the expectation of Eq. (5.31) then yields

$$\mathbb{E}_{\mathbf{x}}\left[\breve{w}_i(\boldsymbol{\Phi})\right] = \sum_{\substack{J \subseteq I \\ |J| \leq m}} \left(\sum_{\substack{\mathbf{k} \in \mathbb{N}^J \\ |\mathbf{k}| \leq m}} \frac{\partial^{\mathbf{k}} \breve{w}_i(\boldsymbol{\Phi}_0)}{\mathbf{k}!} \epsilon^{|\mathbf{k}|}\right) \prod_{j \in J} \varphi_j(\mathbf{x}_{|G_j}) + \mathcal{O}\left(\epsilon^{m+1}\right). \tag{5.34}$$

Comparing to Eq. (5.28), we observe that the fitness effect of collective $J \subseteq I$ on individual $i \in I$ can be approximated as

$$\breve{c}_{J,i} = \sum_{\substack{\mathbf{k} \in \mathbb{N}^J \\ |\mathbf{k}| \leq m}} \frac{\partial^{\mathbf{k}} \breve{w}_i(\boldsymbol{\Phi}_0)}{\mathbf{k}!} \epsilon^{|\mathbf{k}|} + \mathcal{O}\left(\epsilon^{m+1}\right). \tag{5.35}$$

Applying Theorem 5.5 and reducing by a factor of $\epsilon$, we obtain an $m$th-order approximation to the condition for $A$ to be selected:

$$\sum_{i \in I} n_i \sum_{\substack{J \subseteq I \\ |J| \leq m}} \left(\sum_{\substack{\mathbf{k} \in \mathbb{N}^J \\ |\mathbf{k}| \leq m}} \frac{\partial^{\mathbf{k}} \breve{w}_i(\boldsymbol{\Phi}_0)}{\mathbf{k}!} \epsilon^{|\mathbf{k}|-1}\right) r_{J,i} > 0. \tag{5.36}$$

Above, the collective phenotypic relatedness $r_{J,i}$ is defined as

$$r_{J,i} = \frac{\left\langle \prod_{j \in J} \varphi_j(\mathbf{x}_{|G_j})\, (\bar{x}_i - \bar{x}) \right\rangle}{\langle \bar{x}(1-\bar{x}) \rangle}, \tag{5.37}$$

in accordance with Eq. (4.19a). We observe that this $m$th-order approximation, Condition (5.36) involves fitness effects due only to collectives of size at most $m$.

Setting $m = 1$ yields a linear approximation, which determines the direction of adaptive change when the mutation effect $\epsilon$ is very small [53, 54]:

$$\sum_{i,j \in I} n_i \frac{\partial \breve{w}_i}{\partial \Phi_j}(\mathbf{\Phi}_0)\, r_{\{j\},i} > 0. \tag{5.38}$$

In this linear approximation, the condition for success involves only fitness effects due to single individuals. The effect of individual $j$ on $i$'s fitness is quantified by the partial derivative $\frac{\partial \breve{w}_i}{\partial \Phi_j}(\mathbf{\Phi}_0)$, and is weighted by the relatedness $r_{\{j\},i}$ between them. If the ploidy $n_i$ is constant (e.g., all haploid or all diploid), then Condition (5.38) can be written as a sum of inclusive fitness effects, $\sum_{j \in I} w_j^{\text{IF}} > 0$, where $w_j^{\text{IF}} = \sum_{i \in I} \frac{\partial \breve{w}_i}{\partial \Phi_j}(\mathbf{\Phi}_0)\, r_{\{j\},i}$ is the inclusive fitness effect of individual $j$, as defined, for example, in Eq. (11) of Rousset and Billiard [21]. Condition (5.38) was also obtained, in the case of a homogeneous haploid population, as Theorem 2 of Wakano et al. [19].

A quadratic approximation—which is relevant to evolutionary stability and branching [28, 55]—is obtained by setting $m = 2$:

$$\sum_{i,j \in I} n_i \left( \frac{\partial \breve{w}_i}{\partial \Phi_j}(\mathbf{\Phi}_0) + \frac{\epsilon}{2} \frac{\partial^2 \breve{w}_i}{\partial \Phi_j^2}(\mathbf{\Phi}_0) \right) r_{\{j\},i} + \frac{\epsilon}{2} \sum_{\substack{i,j,k \in I \\ k \neq j}} n_i \frac{\partial^2 \breve{w}_i}{\partial \Phi_j \partial \Phi_k}(\mathbf{\Phi}_0)\, r_{\{j,k\},i} > 0. \tag{5.39}$$

The first term in Condition (5.39) quantifies effects due to individuals $j$, while the second term quantifies effects due to pairs $\{j, k\}$. Related results, in populations comprised of many identical islands, were obtained by Ajar [56], Wakano and Lehmann [57], and Mullon et al. [28].

## 6. Collective Action Dilemma

Having derived conditions for selection in arbitrary social interactions, we now introduce the Collective Action Dilemma as a simple model of costly collective action towards an individual. We describe this dilemma here in the context of a haploid population; we will extend to diploids in Section 8.

A set $S \subseteq G$, of size $m$, may (or may not) pay total cost $c$, divided evenly among $S$'s members, to help or harm a target site $g \in G$, inside or outside of $S$. Individuals with allele $A$ contribute to the action, while those with allele $a$ do not.

### 6.1. Unconditional costs.
We first suppose that costs are unconditional, meaning each individual in $S$ with an $A$ allele pays cost $c/m$. Those with allele $a$ do not pay any costs.

The benefit to site $g$ depends on the fraction $\bar{x}_S = \frac{1}{m}\sum_{h \in S} x_h$ of $S$ that contributes, and is represented as a function $b(\bar{x}_S)$. Positive values of $b(\bar{x}_S)$ represent help to $g$, while negative values represent harm. We assume $b(0) = 0$ (no contribution leads to no help or harm), and use the shorthand $b(1) = b$. The main text focuses on *all-or-nothing* benefits, $b(x) = 0$ for $0 \leq x < 1$, meaning that the benefit $b$ is achieved only if all members of $S$ contribute. Here, we allow for arbitrary benefit function $b(x)$, subject only to $b(0) = 0$ and $b(1) = b$.

The outcome of this scenario, for each site $h$ in each state $\mathbf{x}$, is represented by a payoff function $f_h(\mathbf{x})$. In the case that $g \notin S$, this payoff function can be expressed as

$$f_h(\mathbf{x}) = \begin{cases} b(\bar{x}_S) & h = g \\ -\dfrac{c}{m} x_h & h \in S \\ 0 & h \notin S \cup \{g\}. \end{cases} \tag{6.1a}$$

For $g \in S$, payoffs are instead given by

$$f_h(\mathbf{x}) = \begin{cases} -\dfrac{c}{m} x_h + b(\bar{x}_S) & h = g \\ -\dfrac{c}{m} x_h & h \in S - \{g\} \\ 0 & h \notin S. \end{cases} \tag{6.1b}$$

**6.2. Conditional costs.** Conditional costs are only paid if the benefit is achieved. In this case, we assume benefits are all-or-nothing. Thus, if all members of $S$ have allele $A$, they each pay cost $c/m$ and $g$ receives benefit $b$; otherwise, no costs are paid nor benefits received. For $g \notin S$, the payoff function is given by

$$f_h(\mathbf{x}) = \begin{cases} b & \text{if } h = g \text{ and } \bar{x}_S = 1 \\ -\dfrac{c}{m} & \text{if } h \in S \text{ and } \bar{x}_S = 1 \\ 0 & \text{otherwise.} \end{cases} \tag{6.2a}$$

For $g \in S$, payoffs are given by

$$f_h(\mathbf{x}) = \begin{cases} -\dfrac{c}{m} + b & \text{if } h = g \text{ and } \bar{x}_S = 1 \\ -\dfrac{c}{m} & \text{if } h \in S - \{g\} \text{ and } \bar{x}_S = 1 \\ 0 & \text{otherwise.} \end{cases} \tag{6.2b}$$

The conditions for selection in the Collective Action Dilemma depend on how the payoffs, $f_h(\mathbf{x})$, translate into fitness increments, $w_h(\mathbf{x})$. This in turn depends on the selection process in question. We therefore derive conditions separately for a haploid well-mixed population (Section 7), diploid siblings (Section 8), and network-structured populations (Section 9).

## 7. Collective action in well-mixed populations

Our first example of a selection process is the Moran [58] process with frequency-dependent selection [59]. This model describes a haploid well-mixed population of size $N$. We first consider arbitrary payoff functions, and then specialize to the Collective Action Dilemma.

**7.1. Model.** For the frequency-dependent Moran process [59], in each state $\mathbf{x} \in \{0,1\}^G$, each site $g \in G$ receives a payoff $f_g(\mathbf{x})$. A site $g \in G$ is then chosen to reproduce, with probability proportional to the rescaled payoff $1 + \delta f_g(\mathbf{x})$. Here, the selection intensity parameter, $\delta$, quantifies the extent to which payoffs affect reproduction. The offspring displaces an individual (possibly the parent) chosen uniformly at random from the population.

## 7.2. Analysis of selection.
By symmetry, each site $g \in G$ has reproductive value $v_g = 1$. The fitness increment of site $g$ in state $\mathbf{x}$ is given by

$$w_g(\mathbf{x}) = \frac{N-1}{N} + \frac{1}{N}\left(\frac{1+\delta f_g(\mathbf{x})}{1+\delta \bar{f}(\mathbf{x})}\right) - 1 = \frac{\delta}{N}\left(\frac{f_g(\mathbf{x}) - \bar{f}(\mathbf{x})}{1+\delta \bar{f}(\mathbf{x})}\right), \tag{7.1}$$

where $\bar{f}(\mathbf{x}) = \frac{1}{N}\sum_{h \in G} f_h(\mathbf{x})$ is the population average payoff. The first two terms in the middle expression of Eq. (7.1) correspond to survival and reproduction, respectively, while the third subtracts $g$'s own reproductive value.

Taking the $\delta$-derivative of Eq. (7.1) at $\delta = 0$, we obtain the weak selection fitness increments:

$$w'_g(\mathbf{x}) = \frac{1}{N}\left(f_g(\mathbf{x}) - \bar{f}(\mathbf{x})\right). \tag{7.2}$$

The selection increment, for weak selection, can then be computed according to Eq. (3.9):

$$\Delta'(\mathbf{x}) = \frac{1}{N^2}\sum_{g \in G}(x_g - \bar{x})\left(f_g(\mathbf{x}) - \bar{f}(\mathbf{x})\right) = \frac{1}{N^2}\sum_{g \in G}(x_g - \bar{x})f_g(\mathbf{x}). \tag{7.3}$$

The second equality arises because $\sum_{g \in G}(x_g - \bar{x}) = 0$.

We proceed to derive a version of our main condition specialized to this process, using coefficients based on payoffs rather than fitness. To do this, we write the payoff to each site $g$ uniquely as a polynomial function of the state $\mathbf{x}$:

$$f_g(\mathbf{x}) = \sum_{S \subseteq G} C_{S,g}\, \imath_S^A(\mathbf{x}). \tag{7.4}$$

Above, $C_{S,g}$ quantifies the synergistic effect of set $S$ all having allele $A$ on $h$'s payoff. The selection increment, Eq. (7.3), can then be written as

$$\Delta'(\mathbf{x}) = \frac{1}{N^2}\sum_{g \in G}\sum_{S \subseteq G} \imath_S^A(\mathbf{x})(x_g - \bar{x}). \tag{7.5}$$

Applying Theorem 3.4, weak selection favors allele $A$ if

$$\sum_{g \in G}\sum_{S \subseteq G} C_{S,g}\langle \imath_S^A(\mathbf{x})(x_g - \bar{x})\rangle > 0. \tag{7.6}$$

Dividing by $\langle \bar{x}(1-\bar{x})\rangle^{\circ}$, and applying the definition of collective relatedness in Eq. (4.10a), we arrive at the condition

$$\sum_{g \in G}\sum_{S \subseteq G} C_{S,g}\, r_{S,g}^{\circ} > 0. \tag{7.7}$$

This has the same form as our main result in Theorem 5.2, but with coefficients $C_{S,g}$ that describe effects on payoff rather than on fitness.

## 7.3. Coalescence and relatedness.
To evaluate Eq. (7.7) for the collective action scenarios introduced in Section 6, we must first compute collective relatedness via coalescence lengths using Eq. (4.12). The coalescent properties of the Moran process are well-known [18]. Specifically, the coalescence length of any set of size $m$ is

$$\ell_m = \sum_{j=1}^{m-1}\frac{1}{j}. \tag{7.8}$$

Combining this result with the expression for collective relatedness in Eq. (4.12), the relatedness of a collective of size $m$ to each of its members is

$$\begin{aligned}
r_m^\circ &= \frac{\frac{1}{N}\left((N-m)\ell_{m+1} + m\ell_m\right) - \ell_m}{\frac{N-1}{N}\ell_2} \\
&= \frac{(N-m)(\ell_{m+1} - \ell_m)}{(N-1)\ell_2} \\
&= \frac{N-m}{m(N-1)}.
\end{aligned} \tag{7.9}$$

The relatedness of a collective to a site outside the collective is

$$\begin{aligned}
r_{m,1}^\circ &= \frac{\frac{1}{N}\left((N-m)\ell_{m+1} + m\ell_m\right) - \ell_{m+1}}{\frac{N-1}{N}\ell_2} \\
&= \frac{-m(\ell_{m+1} - \ell_m)}{(N-1)\ell_2} \\
&= \frac{-1}{N-1}.
\end{aligned} \tag{7.10}$$

**7.4. Conditions for collective action.** Finally, combining Condition (7.7) and the relatedness values computed in the previous subsection, we obtain conditions for allele $A$ to be favored in the Collective Action Dilemma introduced in Section 6.

**7.4.1. Unconditional costs.** We first consider unconditional costs with all-or-nothing benefits. When written in the polynomial form Eq. (7.4), the payoff function Eq. (6.1) (in the case $g \notin S$) has $C_{\{h\},h} = -c/m$ for all $h \in S$, $C_{S,g} = b$ and $C_{T,h} = 0$ for all other combinations of $T \subseteq G$ and $h \in G$. Condition (7.7) then becomes

$$-\frac{c}{m}\sum_{h \in S} r_{\{h\},h}^\circ + b r_{S,g}^\circ > 0, \tag{7.11}$$

which is the first case of the collective Hamilton's rule, Eq. (19) of the main text. A similar argument shows that this condition also applies when $g \in S$.

From Eq. (7.9), $r_{\{h\},h}^\circ = 1$ for all $h \in S$; consequently, the cost term in Eq. (7.11) simplifies to $-c$. For the benefit term, if $g \in S$ (meaning the target is a member of the collective), applying Eq. (7.9) yields the condition

$$-c + \left(\frac{N-m}{m(N-1)}\right)b > 0. \tag{7.12}$$

For $g \notin S$ (target outside the collective), applying Eq. (7.10) gives the condition

$$-c + \left(\frac{-1}{N-1}\right)b > 0. \tag{7.13}$$

These correspond to the results for well-mixed populations reported in the main text.

Interestingly, Conditions Eq. (7.12) and Eq. (7.13) hold not only for all-or-nothing benefits, but for *any* benefit function $b(\bar{x}_S)$ as described in Section 6. We state this as a theorem:

**Theorem 7.1.** *In the frequency-dependent Moran process described in Section 7.1, for the Collective Action Dilemma with unconditional costs and any benefit function $b(\bar{x}_S)$, allele $A$ is favored if*

$$\begin{cases} (N-m)\,b > m(N-1)c & \text{if } g \in S \\ -b > (N-1)c & \text{if } g \notin S. \end{cases} \tag{7.14}$$

*Proof.* Our first task is to express the payoff function, Eq. (6.1), in polynomial form. We can build such a polynomial out of functions of the form

$$\imath_T^A(\mathbf{x})\, \imath_{S-T}^a(\mathbf{x}) = \begin{cases} 1 & \text{if } x_h = A \text{ for all } h \in T \text{ and } x_h = a \text{ for all } h \in S - T \\ 0 & \text{otherwise.} \end{cases}$$

The idea is that, if $\imath_T^A(\mathbf{x})\, \imath_{S-T}^a(\mathbf{x}) = 1$, then $T$ is exactly the subset of $S$ containing $A$ alleles, and the benefit in this case is $b(|T|/m)$. So overall, the benefit that site $g$ receives in state $\mathbf{x}$ can be expressed as

$$\sum_{T \subseteq S} b\left(\frac{|T|}{m}\right) \imath_T^A(\mathbf{x})\, \imath_{S-T}^a(\mathbf{x}). \tag{7.15}$$

Applying Eq. (4.3) and expanding algebraically, we obtain the identity

$$\imath_T^A(\mathbf{x})\, \imath_{S-T}^a(\mathbf{x}) = \sum_{R;\, T \subseteq R \subseteq S} (-1)^{|R|-|T|} \imath_R^A(\mathbf{x}). \tag{7.16}$$

Now applying Condition (7.7), weak selection favors allele $A$ if and only if

$$-\frac{c}{m} \sum_{h \in S} r_h^\circ + \sum_{T \subseteq S} b\left(\frac{|T|}{m}\right) \sum_{R;\, T \subseteq R \subseteq S} (-1)^{|R|-|T|} r_{R,g}^\circ > 0. \tag{7.17}$$

As before, the cost term simplifies to $-c$. We now separate into cases depending on whether $g \in S$. For $g \notin S$, $r_{R,g}^\circ = -1/(N-1)$ for all $R \subseteq S$ by Eq. (7.10), so $A$ is favored if

$$-c - \frac{1}{N-1} \sum_{T \subseteq S} b\left(\frac{|T|}{m}\right) \sum_{R;\, T \subseteq R \subseteq S} (-1)^{|R|-|T|} > 0. \tag{7.18}$$

The inner sum can be simplified as follows:

$$\sum_{R;\, T \subseteq R \subseteq S} (-1)^{|R|-|T|} = \sum_{V \subseteq S-T} (-1)^{|V|} \qquad \text{letting } V = R - T$$

$$= \sum_{j=0}^{m-|T|} \binom{m-|T|}{j} (-1)^j \qquad \text{letting } j = |V|$$

$$= (1 + (-1))^{m-|T|} \qquad \text{by the Binomial Theorem}$$

$$= \begin{cases} 1 & \text{if } |T| = m \\ 0 & \text{otherwise} \end{cases}$$

$$= \begin{cases} 1 & \text{for } T = S \\ 0 & \text{otherwise.} \end{cases} \tag{7.19}$$

Combining with Eq. (7.18) and recalling that $b(1) = b$, weak selection favors $A$ for $g \notin S$ if

$$-c - \frac{b}{N-1} > 0, \tag{7.20}$$

proving the desired result in this case.

For $g \in S$, we first observe that, by the symmetry of Condition (7.17) and of the Moran process, the condition for selection does not depend on which member of $S$ is the target. Consequently, Condition (7.17) can be averaged over all $g \in S$ and still give the correct condition. It follows that weak selection favors allele $A$ if and only if

$$-c + \sum_{T \subseteq S} b\left(\frac{|T|}{m}\right) \sum_{R;T \subseteq R \subseteq S} (-1)^{|R|-|T|} \left(\frac{1}{m} \sum_{g \in S} r^\circ_{R,g}\right) > 0. \qquad (7.21)$$

Evaluating the inner sum in parentheses by means of Eq. (7.9) and Eq. (7.10), this condition becomes

$$-c + \sum_{T \subseteq S} b\left(\frac{|T|}{m}\right) \sum_{R;T \subseteq R \subseteq S} (-1)^{|R|-|T|} \left(\frac{|R|}{m}\left(\frac{N-|R|}{|R|(N-1)}\right) + \frac{m-|R|}{m}\left(\frac{-1}{N-1}\right)\right) > 0. \qquad (7.22)$$

This simplifies to

$$-c + \frac{N-m}{m(N-1)} \sum_{T \subseteq S} b\left(\frac{|T|}{m}\right) \sum_{R;T \subseteq R \subseteq S} (-1)^{|R|-|T|} > 0, \qquad (7.23)$$

whereupon applying Eq. (7.19), we find that weak selection favors $A$ for $g \in S$ if

$$-c + \left(\frac{N-m}{m(N-1)}\right) b > 0, \qquad (7.24)$$

as desired. □

### 7.4.2. Conditional costs.
For conditional costs, with payoffs given by Eq. (6.2), Condition (7.7) becomes

$$-\frac{c}{m} \sum_{h \in S} r^\circ_{S,h} + b r^\circ_{S,g} > 0. \qquad (7.25)$$

This is the second case of the collective Hamilton's rule, Eq. (19) of the main text. For $g \notin S$, applying the relatedness values in Eq. (7.9) and Eq. (7.10), we obtain

$$-\left(\frac{N-m}{m(N-1)}\right) c + \left(\frac{-1}{N-1}\right) b > 0, \qquad (7.26)$$

which simplifies to

$$-mb > (N-m)c. \qquad (7.27)$$

For $g \in S$, the condition reduces to simply $b > c$.

## 8. Diploid relatives

Here we apply our results to social behavior among relatives in a diploid population. This model uses the formalism for individual phenotypes, as introduced in Section 1.7 and further developed in Sections 4.3.3 and 5.5.

### 8.1. Model.
We consider a diploid population with discrete generations. We first describe the model for a hermaphroditic (one sex) population, and then extend it to two sexes.

#### 8.1.1. One sex.
We focus on the population of juveniles in each generation. There is a set $I$ of juvenile individuals, with size $N = |I|$. These individuals are partitioned into $N/M$ families of size $M$ each. Each individual $i \in I$ has genetic sites $G_i = \{i_1, i_2\}$.

There are two phenotypes, numbered 1 and 0. $AA$ individuals have phenotype 1, $aa$ individuals have phenotype 0, and heterozygotes have phenotype 1 or 0 with probabilities $h$ and $1-h$, respectively, where $0 \leq h \leq 1$ represents the degree of genetic dominance. Overall, the probability that individual $i \in I$ has phenotype $\Phi_i = 1$ in state **x** is given by Eq. (1.4). The overall phenotypic state of the population is represented by the vector $\mathbf{\Phi} = (\Phi_i)_{i \in I}$.

In each phenotypic state $\boldsymbol{\Phi}$, each individual $i$ receives payoff $f_i(\boldsymbol{\Phi})$, summarizing the results of all interactions. For now we allow the payoff functions $f_i(\boldsymbol{\Phi})$ to be arbitrary; we will specialize to the Collective Action Dilemma in Section 8.5 below. This payoff is then rescaled to a junvenile survival function $F_i(\boldsymbol{\Phi}) = 1 + \delta f_i(\boldsymbol{\Phi})$.

The population follows a three-stage lifecycle. In the one-sex case, the stages proceed as follwos:

1. *Survival:* A fixed number $N_A \leq N$ of juveniles survive to adulthood. These $N_A$ surviving adults are sampled in sequence (without replacement) from $I$, each proportionally to $F_i(\boldsymbol{\Phi})$.

2. *Mating:* From these $N_A$ surviving adults, $N/M$ ordered pairs of individuals are sampled, uniformly and independently (with replacement).

3. *Reproduction:* Each ordered pair produces $M$ juvenile offspring to fill a single family group. Alleles are transmitted according to Mendelian inheritance: For each new individual $i$ in a family group with parents $(j, k)$, site $i_1$ randomly inherits the allele in $j_1$ or $j_2$, and site $i_2$ randomly inherits the allele in $k_1$ or $k_2$, each chosen with equal probability independently of all other such choices. Mutation occurs independently with probability $u$ at each site.

**8.1.2. Two sexes.** The model for two sexes has the same lifecycle, with the following amendments: The set $I$ of juveniles is partitioned into equally-sized subsets $I_M$ and $I_F$ for males and females, respectively. In the survival stage, $N_A/2$ males and $N_A/2$ females are sampled from sets $I_M$ and $I_F$ respectively, again proportionally to $F_i(\boldsymbol{\Phi})$. In the mating stage, the first entry of each mating pair $(j, k)$ is sampled from the surviving males, and the second from the surviving females. In the reproduction stage, each mating pair $(j, k)$ produces $M/2$ male offspring and $M/2$ female offspring. For each juvenile offspring $i$, site $i_1$ inherits an allele from one of the paternal sites, $j_1$ or $j_2$, and $i_2$ inherits an allele from one of the maternal sites, $k_1$ or $k_2$, each independently with equal probability.

**8.2. Analysis of selection.** To make analytical progress, we first assume that $M \ll N$. Under this assumption, the probability that a given individual $i \in I$ survives to adulthood is asymptotically $(N_A F_i(\boldsymbol{\Phi}))/(N \bar{F}(\boldsymbol{\Phi}))$, where $\bar{F}(\boldsymbol{\Phi}) = \frac{1}{N} \sum_{j \in I} F_j(\boldsymbol{\Phi})$. Each adult individual has, on expectation, $2N/N_A$ offspring (regardless of family size $M$). Therefore, each juvenile individual $i \in I$ produces $2F_i(\boldsymbol{\Phi})/\bar{F}(\boldsymbol{\Phi})$ offspring on expectation. Since each parental allele has $1/2$ probability to be transmitted to each offspring, each allele in individual $i$ produces an expected $F_i(\boldsymbol{\Phi})/\bar{F}(\boldsymbol{\Phi})$ copies in the next generation. By symmetry, each site $g \in G$ has reproductive value $v_g = 1$. Therefore, applying Eq. (5.21), the fitness increment of each site in a given individual $i \in I$ is

$$\breve{w}_i(\boldsymbol{\Phi}) = \frac{F_i(\boldsymbol{\Phi})}{\bar{F}(\boldsymbol{\Phi})} - 1 = \frac{F_i(\boldsymbol{\Phi}) - \bar{F}(\boldsymbol{\Phi})}{\bar{F}(\boldsymbol{\Phi})}. \tag{8.1}$$

For weak selection, taking the $\delta$-derivative at $\delta = 0$ yields

$$\breve{w}'_i(\boldsymbol{\Phi}) = f_i(\boldsymbol{\Phi}) - \bar{f}(\boldsymbol{\Phi}), \tag{8.2}$$

where $\bar{f}(\boldsymbol{\Phi}) = \frac{1}{N} \sum_{j \in I} f_j(\boldsymbol{\Phi})$. Applying Lemma 5.4, the weak selection increment in each state $\mathbf{x}$ is

$$\Delta'(\mathbf{x}) = \sum_{i \in I} \mathbb{E}_{\mathbf{x}} \left[ f_i(\boldsymbol{\Phi}) - \bar{f}(\boldsymbol{\Phi}) \right] (\bar{x}_i - \bar{x}) = \sum_{i \in I} \mathbb{E}_{\mathbf{x}} \left[ f_i(\boldsymbol{\Phi}) \right] (\bar{x}_i - \bar{x}). \tag{8.3}$$

In the second equality above, the terms involving $\bar{f}(\boldsymbol{\Phi})$ cancel because $\sum_{i \in I}(\bar{x}_i - \bar{x}) = 0$. Eqs. (8.1)–(8.3) apply to both the one-sex and two-sex models.

To obtain a condition for selection, we specialize Condition (5.27b) to this process. Following the approach of Section 7.2, we write $f_i(\boldsymbol{\Phi})$ uniquely in polynomial form:

$$f_i(\boldsymbol{\Phi}) = \sum_{J \subseteq I} C_{J,i} \prod_{j \in J} \Phi_j. \tag{8.4}$$

The expected payoff in population state $\mathbf{x}$ can then be written as

$$\mathbb{E}_{\mathbf{x}}[f_i(\mathbf{\Phi})] = \sum_{J \subseteq I} C_{J,i} \prod_{j \in J} \varphi_j(x_{j_1}, x_{j_2}), \tag{8.5}$$

and Eq. (8.3) for the weak selection increment becomes

$$\Delta'(\mathbf{x}) = \sum_{J \subseteq I} C_{J,i} \prod_{j \in J} \varphi_j(x_{j_1}, x_{j_2})(\bar{x}_i - \bar{x}). \tag{8.6}$$

Applying Theorem 3.4, dividing by $\langle \bar{x}(1-\bar{x}) \rangle^{\circ}$, and using the phenotypic relatedness definition in Eq. (4.19), we obtain that weak selection favors allele $A$ if and only if

$$\sum_{J \subseteq I} C_{J,i} r_{J,i}^{\circ} > 0. \tag{8.7}$$

**8.3. Coalescence lengths.** To evaluate Eq. (8.7), we must compute coalescence lengths for the neutral ($\delta = 0$) case of this model. For this, we introduce the further assumption that $M \ll N_A \ll N$.

We first consider the population of adults at each generation. In the one-sex case, this adult population is asymptotically described by the neutral Wright-Fisher process for $N_A$ diploid individuals. The two-sex case was formally characterized, in the context of our framework, in Section 8.2 of Allen and McAvoy [5] (see also Möhle [60]). In either case, coalescence lengths for alleles in adults are asymptotically characterized by the standard Kingman coalescent [17, 18] on $2N_A$ alleles.

Let $\lambda_k$ denote the coalescence length of a set of $k \geq 1$ alleles in adults, and let $L_k = \lim_{N_A \to \infty} \lambda_k/(4N_A)$. A classical result [61] gives

$$L_k = \sum_{j=1}^{k-1} \frac{1}{j}, \tag{8.8}$$

which is the expected total branch length for $k$ sites in the Kingman coalescent [17, 18].

Now we return to the juvenile population. For any nonempty $S \subseteq G$, let $P_S$ be a random variable representing the number of distinct sites among the parents of alleles in $S$, within the adult population, in the $N \to \infty$ limit. (This requires $S$ to be defined in such a way as to be independent of $N$, which will be the case for all sets we examine.) Note that $1 \leq P_S \leq |S|$. Then neutral expected coalescence lengths obey $\lim_{N \to \infty} \ell_S^{\circ} = |S| + \mathbb{E}[\lambda_{P_S}]$. Normalizing by $4N_A$ and applying Eq. (8.8), we obtain

$$\lim_{N_A \to \infty} \lim_{N \to \infty} \frac{\ell_S^{\circ}}{4N_A} = \lim_{N_A \to \infty} \frac{|S| + \mathbb{E}[\lambda_{P_S}]}{4N_A} = \mathbb{E}[L_{P_S}]. \tag{8.9}$$

In particular, let us apply this result to $\bar{\ell} = \frac{1}{n^2} \sum_{h,k \in G} \ell_{\{h,k\}}$, the average coalescence length among all pairs of (juvenile) sites. For two sites sampled uniformly from the juvenile population, their likelihood of having the same parent allele is negligible, since they are unlikely to belong to the same family ($N \gg M$) and distinct families are unlikely to have the same parents ($N_A \gg 1$). Therefore, in the relevant limits (first $N \to \infty$, then $N_A \to \infty$), $P_{\{h,k\}} = 2$ almost surely for almost all pairs $h, k \in G$. Eq. (8.9) then gives

$$\lim_{N_A \to \infty} \lim_{N \to \infty} \frac{\bar{\ell}^{\circ}}{4N_A} = L_2 = 1. \tag{8.10}$$

It will be useful to define an operator $[\![\,]\!]$ on state functions $f(\mathbf{x})$ with $f(\mathbf{A}) = 1$ and $f(\mathbf{a}) = 0$ by

$$[\![f]\!] = \lim_{N_A \to \infty} \lim_{N \to \infty} \frac{\langle \bar{x} - f(\mathbf{x}) \rangle^{\circ}}{\langle \bar{x}(1-\bar{x}) \rangle^{\circ}} = \lim_{N_A \to \infty} \lim_{N \to \infty} \frac{2 \langle \bar{x} - f(\mathbf{x}) \rangle^{\circ}}{\bar{\ell}^{\circ}}. \tag{8.11}$$

This operator is affine in the sense that, for state functions $f(\mathbf{x})$ and $g(\mathbf{x})$ with $f(\mathbf{A}) = g(\mathbf{A}) = 1$ and $f(\mathbf{a}) = g(\mathbf{a}) = 0$, we have $[\![af + bg]\!] = a[\![f]\!] + b[\![g]\!]$ for all scalars $a, b \in \mathbb{R}$ with $a + b = 1$.

As an example, let $f(\mathbf{x})$ be the identity-by-state function $\iota_S^A(\mathbf{x}) = \prod_{g \in S} x_g$ for some nonempty $S \subseteq G$. By Eq. (4.5) and symmetry of $A$ and $a$ under neutral drift, we have

$$\left\langle \bar{x} - \iota_S^A \right\rangle^\circ = \langle 1 - \bar{x} - \iota_S^a \rangle^\circ = \frac{1}{2} \langle 1 - \iota_S \rangle^\circ = \frac{\ell_S^\circ}{2}. \tag{8.12}$$

Combining with Eq. (8.9) and Eq. (8.11) gives

$$[\![\iota_S^A]\!] = \lim_{N_A \to \infty} \lim_{N \to \infty} \frac{\left\langle \bar{x} - \iota_S^A \right\rangle^\circ}{\langle \bar{x}(1 - \bar{x}) \rangle^\circ} = \lim_{N_A \to \infty} \lim_{N \to \infty} \frac{\ell_S^\circ}{\bar{\ell}^\circ} = \frac{\mathbb{E}\left[L_{P_S}\right]}{L_2} = \mathbb{E}\left[L_{P_S}\right]. \tag{8.13}$$

This means that, for any set $S$ of sites in the juvenile population, $[\![\iota_S^A]\!]$ is equal to the expected total branch length, under the Kingman coalescent, of the parental sites from which the alleles in $S$ are inherited. Eq. (8.13) will be instrumental in computing collective relatedness.

**8.4. Collective relatedness among siblings.** We now begin computing collective relatedness quantities, in the limits of first $N \to \infty$ and then $N_A \to \infty$.

***8.4.1. Self-relatedness.*** We begin with the relatedness of any individual $j \in J$ to itself, using $r_j^\circ$ as shorthand for $r_{\{j\},j}^\circ$. Combining Eq. (4.19) and Eq. (1.4), and applying the limits $N \to \infty$ and then $N_A \to \infty$, we obtain

$$r_j^\circ = \lim_{N_A \to \infty} \lim_{N \to \infty} \frac{\left\langle (h(x_{j_1} + x_{j_2}) + (1 - 2h)x_{j_1}x_{j_2})\left(\frac{x_{j_1} + x_{j_2}}{2} - \bar{x}\right) \right\rangle^\circ}{\langle \bar{x}(1 - \bar{x}) \rangle^\circ}.$$

Recalling that $x_g^2 = x_g$ for each site $g$, the bracketed quantity in the numerator can be expanded as

$$h\left(\frac{x_{j_1} + x_{j_2}}{2} + x_{j_1}x_{j_2} - 2\left(\frac{x_{j_1} + x_{j_2}}{2}\right)\bar{x}\right) + (1 - 2h)\left(x_{j_1}x_{j_2} - x_{j_1}x_{j_2}\bar{x}\right). \tag{8.14}$$

Applying Eq. (8.11), we can express $r_j^\circ$ in terms of the $[\![\ ]\!]$ operator as

$$r_j^\circ = h\left(2\left[\!\left[\left(\frac{x_{j_1} + x_{j_2}}{2}\right)\bar{x}\right]\!\right] - \left[\!\left[\frac{x_{j_1} + x_{j_2}}{2}\right]\!\right] - [\![x_{j_1}x_{j_2}]\!]\right) + (1 - 2h)\left([\![x_{j_1}x_{j_2}\bar{x}]\!] - [\![x_{j_1}x_{j_2}]\!]\right). \tag{8.15}$$

Invoking Eq. (8.13), and noting that (in the $N_A \to \infty$ limit), $j_1$ and $j_2$ must come from distinct parental sites, we have

$$[\![x_{j_1}]\!] = [\![x_{j_2}]\!] = L_1$$
$$[\![x_{j_1}x_{j_2}]\!] = [\![x_{i_1}\bar{x}]\!] = [\![x_{i_2}\bar{x}]\!] = L_2$$
$$[\![x_{j_1}x_{j_2}\bar{x}]\!] = L_3.$$

Substituting into Eq. (8.15) and evaluating via Eq. (8.8), we obtain

$$\begin{aligned} r_j^\circ &= h\left(2L_2 - L_1 - L_2\right) + (1 - 2h)(L_3 - L_2) \\ &= h(1) + (1 - 2h)\left(\frac{1}{2}\right) \\ &= \frac{1}{2}. \end{aligned} \tag{8.16}$$

Thus the neutral relatedness of any individual $j$ to itself is $r_j^\circ = \frac{1}{2}$ in this model.

### 8.4.2. Collective relatedness to another sibling.

We now consider the relatedness of a set $J$ of $m$ siblings to another sibling $i \notin J$. Again combining Eq. (4.19) and Eq. (1.4), we obtain

$$r^\circ_{J,i} = \lim_{N_A \to \infty} \lim_{N \to \infty} \frac{\left\langle \left(\prod_{j \in J}(hx_{j_1} + hx_{j_2} + (1-2h)x_{j_1}x_{j_2})\right)\left(\frac{x_{i_1}+x_{i_2}}{2} - \bar{x}\right)\right\rangle^\circ}{\langle \bar{x}(1-\bar{x})\rangle^\circ}. \qquad (8.17)$$

Let us define the notation $\underline{x}_{(a,b)}$, for $0 \le a \le m$ and $0 \le b \le m$, to be the average value of all products involving $a$ distinct factors of the form $x_{j_1}$ and $b$ distinct factors of the form $x_{j_2}$, for $j$ varying over $J$. Expanding the first factor in the numerator of Eq. (8.17) according to the multinomial theorem, we have

$$\prod_{j \in J}(hx_{j_1} + hx_{j_2} + (1-2h)x_{j_1}x_{j_2}) = \sum_{\substack{k_1,k_2 \ge 0 \\ k_1+k_2 \le m}} \frac{m!\, h^{k_1+k_2}(1-2h)^{m-k_1-k_2}}{k_1!\,k_2!\,(m-k_1-k_2)!}\, \underline{x}_{(m-k_2,m-k_1)}. \qquad (8.18)$$

Each term on the right-hand side corresponds to choosing $k_1$ factors of the form $hx_{j_1}$, $k_2$ factors of the form $hx_{j_2}$, and $m - k_1 - k_2$ factors of the form $(1-2h)x_{j_1}x_{j_2}$.

Using this result, we can express Eq. (8.17) as

$$r^\circ_{J,i} = \sum_{\substack{k_1,k_2 \ge 0 \\ k_1+k_2 \le m}} \frac{m!\, h^{k_1+k_2}(1-2h)^{m-k_1-k_2}}{k_1!\,k_2!\,(m-k_1-k_2)!}$$

$$\times \left(\llbracket \underline{x}_{(m-k_2,m-k_1)}\bar{x}\rrbracket - \llbracket \frac{\underline{x}_{(m-k_2+1,m-k_1)} + \underline{x}_{(m-k_2,m-k_1+1)}}{2}\rrbracket\right). \qquad (8.19)$$

We now evaluate the $\llbracket\ \rrbracket$ operations in Eq. (8.19), by means of Eq. (8.13). We start with terms of the form $\underline{x}_{(a,b)}$. For $b = 0$, we have

$$\llbracket \underline{x}_{(a,0)}\rrbracket = \frac{1}{2^{a-1}}L_1 + \left(1 - \frac{1}{2^{a-1}}\right)L_2 = 1 - \frac{1}{2^{a-1}}. \qquad (8.20)$$

Similarly, for $a = 0$,

$$\llbracket \underline{x}_{(0,b)}\rrbracket = 1 - \frac{1}{2^{b-1}}. \qquad (8.21)$$

For $a, b \ne 0$ we have

$$\llbracket \underline{x}_{(a,b)}\rrbracket = \left(\frac{1}{2^{a-1}}\right)\left(\frac{1}{2^{b-1}}\right)L_2$$
$$+ \left(\frac{1}{2^{a-1}}\left(1 - \frac{1}{2^{b-1}}\right) + \left(1 - \frac{1}{2^{a-1}}\right)\frac{1}{2^{b-1}}\right)L_3$$
$$+ \left(1 - \frac{1}{2^{a-1}}\right)\left(1 - \frac{1}{2^{b-1}}\right)L_4$$
$$= \frac{11}{6} - \frac{1}{3}\left(\frac{1}{2^{a-1}} + \frac{1}{2^{b-1}}\right) - \frac{1}{6}\left(\frac{1}{2^{a+b-2}}\right). \qquad (8.22)$$

We next consider combined terms of the form $\frac{1}{2}\left(\underline{x}_{(a+1,b)} + \underline{x}_{(a,b+1)}\right)$. For $b = 0$, using Eq. (8.20) and Eq. (8.22), we have

$$\llbracket \frac{\underline{x}_{(a+1,0)} + \underline{x}_{(a,1)}}{2}\rrbracket = \frac{1}{2}\left(1 - \frac{1}{2^a} + \frac{11}{6} - \frac{1}{3}\left(\frac{1}{2^{a-1}} + 1\right) - \frac{1}{6}\left(\frac{1}{2^{a-1}}\right)\right)$$
$$= \frac{1}{2}\left(\frac{5}{2} - \frac{1}{2^{a-1}}\right)$$
$$= \frac{5}{4} - \frac{1}{2^a}. \qquad (8.23)$$

Similarly, for $a = 0$,

$$\left[\!\left[\frac{x_{(1,b)} + x_{(0,b+1)}}{2}\right]\!\right] = \frac{5}{4} - \frac{1}{2^b}. \tag{8.24}$$

For $a, b \neq 0$, applying Eq. (8.22) gives

$$\left[\!\left[\frac{x_{(a+1,b)} + x_{(a,b+1)}}{2}\right]\!\right] = \frac{11}{6} - \frac{1}{6}\left(\frac{1}{2^{a-1}} + \frac{1}{2^a} + \frac{1}{2^{b-1}} + \frac{1}{2^b}\right) - \frac{1}{6}\left(\frac{1}{2^{a+b-1}}\right)$$

$$= \frac{11}{6} - \frac{1}{2}\left(\frac{1}{2^a} + \frac{1}{2^b}\right) - \frac{1}{3}\left(\frac{1}{2^{a+b}}\right). \tag{8.25}$$

We also require terms of the form $\underline{x}_{(a,b)}\bar{x}$. For $b = 0$,

$$[\![\underline{x}_{(a,0)}\bar{x}]\!] = \frac{1}{2^{a-1}} L_2 + \left(1 - \frac{1}{2^{a-1}}\right) L_3 = \frac{3}{2} - \frac{1}{2^a}. \tag{8.26}$$

Similarly, for $a = 0$,

$$[\![\underline{x}_{(0,b)}\bar{x}]\!] = \frac{3}{2} - \frac{1}{2^b}. \tag{8.27}$$

For $a, b \neq 0$,

$$[\![\underline{x}_{(a,b)}\bar{x}]\!] = \frac{1}{2^{a+b-2}} L_3 + \left(\frac{1}{2^{a-1}}\left(1 - \frac{1}{2^{b-1}}\right) + \left(1 - \frac{1}{2^{a-1}}\right)\frac{1}{2^{b-1}}\right) L_4$$

$$+ \left(1 - \frac{1}{2^{a-1}}\right)\left(1 - \frac{1}{2^{b-1}}\right) L_5$$

$$= \frac{3}{2} + \frac{1}{3}\left(1 - \frac{1}{2^{a+b-2}}\right) + \frac{1}{4}\left(1 - \frac{1}{2^{a-1}}\right)\left(1 - \frac{1}{2^{b-1}}\right)$$

$$= \frac{25}{12} - \frac{1}{4}\left(\frac{1}{2^{a-1}} + \frac{1}{2^{b-1}}\right) - \frac{1}{12}\left(\frac{1}{2^{a+b-2}}\right)$$

$$= \frac{25}{12} - \frac{1}{2}\left(\frac{1}{2^a} + \frac{1}{2^b}\right) - \frac{1}{3}\left(\frac{1}{2^{a+b}}\right). \tag{8.28}$$

Combining Eqs. (8.23)–(8.28), we find that for any $a, b \geq 0$ with $a + b \geq 1$,

$$[\![\underline{x}_{(a,b)}\bar{x}]\!] - \left[\!\left[\frac{x_{(a+1,b)} + x_{(a,b+1)}}{2}\right]\!\right] = \frac{1}{4}. \tag{8.29}$$

It then follows from Eq. (8.17) and Eq. (8.19) that the collective relatedness of $J$ to $i$ is

$$r_{J,i}^\circ = 1/4. \tag{8.30}$$

Interestingly, this result does not depend on the number of siblings, $m$, nor on the degree of dominance, $h$.

**8.4.3. Collective intra-relatedness of siblings.** Finally, we compute the intra-relatedness $r_J^\circ$ of a set $J$ of siblings, given by

$$r_J^\circ = \lim_{N_A \to \infty} \lim_{N \to \infty} \frac{\left\langle \left(\prod_{j \in J}(hx_{j_1} + hx_{j_2} + (1-2h)x_{j_1}x_{j_2})\right)\left(\frac{1}{2m}\sum_{j \in J}(x_{j_1} + x_{j_2}) - \bar{x}\right)\right\rangle^\circ}{\langle \bar{x}(1 - \bar{x})\rangle^\circ}. \tag{8.31}$$

Using the multinomial expansion in Eq. (8.18), and the identity

$$\underline{x}_{(m-k_2, m-k_1)}\left(\frac{\sum_{j \in J}(x_{j_1} + x_{j_2})}{2m}\right)$$

$$= \frac{k_2}{2m}\underline{x}_{(m-k_2+1, m-k_1)} + \frac{k_1}{2m}\underline{x}_{(m-k_2, m-k_1+1)} + \frac{2m - k_1 - k_2}{2m}\underline{x}_{(m-k_2, m-k_1)}, \tag{8.32}$$

we can express the intra-relatedness $r_J^\circ$ in terms of the $[\![\,]\!]$ operator as

$$r_J^\circ = \sum_{\substack{k_1,k_2 \geq 0 \\ k_1+k_2 \leq m}} \frac{m!\, h^{k_1+k_2}\,(1-2h)^{m-k_1-k_2}}{k_1!\,k_2!\,(m-k_1-k_2)!}$$
$$\times \left( [\![\underline{x}_{(m-k_2,m-k_1)}\bar{x}]\!] - \frac{k_2}{2m}[\![\underline{x}_{(m-k_2+1,m-k_1)}]\!] - \frac{k_1}{2m}[\![\underline{x}_{(m-k_2,m-k_1+1)}]\!] \right.$$
$$\left. - \frac{2m-k_1-k_2}{2m}[\![\underline{x}_{(m-k_2,m-k_1)}]\!] \right). \quad (8.33)$$

Unlike for $r_{J,i}^\circ$, the expression for $r_J^\circ$ does not appear to simplify in general. However, by evaluating the terms in Eq. (8.33) according to Eq. (8.20)–Eq. (8.27), one can compute $r_J^\circ$ for any number $m = |J|$ of siblings. The first five values are:

$$r_J^\circ = \begin{cases} \frac{1}{2} & m = 1 \\ \frac{17+2h}{48} & m = 2 \\ \frac{57+12h-4h^2}{192} & m = 3 \\ \frac{209+38h+12h^2-24h^3}{768} & m = 4 \\ \frac{801+104h+88h^2-32h^3-80h^4}{3072} & m = 5. \end{cases} \quad (8.34)$$

Evaluating these expressions for different values of $h$ leads to the intra-relatedness values given in Table 1 of the main text.

In the recessive case ($h = 0$), a closed-form expression for $r_J^\circ$ can be obtained. Eq. (8.31) simplifies in the recessive case to

$$r_J^\circ = \lim_{N_A \to \infty} \lim_{N \to \infty} \frac{\left\langle \left(\prod_{j \in J} x_{j_1} x_{j_2}\right)(1-\bar{x}) \right\rangle^\circ}{\langle \bar{x}(1-\bar{x}) \rangle^\circ}$$
$$= [\![\underline{x}_{(m,m)}\bar{x}]\!] - [\![\underline{x}_{(m,m)}]\!]. \quad (8.35)$$

From Eq. (8.22) and Eq. (8.28), we have

$$[\![\underline{x}_{(m,m)}]\!] = \frac{11}{6} - \frac{2}{3}\left(\frac{1}{2^{m-1}}\right) - \frac{1}{6}\left(\frac{1}{2^{2m-2}}\right)$$
$$= \frac{22}{12} - \frac{4}{3}\left(\frac{1}{2^m}\right) - \frac{2}{3}\left(\frac{1}{2^{2m}}\right)$$
$$[\![\underline{x}_{(m,m)}\bar{x}]\!] = \frac{25}{12} - \frac{1}{2^m} - \frac{1}{3}\left(\frac{1}{2^{2m}}\right).$$

Substituting in Eq. (8.35) gives the intra-relatedness in the recessive case:

$$r_J^\circ = \frac{1}{4} + \frac{1}{3}\left(\frac{1}{2^m} + \frac{1}{2^{2m}}\right). \quad (8.36)$$

**8.5. Conditions for collective action.** We now adapt the Collective Action Dilemma introduced in Section 6 to diploids, and obtain conditions for action to be favored. We seek to determine whether selection favors a set $J \subseteq I$ of $m$ siblings to help a target sibling $i \in I$. Phenotype 1 contributes to collective action, while phenotype 0 does not. As before, the benefit depends on the fraction of $J$ that contributes; we write this benefit as a function $b(\bar{\Phi}_J)$, where $\bar{\Phi}_J = \frac{1}{m}\sum_{j \in J} \Phi_j$ is the fraction that contributes. We again require that $b(0) = 0$ and shorten $b(1) = b$.

### 8.5.1. Unconditional costs, target outside collective.
We start with the case of unconditional costs—meaning every phenotype-1 individual in $J$ pays cost $c/m$—and the target sibling $i$ being outside the collective $J$. In this case, the payoff function is given by

$$f_j(\mathbf{\Phi}) = \begin{cases} b\left(\bar{\Phi}_J\right) & j = i \\ -\dfrac{c}{m} \Phi_j & j \in J \\ 0 & j \notin J \cup \{i\}. \end{cases} \tag{8.37}$$

For all-or-nothing benefits ($b(x) = 0$ for $0 \leq x < 1$), Condition (8.7) becomes

$$-\frac{c}{m} \sum_{j \in J} r_j^\circ + b r_{J,i}^\circ > 0. \tag{8.38}$$

Substituting the relatedness values from Eq. (8.16) and Eq. (8.30), we obtain $-\frac{1}{2}c + \frac{1}{4}b > 0$, or $b > 2c$. Thus, as reported in the main text, collective help to $i$ is favored if the benefit exceeds twice the cost.

It turns out that, since $r_{J,i} = 1/4$ is independent of the size of $J$, the $b > 2c$ condition applies not only to all-or-nothing benefits, but to *any* benefit function:

**Theorem 8.1.** *In the model described in Section 8.1, for the Collective Action Dilemma played by a collective $J$ of siblings and target sibling $i \notin J$, with any benefit function $b(\bar{\Phi}_J)$, weak selection favors allele $A$ if and only if $b > 2c$.*

*Proof.* Following the same approach as the proof of Theorem 7.1, we express the benefit function in the form

$$\begin{aligned} b\left(\bar{\Phi}_J\right) &= \sum_{K \subseteq J} b\left(\frac{|K|}{m}\right) \prod_{k \in K} \Phi_k \prod_{h \in J-K} (1 - \Phi_h) \\ &= \sum_{K \subseteq J} b\left(\frac{|K|}{m}\right) \sum_{H; K \subseteq H \subseteq J} (-1)^{|H|-|K|} \prod_{h \in H} \Phi_h. \end{aligned} \tag{8.39}$$

Applying Condition (8.7), weak selection favors allele $A$ if and only if

$$-\frac{c}{m} \sum_{j \in J} r_j^\circ + \sum_{K \subseteq J} b\left(\frac{|K|}{m}\right) \sum_{H; K \subseteq H \subseteq J} (-1)^{|H|-|K|} r_{H,i}^\circ > 0. \tag{8.40}$$

Substituting the relatedness values from Eq. (8.16) and Eq. (8.30), this condition becomes

$$-c\left(\frac{1}{2}\right) + \frac{1}{4} \sum_{K \subseteq J} b\left(\frac{|K|}{m}\right) \sum_{H; K \subseteq H \subseteq J} (-1)^{|H|-|K|} > 0. \tag{8.41}$$

Now using the identity Eq. (7.19), the benefit term reduces to $\frac{1}{4}b$, and the condition simplifies to $b > 2c$. ☐

### 8.5.2. Unconditional costs, target inside collective.
If the target sibling $i$ belongs to the collective $J$, the payoffs are given by

$$f_j(\mathbf{\Phi}) = \begin{cases} -\dfrac{c}{m} + b\left(\bar{\Phi}_J\right) & j = i \\ -\dfrac{c}{m} \Phi_j & j \in J - \{i\} \\ 0 & j \notin J. \end{cases} \tag{8.42}$$

In this case, we only obtain results for all-or-nothing benefits. Applying Condition (8.7), weak selection favors allele $A$ if and only if

$$-\frac{c}{m} \sum_{j \in J} r_j^\circ + b r_J^\circ > 0. \tag{8.43}$$

Using $r_j^\circ = 1/2$, we obtain that help to a member of the collective is favored if $2br_J^\circ > c$.

### 8.5.3. Conditional costs.
For conditional costs, with the target $i$ outside the collective $J$ and all-or-nothing benefits, the payoff function is

$$f_j(\mathbf{\Phi}) = \begin{cases} b \prod_{j \in J} \Phi_j & j = i \\ -\dfrac{c}{m} \prod_{j \in J} \Phi_j & j \in J \\ 0 & j \notin J \cup \{i\}. \end{cases} \quad (8.44)$$

Applying Condition (8.7), weak selection favors allele $A$ if and only if

$$-cr_J^\circ + br_{J,i}^\circ > 0. \quad (8.45)$$

Noting that $r_{J,i}^\circ = \frac{1}{4}$ by Eq. (8.30), we obtain the condition $b > 4cr_J^\circ$.

### 8.6. Two arbitrary relatives.
We now extend to relationships beyond full siblings. We imagine a model similar to the model described above, but with individuals grouped by a relationship other than full siblings (half-siblings, cousins, etc.). We do not attempt to explicitly construct such a model, as the setup would depend on the kin relationship in question. Instead, we suppose that such a model will share the following features with the full-sibling model:

1. The juvenile population is represented a set $I$ of $N$ diploid individuals, with sites $i_1$ and $i_2$ for each individual $i \in I$.

2. The effects of phenotypic state $\mathbf{\Phi}$ on the survival of each individual $i \in I$ to adulthood is captured by an arbitrary function $f_i(\mathbf{\Phi})$.

3. At each time-step, $N_A$ surviving adults are sampled from $I$, in such a manner that, as $N \to \infty$, the probability individual $i$ survives becomes proportional to $F_i(\mathbf{\Phi}) = 1 + \delta f_i(\mathbf{\Phi})$.

4. Each of the $2N_A$ sites in the adult population produces, on expectation, $N/N_A$ copies in the next juvenile population.

5. For $k \geq 1$, let $\bar{\ell}_k^\circ$ denote the average of $\ell_S^\circ$ over all sets $S \subseteq G$ of size $k$. Then there exists a constant $C$ such that

$$\lim_{N_A \to \infty} \lim_{N \to \infty} \frac{\bar{\ell}_k^\circ}{N_A} = CL_k,$$

where $L_k = \sum_{j=1}^{k-1} 1/j$.

The above assumptions are quite flexible in allowing for various kinds of family relationships to be described. In particular, Property 5 holds for any model for which the adult population converges (in the specified limits, under some rescaling of time) to the Kingman coalescent [17]. Many population models satisfy this property, including models with two distinct sexes [5, 18, 60]. We also note that Properties 2 and 4 do not require independence across individuals in the survival of juveniles to adulthoood, nor in the production of offspring by adults. Properties 1-5 suffice to ensure that Eqs. (8.3)–(8.7) of Section 8.2 remain valid.

Within such a model, we consider two individuals $i, j \in I$ with a particular kin relationship (siblings, cousins, etc), quantified by two probabilities $p_1$ and $p_2$. With probability $p_1$, site $i_1$ shares a recent common ancestor with $j_1$. Likewise, with probability $p_2$, site $i_2$ shares a recent common ancestor with $j_2$. For example, if we suppose that sites $i_1$ and $j_1$ are maternally inherited, then maternal half-siblings have $p_1 = 1/2$ and $p_2 = 0$. To state this formally, we assume there is some fixed $T \geq 1$ such that, under neutral drift ($\delta = 0$), asymptotically as first $N \to \infty$ and then $N_A \to \infty$, the ancestral map $A_0^T$ has the following properties:

(i) With probability $p_1$, $A_0^T(i_1) = A_0^T(j_1)$, otherwise $A_0^T(i_1)$ and $A_0^T(j_1)$ are distributed uniformly and independently over $G$;

(ii) With probability $p_2$, $A_0^T(i_2) = A_0^T(j_2)$, otherwise $A_0^T(i_2)$ and $A_0^T(j_2)$ are distributed uniformly and independently over $G$;

(iii) The events described in (i) and (ii) are independent of each other.

$T$ represents the number of generations needed to characterize the relationship: $T = 1$ for siblings, $T = 2$ for cousins, and so on.

**8.6.1. Relatedness of one relative to another.** We first compute the relatedness $r^\circ_{\{i\},j}$ of (the set containing) one relative to the other:

$$r^\circ_{\{i\},j} = \lim_{N_A \to \infty} \lim_{N \to \infty} \frac{\left\langle (h(x_{i_1} + x_{i_2}) + (1-2h)x_{i_1}x_{i_2}) \left(\frac{1}{2}(x_{j_1} + x_{j_2}) - \bar{x}\right) \right\rangle^\circ}{\langle \bar{x}(1-\bar{x}) \rangle^\circ}. \tag{8.46}$$

The bracketed quantity in the numerator can be expanded as

$$h\left(\frac{x_{i_1}x_{j_1} + x_{i_2}x_{j_2}}{2} + \frac{x_{i_1}x_{j_2} + x_{i_2}x_{j_1}}{2} - 2\left(\frac{x_{i_1}\bar{x} + x_{i_2}\bar{x}}{2}\right)\right)$$
$$+ (1-2h)\left(\frac{x_{i_1}x_{i_2}x_{j_1} + x_{i_1}x_{i_2}x_{j_2}}{2} - x_{i_1}x_{i_2}\bar{x}\right).$$

We again make use of the $[\![\,]\!]$ operation, defined in Eq. (8.11), to express this relatedness:

$$h\left(2\left[\!\left[\frac{x_{i_1}\bar{x} + x_{i_2}\bar{x}}{2}\right]\!\right] - \left[\!\left[\frac{x_{i_1}x_{j_1} + x_{i_2}x_{j_2}}{2}\right]\!\right] - \left[\!\left[\frac{x_{i_1}x_{j_2} + x_{i_2}x_{j_1}}{2}\right]\!\right]\right)$$
$$+ (1-2h)\left([\![x_{i_1}x_{i_2}\bar{x}]\!] - \left[\!\left[\frac{x_{i_1}x_{i_2}x_{j_1} + x_{i_1}x_{i_2}x_{j_2}}{2}\right]\!\right]\right). \tag{8.47}$$

Using Eq. (8.13) and the given relationship between $i$ and $j$, we compute

$$[\![x_{i_1}\bar{x}]\!] = [\![x_{i_2}\bar{x}]\!] = [\![x_{i_1}x_{j_2}]\!] = [\![x_{i_2}x_{j_1}]\!] = L_2 = 1$$
$$[\![x_{i_1}x_{j_1}]\!] = p_1 L_1 + (1-p_1)L_2 = 1 - p_1$$
$$[\![x_{i_2}x_{j_2}]\!] = p_2 L_1 + (1-p_2)L_2 = 1 - p_2$$
$$[\![x_{i_1}x_{i_2}\bar{x}]\!] = L_3 = \frac{3}{2}$$
$$[\![x_{i_1}x_{i_2}x_{j_1}]\!] = p_1 L_2 + (1-p_1)L_3 = \frac{3-p_1}{2}$$
$$[\![x_{i_1}x_{i_2}x_{j_2}]\!] = p_2 L_2 + (1-p_2)L_3 = \frac{3-p_2}{2}.$$

Substituting in Eq. (8.47) yields

$$r^\circ_{\{i\},j} = h\left(2(1) - \left(1 - \frac{p_1+p_2}{2}\right) - 1\right) + (1-2h)\left(\frac{3}{2} - \frac{6-p_1-p_2}{4}\right)$$
$$= h\left(\frac{p_1+p_2}{2}\right) + (1-2h)\left(\frac{p_1+p_2}{4}\right)$$
$$= \frac{p_1+p_2}{4}.$$

Letting $r = (p_1+p_2)/2$ denote Wright's coefficient of relationship [62], we conclude that

$$r^\circ_{\{i\},j} = \frac{r}{2}. \tag{8.48}$$

***8.6.2. Intra-relatedness of two relatives.*** Next we compute the intra-relatedness $r^\circ_{\{i,j\}}$ of the set containing the two relatives together:

$$r^\circ_{\{i,j\}} = \lim_{N_A \to \infty} \lim_{N \to \infty} \frac{\left\langle \varphi_i(x_{i_1}, x_{i_2})\, \varphi_j(x_{j_1}, x_{j_2}) \left( \frac{x_{i_1}+x_{i_2}+x_{j_1}+x_{j_2}}{4} - \bar{x} \right) \right\rangle^\circ}{\left\langle \sum_{g \in G} x_g (x_g - \bar{x}) \right\rangle^\circ}, \tag{8.49}$$

where $\varphi_i(x_{i_1}, x_{i_2})$ and $\varphi_j(x_{j_1}, x_{j_2})$ are as given in Eq. (1.4). For the sake of symmetry, $r^\circ_{\{i,j\}}$ is expressed in Eq. (8.49) as the average of $r^\circ_{\{i,j\},i}$ and $r^\circ_{\{i,j\},j}$; we recall from Section 4.3.2 that $r^\circ_{\{i,j\},i} = r^\circ_{\{i,j\},j} = r^\circ_{\{i,j\}}$.

The bracketed quantity in the numerator can be expanded as follows:

$$\begin{aligned}
&\varphi_i(x_{i_1}, x_{i_2})\, \varphi_j(x_{j_1}, x_{j_2}) \left( \frac{x_{i_1} + x_{i_2} + x_{j_1} + x_{j_2}}{4} - \bar{x} \right) \\
&= h^2 (x_{i_1} + x_{i_2})(x_{j_1} + x_{j_2}) \left( \frac{x_{i_1} + x_{i_2} + x_{j_1} + x_{j_2}}{4} - \bar{x} \right) \\
&\quad + h(1-2h) \left( x_{i_1} x_{i_2}(x_{j_1} + x_{j_2}) + x_{j_1} x_{j_2}(x_{i_1} + x_{i_2}) \right) \left( \frac{x_{i_1} + x_{i_2} + x_{j_1} + x_{j_2}}{4} - \bar{x} \right) \\
&\quad + (1-2h)^2 x_{i_1} x_{i_2} x_{j_1} x_{j_2} \left( \frac{x_{i_1} + x_{i_2} + x_{j_1} + x_{j_2}}{4} - \bar{x} \right) \\
&= h^2 \left( \frac{x_{i_1} x_{j_1} + x_{i_2} x_{j_2}}{2} + \frac{x_{i_1} x_{j_2} + x_{i_2} x_{j_1}}{2} \right. \\
&\quad + 2 \left( \frac{x_{i_1} x_{i_2} x_{j_1} + x_{i_1} x_{j_1} x_{j_2} + x_{i_1} x_{i_2} x_{j_2} + x_{i_2} x_{j_1} x_{j_2}}{4} \right) \\
&\quad \left. -2 \left( \frac{x_{i_1} x_{j_1} \bar{x} + x_{i_2} x_{j_2} \bar{x}}{2} \right) - 2 \left( \frac{x_{i_1} x_{j_2} \bar{x} + x_{i_2} x_{j_1} \bar{x}}{2} \right) \right) \\
&\quad + h(1-2h) \left( 3 \left( \frac{x_{i_1} x_{i_2} x_{j_1} + x_{i_1} x_{j_1} x_{j_2} + x_{i_1} x_{i_2} x_{j_2} + x_{i_2} x_{j_1} x_{j_2}}{4} \right) + x_{i_1} x_{i_2} x_{j_1} x_{j_2} \right. \\
&\quad \left. -4 \left( \frac{x_{i_1} x_{i_2} x_{j_1} \bar{x} + x_{i_1} x_{j_1} x_{j_2} \bar{x} + x_{i_1} x_{i_2} x_{j_2} \bar{x} + x_{i_2} x_{j_1} x_{j_2} \bar{x}}{4} \right) \right) \\
&\quad + (1-2h)^2 \left( x_{i_1} x_{i_2} x_{j_1} x_{j_2} - x_{i_1} x_{i_2} x_{j_1} x_{j_2} \bar{x} \right).
\end{aligned}$$

We can therefore express the relatedness $r^\circ_{\{i,j\}}$ as

$$\begin{aligned}
r^\circ_{\{i,j\}} = h^2 &\left( 2 \left[\!\!\left[ \frac{x_{i_1} x_{j_1} \bar{x} + x_{i_2} x_{j_2} \bar{x}}{2} \right]\!\!\right] + 2 \left[\!\!\left[ \frac{x_{i_1} x_{j_2} \bar{x} + x_{i_2} x_{j_1} \bar{x}}{2} \right]\!\!\right] \right. \\
&\quad - \left[\!\!\left[ \frac{x_{i_1} x_{j_1} + x_{i_2} x_{j_2}}{2} \right]\!\!\right] - \left[\!\!\left[ \frac{x_{i_1} x_{j_2} + x_{i_2} x_{j_1}}{2} \right]\!\!\right] \\
&\quad \left. -2 \left[\!\!\left[ \frac{x_{i_1} x_{i_2} x_{j_1} + x_{i_1} x_{j_1} x_{j_2} + x_{i_1} x_{i_2} x_{j_2} + x_{i_2} x_{j_1} x_{j_2}}{4} \right]\!\!\right] \right) \\
&+ h(1-2h) \left( 4 \left[\!\!\left[ \frac{x_{i_1} x_{i_2} x_{j_1} \bar{x} + x_{i_1} x_{j_1} x_{j_2} \bar{x} + x_{i_1} x_{i_2} x_{j_2} \bar{x} + x_{i_2} x_{j_1} x_{j_2} \bar{x}}{4} \right]\!\!\right] \right. \\
&\quad -3 \left[\!\!\left[ \frac{x_{i_1} x_{i_2} x_{j_1} + x_{i_1} x_{j_1} x_{j_2} + x_{i_1} x_{i_2} x_{j_2} + x_{i_2} x_{j_1} x_{j_2}}{4} \right]\!\!\right] \\
&\quad \left. - [\![ x_{i_1} x_{i_2} x_{j_1} x_{j_2} ]\!] \right) \\
&+ (1-2h)^2 \left( [\![ x_{i_1} x_{i_2} x_{j_1} x_{j_2} \bar{x} ]\!] - [\![ x_{i_1} x_{i_2} x_{j_1} x_{j_2} ]\!] \right).
\end{aligned}$$

Using Eq. (8.13) and the given relationship between $i$ and $j$, we compute

$$[\![x_{i_1}x_{j_2}]\!] = [\![x_{i_2}x_{j_1}]\!] = L_2 = 1$$
$$[\![x_{i_1}x_{j_1}]\!] = p_1 L_1 + (1-p_1)L_2 = 1 - p_1$$
$$[\![x_{i_2}x_{j_2}]\!] = p_2 L_1 + (1-p_2)L_2 = 1 - p_2$$
$$[\![x_{i_1}x_{i_2}x_{j_1}]\!] = [\![x_{i_1}x_{j_1}x_{j_2}]\!] = [\![x_{i_1}x_{j_1}\bar{x}]\!] = p_1 L_2 + (1-p_1)L_3 = \frac{3-p_1}{2}$$
$$[\![x_{i_1}x_{i_2}x_{j_2}]\!] = [\![x_{i_2}x_{j_1}x_{j_2}]\!] = [\![x_{i_2}x_{j_2}\bar{x}]\!] = p_2 L_2 + (1-p_2)L_3 = \frac{3-p_2}{2}$$
$$[\![x_{i_1}x_{j_2}\bar{x}]\!] = [\![x_{i_2}x_{j_1}\bar{x}]\!] = L_3 = \frac{3}{2}$$
$$[\![x_{i_1}x_{i_2}x_{j_1}\bar{x}]\!] = [\![x_{i_1}x_{j_1}x_{j_2}\bar{x}]\!] = p_1 L_3 + (1-p_1)L_4 = \frac{11-2p_1}{6}$$
$$[\![x_{i_1}x_{i_2}x_{j_2}\bar{x}]\!] = [\![x_{i_2}x_{j_1}x_{j_2}\bar{x}]\!] = p_2 L_3 + (1-p_2)L_4 = \frac{11-2p_2}{6}$$
$$[\![x_{i_1}x_{i_2}x_{j_1}x_{j_2}]\!] = p_1 p_2 L_2 + (p_1 + p_2 - 2p_1 p_2) L_3$$
$$+ (1-p_1)(1-p_2) L_4$$
$$= \frac{11 - 2p_1 - 2p_2 - p_1 p_2}{6}$$
$$[\![x_{i_1}x_{i_2}x_{j_1}x_{j_2}\bar{x}]\!] = p_1 p_2 L_3 + (p_1 + p_2 - 2p_1 p_2) L_4$$
$$+ (1-p_1)(1-p_2) L_5$$
$$= \frac{25 - 3p_1 - 3p_2 - p_1 p_2}{12}.$$

Subsituting, and using $r = (p_1 + p_2)/2$ for Wright's coefficient of relationship, we obtain

$$r^\circ_{\{i,j\}} = h^2 \left( 2\left(\frac{3-r}{2}\right) + 2\left(\frac{3}{2}\right) - (1-r) - 1 - 2\left(\frac{3-r}{2}\right) \right)$$
$$+ h(1-2h)\left( 4\left(\frac{11-2r}{6}\right) - 3\left(\frac{3-r}{2}\right) - \frac{11 - 4r - p_1 p_2}{6} \right)$$
$$+ (1-2h)^2 \left( \frac{25 - 6r - p_1 p_2}{12} - \frac{11 - 4r - p_1 p_2}{6} \right).$$

Simplifying, we arrive at

$$r^\circ_{\{i,j\}} = \frac{1+r}{4} + \frac{\left(h - \frac{1}{2}\right)(r - p_1 p_2)}{6}. \quad (8.50)$$

***8.6.3. Two-player game.*** We now suppose $i$ and $j$ interact according to the matrix game

$$\begin{array}{c} \\ \text{C} \\ \text{D} \end{array} \begin{pmatrix} \text{C} & \text{D} \\ b - c + d & -c \\ b & d \end{pmatrix}, \quad (8.51)$$

with phenotypes 1 and 0 corresponding to strategies C and D, respectively. Any $2 \times 2$ matrix game can be written in this form, up to an additive constant. Specifically, let $\pi_{XY}$ denote the payoff to strategy $X$ interacting with strategy $Y$ in an arbitrary $2 \times 2$ matrix game, with $X, Y \in \{\text{C}, \text{D}\}$. Then the cost, benefit, and synergy parameters can be defined, respectively, as

$$c = \frac{1}{2}(\pi_{\text{DC}} + \pi_{\text{DD}}) - \frac{1}{2}(\pi_{\text{CC}} + \pi_{\text{CD}}) \tag{8.52a}$$

$$b = \frac{1}{2}(\pi_{\text{CC}} + \pi_{\text{DC}}) - \frac{1}{2}(\pi_{\text{CD}} + \pi_{\text{DD}}) \tag{8.52b}$$

$$d = \frac{1}{2}(\pi_{\text{CC}} + \pi_{\text{DD}}) - \frac{1}{2}(\pi_{\text{CD}} + \pi_{\text{DC}}), \tag{8.52c}$$

and we have

$$\begin{pmatrix} \pi_{\text{CC}} & \pi_{\text{CD}} \\ \pi_{\text{DC}} & \pi_{\text{DD}} \end{pmatrix} = \frac{-\pi_{\text{CC}} + \pi_{\text{CD}} + \pi_{\text{DC}} + \pi_{\text{DD}}}{2} \begin{pmatrix} 1 & 1 \\ 1 & 1 \end{pmatrix} + \begin{pmatrix} b - c + d & -c \\ b & d \end{pmatrix}. \tag{8.53}$$

The addition of a constant matrix in Eq. (8.53) does not affect weak selection, as can be seen from the subtraction of $\bar{f}(\mathbf{\Phi})$ in Eq. (8.3). Therefore, no generality is lost in using the game matrix in Eq. (8.52).

Given this game matrix, the payoff to a given individual $i$ interacting with its partner, $j$, can be written

$$\begin{aligned} f_i(\mathbf{\Phi}) &= -c\Phi_i + b\Phi_j + d(\Phi_i \Phi_j + (1 - \Phi_i)(1 - \Phi_j)) \\ &= d + (-c - d)\Phi_i + (b - d)\Phi_j + 2d\Phi_i \Phi_j. \end{aligned} \tag{8.54}$$

Now applying Condition (8.7), weak selection favors allele A (corresponding to strategy C) if and only if

$$(-c - d)r_i^\circ + (b - d)r_{\{j\},i}^\circ + 2dr_{\{i,j\}}^\circ > 0. \tag{8.55}$$

Substituting the relatedness coefficients from Eq. (8.16), Eq. (8.48), and Eq. (8.50), this condition becomes

$$-c + br + \frac{2d}{3}\left(h - \frac{1}{2}\right)(r - p_1 p_2) > 0, \tag{8.56}$$

which is Condition (18) of the main text.

**8.7. Relationship to prior results.** We compare our results to those of prior works that modeled synergistic and collective interactions among relatives.

***8.7.1. Jones (2000).*** Jones [63] considered the problem of whether selection favors two full siblings to help a third. Their scenario amounts to the Collective Action Dilemma with conditional costs and all-or-nothing benefits, where the acting collective consists of two siblings and the target is another sibling. As in our framework, the helping phenotype is conferred by an allele at a single biallelic locus with genetic dominance $h$. Their scenario can therefore be seen as the $m = 2$ case of the scenario we consider in Section 8.5.3 above. Jones finds that the Cooperator allele increases from a given frequency $x$ if and only if $bR > c$, where the relatedness quantity $R$ is given by

$$R = \frac{2h^2 + x(1 + 4h) + x^2(3 - 12h^2) + x^3(2 - 8h + 8h^2)}{4h^2 + x(2 + 6h - 4h^2) + x^2(4 - 2h - 12h^2) + x^3(2 - 8h + 8h^2)}. \tag{8.57}$$

For comparison, our Condition (8.45), for a collective of $m = 2$ siblings, can be written in the form $bR > c$, where

$$R = \frac{r_J^\circ}{r_{J,i}^\circ} = \frac{12}{17 + 2h}. \tag{8.58}$$

The results in Eq. (8.57) and Eq. (8.58) are not directly comparable, since Jones's result pertains to a particular Cooperator allele frequency $x$, whereas ours reflects the entire process of selection. These results coincide, however, in one case of interest: if there is no genetic dominance ($h = 1/2$) and both alleles are equally abundant ($x = 1/2$), then Eq. (8.57) and Eq. (8.58) both give $R = 2/3$, indicating that helping is favored if $2b > 3c$.

**8.7.2. Garay et al. (2023).** Garay et al. [64] also investigate collective action among some siblings to help another. In the case that the benefits of help scale linearly with the number of helpers, they obtain the condition $b > 2c$ for a collective of full siblings to help another, in agreement with our Theorem 8.1. However, for a nonlinear benefit function, they obtain a more complicated condition. This contrasts with our finding in Theorem 8.1 that the $b > 2c$ condition is valid for all benefit functions, linear or nonlinear.

This apparent difference in results stems from the criteria used to characterize selection. Garay et al. [64] ask whether collective help is evolutionarily stable, i.e. robust to invasion by a non-helper mutation. They assume the helping behavior is recessive, meaning that a single copy of a non-helper mutant allele eliminates the helping behavior.

In contrast, we have characterized selection here in terms of pairwise fixation probabilities, or equivalently, the low-mutation limit of the stationary distribution (see Theorem 3.3). This condition for success is distinct from—albeit related to—the evolutionary stability condition used by Garay et al. [64]. The relationship between these selection criteria is discussed, for example, in Refs. [59, 65]. This difference in selection criteria explains the different conditions obtained.

**8.7.3. Games between haploid relatives.** A number of works [20, 66–68] have analyzed the evolutionary dynamics of two-player games played by haploid relatives. A typical setup involves a large population with two heritable types, corresponding to the strategies in a $2 \times 2$ matrix game. Each individual plays with a large number of partners, a fraction $r$ of which are clonal relatives of the individual (guaranteed to have the same type). The remaining fraction, $1 - r$, are drawn from the population at large.

For the game in Eq. (8.51), Strategy 1 will increase from a particular frequency, $x$, if and only if

$$-c + br + 2d(1-r)\left(x - \frac{1}{2}\right) > 0. \tag{8.59}$$

This is equivalent to Eq. (5a) of Queller [67] and Eq. (3.3) of Ohtsuki [20], although the parameters $b$, $c$, and $d$ have different meanings in these works.

Eq. (8.59) expresses a similar idea to Eq. (8.56)—which is Condition (4) of the main text—in that Hamilton's rule is extended to incorporate synergy between two relatives. The differences are that (i) Eq. (8.59) applies to a haploid population, while Eq. (8.56) applies to diploid relatives with arbitrary genetic dominance, and (ii) Eq. (8.59) pertains to a particular allele frequency, $x$, while Eq. (8.56) pertains to the overall process of selection.

**8.7.4. Altruism between two diploid siblings.** Roze and Rousset [69] consider a game played with altruistic and non-altruistic strategies, played by diploid relatives. They allow for any degree of dominance in the altruistic phenotype (arbitrary $h$) but restrict to additive games ($d = 0$, in our parameterization). They find, in their Eq. (22), that the altruistic allele increases if $br > c$, recovering Hamilton's rule [70]. This coincides with our result, Eq. (8.56), in the case of an additive game ($d = 0$).

The combined effects of nonadditivity and dominance in sibling interactions were explored by Cavalli-Sforza and Feldman [71] and Uyenoyama and Feldman [72]. In their setup, an altruistic act has cost $0 < \gamma < 1$ and benefit $\beta > 0$, and these effects combine multiplicatively rather than additively. This interaction can be represented by the game matrix

$$\begin{array}{c} \\ \text{C} \\ \text{D} \end{array} \begin{pmatrix} \text{C} & \text{D} \\ (1-\gamma)(1+\beta) & 1-\gamma \\ 1+\beta & 1 \end{pmatrix}. \tag{8.60}$$

The cost, benefit, and synergy parameters for this game are

$$c = \gamma + \frac{\beta\gamma}{2} \qquad b = \beta - \frac{\beta\gamma}{2} \qquad d = -\frac{\beta\gamma}{2}. \tag{8.61}$$

In particular, this game is anti-synergistic ($d < 0$). Cavalli-Sforza and Feldman [71] obtain that the C allele can invade D when rare if $-\gamma + \beta r - \gamma\beta rh > 0$, and can resist invasion by D if $-\gamma + \beta r - \gamma\beta(rh+1) > 0$. These conditions are not directly comparable to our Eq. (8.56), as they pertain to evolutionary stability whereas ours measures relative abundance in the stationary distribution. However, they share the feature that, for an anti-synergistic game, the conditions for cooperation become more stringent as the dominance, $h$, of the cooperative trait increases.

## 9. Collective action on networks

We now analyze populations structured as weighted networks. This extends previous analyses of evolutionary games on graphs [73–79] to arbitrary nonlinear interactions.

**9.1. Model.** To construct this model, we endow the set of sites $G$ with the structure of a weighted (undirected) graph. Each vertex houses a haploid individual. The edge weight between sites $g, h \in G$ is denoted $w_{gh} = w_{hg}$. We define the weighted degree of each vertex $g \in G$ as $d_g = \sum_{h \in G} w_{gh}$. Let $p_{gh}^{(n)}$ denote the $n$-step random walk probability from $g$ to $h$, using step probabilities $p_{gh} = w_{gh}/d_g$.

In each state $\mathbf{x}$, each site $g \in G$ is assigned a payoff $f_g(\mathbf{x})$. For now, this payoff may depend arbitrarily on $\mathbf{x}$; we will apply this to the Collective Action Dilemma in Section 9.4 below. State transitions follow the death-Birth rule [74, 79]: At each time-step, one site $h \in G$ is chosen uniformly at random to be replaced. A neighboring site $g \in G$ is then chosen, with probability proportional to $p_{hg}(1 + \delta f_g(\mathbf{x}))$ to produce an offspring, which fills the vacancy in site $h$. With probability $u$, the offspring acquires a mutation; otherwise it inherits the allele of the parent.

For death-Birth updating, the reproductive value $v_g$ of each site $g \in G$ is proportional to its weighted degree [33, 79]:

$$v_g = \frac{n d_g}{\sum_{h \in G} d_h}. \tag{9.1}$$

These reproductive values possess the reversibility property

$$v_g p_{gh}^{(n)} = v_h p_{hg}^{(n)}, \tag{9.2}$$

for all $g, h \in G$ and $n \geq 0$ [79].

**9.2. Analysis of selection.** We compute the fitness increment of site $g \in G$ in state $\mathbf{x}$ according to Eq. (3.5), making use of the reversibility property Eq. (9.2):

$$\begin{aligned}
w_g(\mathbf{x}) &= \left(\frac{N-1}{N}\right) v_g + \frac{1}{N} \sum_{h \in G} \left(\frac{p_{hg}(1 + \delta f_g(\mathbf{x}))}{\sum_{k \in G} p_{hg}(1 + \delta f_g(\mathbf{x}))}\right) v_h - v_g \\
&= \frac{1}{N} \left(\sum_{h \in G} \left(\frac{v_h p_{hg}(1 + \delta f_g(\mathbf{x}))}{1 + \delta \sum_{k \in G} p_{hk} f_k(\mathbf{x})}\right) - v_g\right) \\
&= \frac{1}{N} \left(\sum_{h \in G} \left(\frac{v_g p_{gh}(1 + \delta f_g(\mathbf{x}))}{1 + \delta \sum_{k \in G} p_{hk} f_k(\mathbf{x})}\right) - v_g\right) \\
&= \frac{v_g}{N} \left(\sum_{h \in G} \frac{p_{gh}(1 + \delta f_g(\mathbf{x}))}{1 + \delta \sum_{k \in G} p_{hk} f_k(\mathbf{x})} - 1\right).
\end{aligned}$$

For weak selection, taking the $\delta$-derivative at $\delta = 0$ and using $\sum_{h \in G} p_{gh} = 1$ gives

$$w'_g(\mathbf{x}) = \frac{v_g}{N} \sum_{h \in G} p_{gh} \left( f_g(\mathbf{x}) - \sum_{k \in G} p_{hk} f_k(\mathbf{x}) \right)$$

$$= \frac{v_g}{N} \left( f_g(\mathbf{x}) - \sum_{h \in G} p_{gh}^{(2)} f_h(\mathbf{x}) \right).$$

An equivalent result was obtained in Eq. (78) in the Supplementary Information of Allen et al. [79]. Applying Eq. (3.9), the weak selection increment in state $\mathbf{x}$ is

$$\Delta'(\mathbf{x}) = \frac{1}{N^2} \sum_{g \in G} v_g x_g \left( f_g(\mathbf{x}) - \sum_{h \in G} p_{gh}^{(2)} f_h(\mathbf{x}) \right). \quad (9.3)$$

To specialize our main condition to this model, we follow the procedure of Section 7.2 and write the payoffs to each site in unique polynomial form:

$$f_g(\mathbf{x}) = \sum_{S \subseteq G} C_{S,g} \, \iota_S^A(\mathbf{x}). \quad (9.4)$$

Then substituting in Eq. (9.3) and making use of Eq. (4.3) and Eq. (9.2), we obtain

$$\Delta'(\mathbf{x}) = \frac{1}{N^2} \sum_{g \in G} \sum_{S \subseteq G} v_g x_g \left( C_{S,g} - \sum_{h \in G} p_{gh}^{(2)} C_{S,h} \right) \iota_S^A(\mathbf{x})$$

$$= \frac{1}{N^2} \sum_{g \in G} \sum_{S \subseteq G} v_g \left( C_{S,g} - \sum_{h \in G} p_{gh}^{(2)} C_{S,h} \right) \iota_{S \cup \{g\}}^A(\mathbf{x})$$

$$= \frac{1}{N^2} \sum_{g \in G} \sum_{S \subseteq G} \left( v_g C_{S,g} - \sum_{h \in G} v_h p_{hg}^{(2)} C_{S,h} \right) \iota_{S \cup \{g\}}^A(\mathbf{x})$$

$$= \frac{1}{N^2} \sum_{g \in G} \sum_{S \subseteq G} v_g C_{S,g} \left( \iota_{S \cup \{g\}}^A(\mathbf{x}) - \sum_{h \in G} p_{gh}^{(2)} \iota_{S \cup \{h\}}^A(\mathbf{x}) \right).$$

Applying Theorem 3.4 and Eq. (4.10a), weak selection favors allele $A$ if and only if

$$\sum_{g \in G} \sum_{S \subseteq G} v_g C_{S,g} \left( r_{S,g}^\circ - \sum_{h \in G} p_{gh}^{(2)} r_{S,h}^\circ \right) > 0. \quad (9.5)$$

**9.3. Coalescence lengths.** For purposes of computation, it can be more convenient to write Condition (9.5) in terms of coalescence lengths:

$$\sum_{g \in G} \sum_{S \subseteq G} v_g C_{S,g} \left( \sum_{h \in G} p_{gh}^{(2)} \ell_{S \cup \{h\}}^\circ - \ell_{S \cup \{g\}}^\circ \right) > 0. \quad (9.6)$$

The neutral coalescence lengths can be obtained using Eq. (4.8). Since each site is replaced with probability $1/N$, and mutation occurs with probability $u$ in each new offspring, mutations arise at rate $\nu_g = 1/N$ at each site $g \in G$, and $\nu_S = |S|/N$ for each $S \subseteq G$. Eq. (4.8) then becomes

$$\ell_S^\circ = \begin{cases} \dfrac{|S|}{N} + \dfrac{N - |S|}{N} \ell_S^\circ + \dfrac{1}{N} \sum_{g \in S} \sum_{h \in G} p_{gh} \, \ell_{(S - \{g\}) \cup \{h\}}^\circ & |S| \geq 2 \\ 0 & |S| = 1, \end{cases} \quad (9.7)$$

which simplifies to

$$\ell_S^\circ = \begin{cases} 1 + \dfrac{1}{|S|} \displaystyle\sum_{g \in S} \sum_{h \in G} p_{gh}\, \ell_{(S-\{g\}) \cup \{h\}}^\circ & |S| \geq 2 \\ 0 & |S| = 1. \end{cases} \quad (9.8)$$

**9.4. Conditions for collective action.** We now apply the above results to the Collective Action Dilemma with unconditional costs (conditional costs can be analyzed similarly). The payoff function for an actor set $S$ and target $g$ is given by Eq. (6.1). Applying Condition (9.5), weak selection favors allele $A$ if and only if

$$-\frac{c}{m} \sum_{h \in S} v_h \left( r_{\{h\},h}^\circ - \sum_{k \in G} p_{hk}^{(2)} r_{\{h\},k}^\circ \right) + b\, v_g \left( r_{S,g}^\circ - \sum_{k \in G} p_{gk}^{(2)} r_{S,k}^\circ \right) > 0. \quad (9.9)$$

By Eq. (9.1), the reproductive values, $v_g$, may be replaced by the weighted degrees, $d_g$, recovering Eq. (21) of the main text. We can rewrite Condition (9.9) in the form $b\kappa_{S,g} > c$, where $\kappa_{S,g}$ is the cost-benefit threshold for $S$ to act on $g$:

$$\kappa_{S,g} = \frac{d_g \left( r_{S,g}^\circ - \sum_{k \in G} p_{gk}^{(2)} r_{S,k}^\circ \right)}{\frac{1}{m} \sum_{h \in S} d_h \left( r_{\{h\},h}^\circ - \sum_{k \in G} p_{hk}^{(2)} r_{\{h\},k}^\circ \right)} \quad (9.10a)$$

$$= \frac{d_g \left( \sum_{k \in G} p_{gk}^{(2)} \ell_{S \cup \{k\}}^\circ - \ell_{S \cup \{g\}}^\circ \right)}{\frac{1}{m} \sum_{h \in S} d_h \sum_{k \in G} p_{hk}^{(2)} \ell_{\{h,k\}}^\circ}. \quad (9.10b)$$

The cost-benefit ratio $\kappa_{S,g}$ also corresponds to "scaled relatedness" [26, 80] or "compensated relatedness" [81].

Eq. (9.10b) is particularly useful for computation as well as theoretical results. Indeed, a number of general results follow directly from this equation:

- The denominator in Eq. (9.10b)—and hence in Eq. (9.10a) as well—is always nonnegative. It is zero only for star graphs, as we show in the next bullet. Outside of star graphs, the sign of $\kappa_{S,g}$ is the same as that of $r_{S,g}^\circ - \sum_{k \in G} p_{gk}^{(2)} r_{S,k}^\circ$ (or equivalently, of $\sum_{k \in G} p_{gk}^{(2)} \ell_{S \cup \{k\}}^\circ - \ell_{S \cup \{g\}}^\circ$). If $r_{S,g}^\circ > \sum_{k \in G} p_{gk}^{(2)} r_{S,k}^\circ$ then collective help can be favored for sufficiently large $b > 0$ (relative to fixed $c > 0$). If $r_{S,g}^\circ < \sum_{k \in G} p_{gk}^{(2)} r_{S,k}^\circ$ then collective help cannot be favored, but collective harm can be favored for sufficiently negative $b$. In short, for colletive help to be favored, it is necessary that $S$ be more related to $g$ than to $g$'s two-step neighbors.

- The denominator in Eq. (9.10b)—and hence in Eq. (9.10a) as well—is zero if and only if every $h \in S$ is the only two-step neighbor of itself. Since $G$ is connected, this can only occur if $G$ is a (possibly weighted) star graph, and $S$ is a singleton set containing only the hub. In this case, the numerator also evaluates to zero for any target vertex $g$. The hub of a star graph has the unique property that its reproductive success does not depend on the payoff of any vertex: it reproduces if and only if a leaf vertex dies. Thus, no payoff-affecting action is either favored or disfavored for the hub of a star graph. Aside from star graphs, the denominator is always positive.

- If the graph is vertex-transitive—meaning that for ever pair of vertices $h, k \in G$ there is an automorphism $\sigma$ of $G$ with $\sigma(h) = k$ [75, 82]—then the denominator in Eq. (9.10b) equals $N - 2$. This follows from Eq. (14) of the SI of Ref. [83].

- If $g \in S$, then $\ell_{S \cup \{g\}}^\circ = \ell_S^\circ$ and $\sum_{k \in G} p_{gk}^{(2)} \ell_{S \cup \{k\}}^\circ \geq \ell_S^\circ$, which together imply that $r_{S,g}^\circ \geq \sum_{k \in G} p_{gk}^{(2)} r_{S,k}^\circ$. It follows that a collective cannot be favored to harm any of its own members.

- Suppose that site $g$ and all of its two-step neighbors are in $S$. Then $\sum_{k \in G} p_{gk}^{(2)} \ell_{S \cup \{k\}}^{\circ} = \ell_S^{\circ} = \ell_{S \cup \{g\}}^{\circ}$, and it follows that $r_{S,g}^{\circ} = \sum_{k \in G} p_{gk}^{(2)} r_{S,k}^{\circ}$. From this we conclude that if $g$ and all of its two-step neighbors are in $S$, then $S$ is never favored to collectively help or harm $g$. Any help or harm to $g$ has the opposite effect on $g$'s two-step neighbors; these effects cancel out to yield no net effect on the spread of allele $A$.

**9.5. Cycles.** We are now in a position to compute conditions for collective help or harm on particular graph families, via Eq. (9.10b). Results are presented in Fig. S1. We start with the cycle (main text, Fig. 2b), which consists of $N$ vertices, each joined to exactly two others.

**9.5.1. States.** The cycle is convenient to analyze in that, instead of considering the full state $\mathbf{x} \in \{0,1\}^G$, one need only keep track of the number of $A$ alleles. This is because, in the death-Birth process without mutation (with initial state sampled from the mutant appearance distribution $\mu$) the only possible states are those for which the $A$ and $a$ alleles each form contiguous blocks of adjacent vertices.

We therefore index the possible states as $k = 0, \ldots, N$, where $k$ indicates the number of $A$ alleles. The initial state (sampled from $\mu$) is either $k = 1$ or $k = N - 1$, with probability $1/2$ each. Under neutral drift without mutation ($u = \delta = 0$), a state with $k$ contiguous $A$ alleles will transition to $k + 1$ or $k - 1$ contiguous $A$ alleles with probability $1/N$ each, and will otherwise remain with $k$ contiguous $A$ alleles.

**9.5.2. Sojourn times.** We compute the neutral bracket operation $\langle \ \rangle^{\circ}$ for death-Birth on the cycle via sojourn times, using Lemma 2.2. For $1 \le k \le N - 1$, let $\sigma_k$ denote the expected duration of time (i.e., sojourn time) spent in states with exactly $k$ $A$ alleles, under neutral drift with initial state sampled from $\mu$:

$$\sigma_k = \mathbb{P}_{\mathcal{M}_0}^{\circ} \left[ \sum_{g \in G} X_g^t = k \ \Big| \ \mathbf{X}^0 \sim \mu \right]. \tag{9.11}$$

Allen & McAvoy [5, Appendix A.3] provide recurrence equations that uniquely determine sojourn times from a given initial distribution over states. For dB on the cycle with $u = \delta = 0$, using the transition probabilities from the previous subsection, these recurrence equations become:

$$\sigma_k = \begin{cases} \dfrac{1}{2} + \dfrac{(N-2)\sigma_1 + \sigma_2}{N} & k = 1 \\[2mm] \dfrac{\sigma_{k-1} + (N-2)\sigma_k + \sigma_{k+1}}{N} & 2 \le k \le N-2 \\[2mm] \dfrac{1}{2} + \dfrac{\sigma_{N-2} + (N-2)\sigma_{N-1}}{N} & k = N-1. \end{cases} \tag{9.12}$$

The unique solution is $\sigma_k = N/2$ for all $k = 1, \ldots, N - 1$. This means that the death-Birth process on the cycle, with $u = \delta = 0$ and initial state sampled from $\mu$, spends an expected $N/2$ time-steps having each number $k = 1, \ldots, N - 1$ of $A$ alleles.

These sojourn times can be used to evaluate the neutral bracket operation $\langle \ \rangle^{\circ}$ on any function that depends only on number of $A$ alleles. Indeed, for any function $f : \{0, \ldots, N\} \to \mathbb{R}$ with $f(0) = f(N) = 0$, Lemma 2.2 gives

$$\left\langle f \left( \sum_{g \in G} X_g^t \right) \right\rangle^{\circ} = \nu_G \sum_{k=1}^{N} \sigma(k) f(k) = \frac{N}{2} \sum_{k=1}^{N} f(k), \tag{9.13}$$

since $\nu_G = 1$ for the death-Birth process.

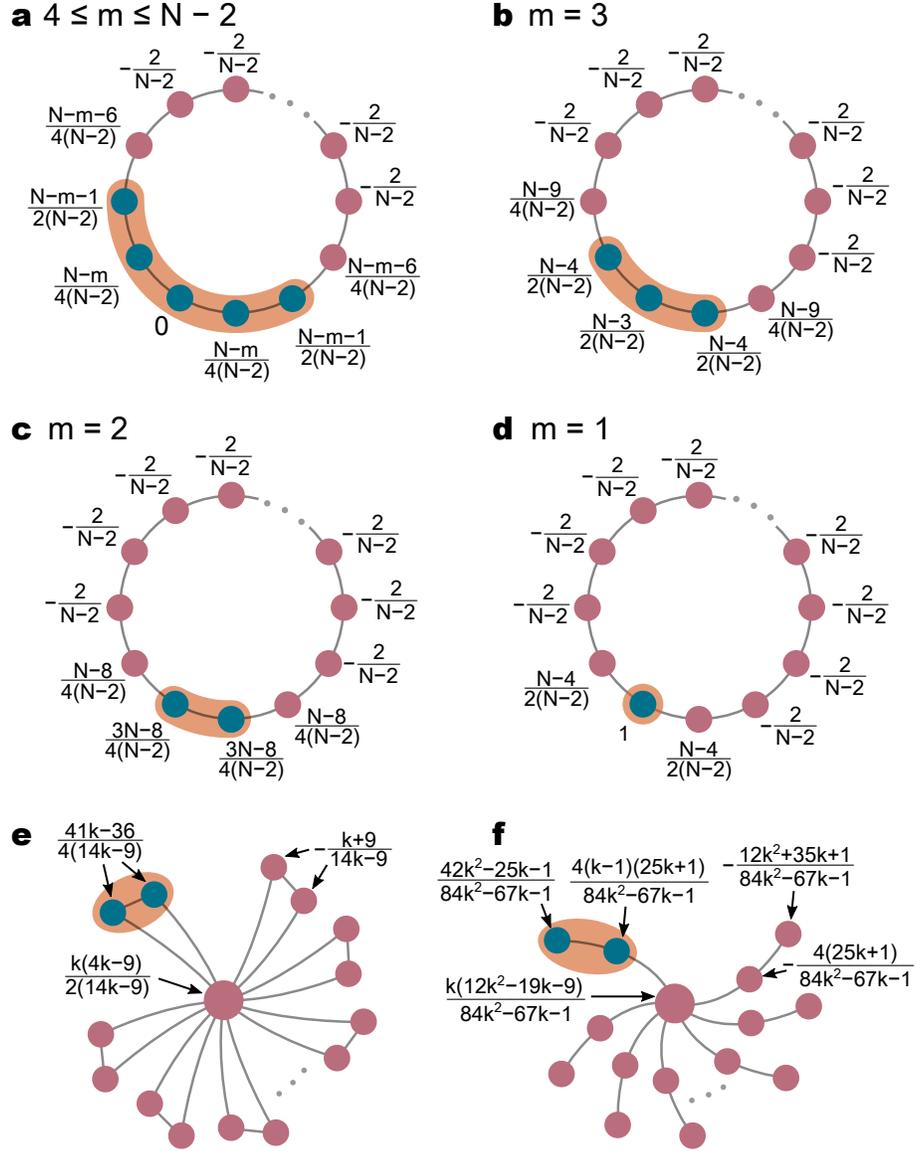

**Fig. S1. Cost-benefit thresholds for collective action on finite cycles, windmills, and spiders** Expressions show the cost-benefit thresholds, $\kappa_{S,g} = m d_g \left( r_{S,g} - r^{(2)}_{S,g} \right) \Big/ \left( \sum_{h \in S} d_h \left( r_h - r^{(2)}_h \right) \right)$, from the highlighted collective $S$, such that collective action toward $g$ is favored if $b \kappa_{S,g} > c$. Panels **a–d** show a cycle of size $N$. Separate cases are needed for collectives of three or fewer. Panels **e** and **f** show the Windmill and Spider graphs, respectively, with the collective $S$ being a single "blade" or "leg". For both the Windmill and Spider graphs, the condition for collective help to the hub is asymptotically $kb > 7c$. Collective help can therefore be favored, for sufficiently large $k$, even if the benefit is very small relative to the cost. Calculations are provided in the Supplementary Information.

***9.5.3. Coalescence lengths.*** We can now compute the necessary coalescence lengths on the cycle, using Eq. (4.5) and Eq. (9.13) rather than the recurrence relations, Eq. (9.8).

Let $\imath_m(k)$, $\imath_m^A(k)$, and $\imath_m^a(k)$ denote the respective averages of $\imath_S(\mathbf{x})$, $\imath_S^A(\mathbf{x})$, and $\imath_S^a(\mathbf{x})$, as $S$ runs over all contiguous blocks of length $m$, in a state $\mathbf{x}$ with exactly $k$ $A$ alleles in a single contiguous block. To compute $\imath_m^A(k)$, we note that there are $k - m + 1$ ways for a block of $m \le k$ contiguous sites to be contained within a block of $k$ contiguous $A$-containing sites. This gives

$$\imath_m^A(k) = \begin{cases} (k - m + 1)/N & m \le k \\ 0 & m > k. \end{cases} \tag{9.14}$$

We now compute $\ell_m^\circ$, the coalescence length for any contiguous contiguous block of $m$ sites. We begin by invoking Eq. (4.5):

$$\begin{aligned}
\ell_m^\circ &= \langle 1 - \imath_S \rangle^\circ & \text{for any set } S \text{ of } m \text{ contiguous sites} \\
&= \left\langle \left( \imath_{\{g\}}^A + \imath_{\{g\}}^a \right) - \left( \imath_S^A + \imath_S^a \right) \right\rangle^\circ & \text{for any } g \in G \\
&= 2 \left\langle \imath_{\{g\}}^A - \imath_S^A \right\rangle^\circ & \text{by symmetry under neutral drift} \\
&= N \sum_{k=1}^{N-1} \left( \imath_1^A(k) - \imath_m^A(k) \right) & \text{by symmetry of cycle and Eq. (9.13)} \\
&= N \left( \sum_{k=1}^{N-1} \frac{k}{N} - \sum_{k=m}^{N-1} \frac{k - m + 1}{N} \right) & \text{by Eq. (9.14)} \\
&= (m - 1) \left( N - \frac{m}{2} \right).
\end{aligned}$$

Therefore, the coalescence length for any contiguous block of $m$ sites on the cycle is

$$\ell_m^\circ = (m-1)(N - m/2). \tag{9.15}$$

We also require the coalescence length for sets of consisting of a block of $m$ contiguous sites together with a site of distance $j \ge 2$ from the block. There are two gaps in such a set, one of size $j - 1$ and the other of size $N - m - j$. Let $\imath_{m,j}^a(k)$ denote the average of $\imath_S^a(\mathbf{x})$ as $S$ runs over all sets of this form, where $\mathbf{x}$ is any state with exactly $k$ contiguous $A$ alleles. There are two ways to have $\imath_S^a(\mathbf{x}) = 1$ for such a set $S$ and state $\mathbf{x}$. First, all of the $A$ alleles in $\mathbf{x}$ may be contained in the gap of size $j - 1$ in $S$; this can happen $j - k$ ways. Second, all of the $A$ alleles in $\mathbf{x}$ may be contained in the gap of size $N - m - j$ in $S$; this can happen $N - m - j - k + 1$ ways. Proceeding similarly to the previous calculation, we compute

$$\begin{aligned}
\ell_S^\circ &= \langle 1 - \imath_S \rangle^\circ \\
&= 2 \left\langle \imath_{\{g\}}^a - \imath_S^a \right\rangle^\circ \\
&= N \sum_{k=1}^{N-1} \left( \imath_1^a(k) - \imath_{m,j}^a(k) \right) \\
&= N \left( \sum_{k=1}^{N-1} \frac{N-k}{N} - \left( \sum_{k=1}^{j-1} \frac{j-k}{N} + \sum_{k=1}^{N-m-j} \frac{N-m-j-k+1}{N} \right) \right) \\
&= \frac{N(N-1)}{2} - \left( \frac{j(j-1)}{2} + \frac{(N-m-j)(N-m-j+1)}{2} \right) \\
&= m \left( N - \frac{m+1}{2} \right) + (N - m - j)(j - 1).
\end{aligned}$$

So the coalescence length for a contiguous block of $m$ sites together with another site distance $j$ away is

$$\ell^\circ_{m,j} = m\left(N - \frac{m+1}{2}\right) + (N - m - j)(j - 1). \tag{9.16}$$

Note that $\ell^\circ_{m,j} = \ell^\circ_{m+1} + (N - m - j)(j - 1)$, where $(N - m - j)(j - 1)$ is the product of the two gap sizes in a set of this form. Although Eq. (9.16) was derived for $j \geq 2$, it also holds in the cases $j = 0$ and $j = 1$, giving

$$\ell^\circ_{m,0} = (m - 1)\left(N - \frac{m}{2}\right) = \ell^\circ_m$$

$$\ell^\circ_{m,1} = m\left(N - \frac{m+1}{2}\right) = \ell^\circ_{m+1}.$$

Overall, Eq. (9.16) is valid for $0 \leq j \leq N - m + 1$.

***9.5.4. Conditions for collective help or harm.*** Using the above coalescence lengths, we compute the conditions for a collective $S$, of $m$ contiguous sites, to help or harm a target site $g$ in the Collective Action game. The cost-benefit thresholds $\kappa_{S,g}$ are computed using Eq. (9.10b) with the coalescence lengths obtained in Eqs. (9.15)–(9.16). We also use the fact that a two-step random walk on the cycle terminates at its starting point with probability $\frac{1}{2}$, and otherwise terminates two spaces to the left or right with probability $\frac{1}{4}$ each. Using this, the denominator of $\kappa_{S,g}$ in Eq. (9.10b) evaluates to

$$\frac{1}{m}\sum_{h \in S} v_h \sum_{k \in G} p^{(2)}_{hk} \ell^\circ_{\{h,k\}} = \frac{1}{2}\left(\ell^\circ_{1,0} + \ell^\circ_{1,2}\right) = N - 2, \tag{9.17}$$

in agreement with the third bullet point in Section 9.4.

We now break into cases according to the target site $g$:

- **To neighbor:** Suppose the target $g$ is an immediate neighbor of the collective $S$. For $2 \leq m \leq N - 2$, evaluating Eq. (9.10b) yields

$$\kappa_{S,g} = \frac{\frac{1}{4}\ell^\circ_{m,3} + \frac{1}{2}\ell^\circ_{m+1} + \frac{1}{4}\ell^\circ_m - \ell^\circ_{m+1}}{N - 2} = \frac{N - m - 6}{4(N - 2)}. \tag{9.18}$$

For $m = 1$,

$$\kappa_{S,g} = \frac{\frac{1}{4}\ell^\circ_2 + \frac{1}{2}\ell^\circ_2 + \frac{1}{4}\ell^\circ_{1,3} - \ell^\circ_2}{N - 2} = \frac{N - 4}{2(N - 2)}. \tag{9.19}$$

Eq. (9.19) recovers a known result for Prisoner's Dilemma games on a cycle, first obtained as Eq. (4.4) of Ohtsuki and Nowak [84].

- **To non-neighbor vertex outside collective:** Now suppose $g$ is distance $j$ from $S$, with $2 \leq j \leq N - m - 2$. Then Eq. (9.10b) becomes

$$\kappa_{S,g} = \frac{\frac{1}{4}\ell^\circ_{m,j-2} + \frac{1}{2}\ell^\circ_{m,j} + \frac{1}{4}\ell^\circ_{m,j+2} - \ell^\circ_{m,j}}{N - 2} = -\frac{2}{N - 2}. \tag{9.20}$$

In this case, collective help is never favored, and harm is favored if $2b < -(N - 2)c$.

- **To internal boundary vertex:** Next, suppose that the target $g$ is a boundary vertex of $S$. For $3 \leq m \leq N - 1$ we have

$$\kappa_{S,g} = \frac{\frac{1}{4}\ell^\circ_m + \frac{1}{2}\ell^\circ_m + \frac{1}{4}\ell^\circ_{m,2} - \ell^\circ_m}{N - 2} = \frac{N - m - 1}{2(N - 2)}. \tag{9.21}$$

For $m = 2$,
$$\kappa_{S,g} = \frac{\frac{1}{4}\ell_3^\circ + \frac{1}{2}\ell_2^\circ + \frac{1}{4}\ell_{2,2}^\circ - \ell_2^\circ}{N-2} = \frac{3N-8}{4(N-2)}. \tag{9.22}$$

For $m = 1$ we have $S = \{g\}$ and $\kappa_{\{g\},g} = 1$.

- **To interior neighbor of boundary vertex:** Now suppose $g$ is in $S$, one space away from the boundary (this requires $m \geq 3$). For $4 \leq m \leq N-1$,
$$\kappa_{S,g} = \frac{\frac{1}{4}\ell_m^\circ + \frac{1}{2}\ell_m^\circ + \frac{1}{4}\ell_{m+1}^\circ - \ell_m^\circ}{N-2} = \frac{N-m}{4(N-2)}. \tag{9.23}$$

For $m = 3$,
$$\kappa_{S,g} = \frac{\frac{1}{4}\ell_4^\circ + \frac{1}{2}\ell_3^\circ + \frac{1}{4}\ell_4^\circ - \ell_3^\circ}{N-2} = \frac{N-3}{2(N-2)}. \tag{9.24}$$

- **To other interior vertex:** Finally, suppose $g$ is within the collective and is more than one space from the boundary (this requires $m \geq 5$). In this case, all two-step neighbors of $g$ are also in $S$. It follows from the final observation in Section 9.4 that $\kappa_{S,g} = 0$ and neither collective help nor harm can be favored.

These results are summarized in Figure S1a–d. Taking $N \to \infty$ for fixed $m \geq 4$ yields the results shown in Figure 3b of the main text.

**9.6. Windmill.** The Windmill graph (Fig. 2c of the main text) has one hub vertex and $2k$ "blade" vertices. Each blade vertex is joined to one exactly other blade vertex, as well as to the hub. We denote the hub vertex by $h$ and the blade vertices by $b$. The weighted degrees are $d_b = 2$ and $d_h = 2k$.

***9.6.1. Coalescence lengths.*** We solve for coalescence lengths using Eq. (9.8). In indexing the coalescence lengths, we use primes ($'$) and double primes ($''$) to indicate vertices on different blades. For example, $\ell_{bbb'}$ is the coalescence length of sets comprising two distinct blade vertices on one blade and one on a different blade, whereas $\ell_{bb'b''}$ corresponds to sets with three blade vertices on three different blades. To reduce clutter, we omit the superscripts $^\circ$, and understand all coalescence lengths to be computed at neutrality ($\delta = 0$).

The recurrence relations, Eq. (9.8), for sets of size two are
$$\ell_{hb} = 1 + \frac{1}{2}\left(\frac{k-1}{k}\ell_{bb'} + \frac{1}{2k}\ell_{bb} + \frac{1}{2}\ell_{hb}\right)$$
$$\ell_{bb} = 1 + \frac{1}{2}\ell_{hb}$$
$$\ell_{bb'} = 1 + \frac{1}{2}\ell_{hb} + \frac{1}{2}\ell_{bb'},$$

giving the solution
$$\ell_{hb} = \frac{16k-6}{2k+3}, \qquad \ell_{bb} = \frac{10k}{2k+3}, \qquad \ell_{bb'} = \frac{20k}{2k+3}. \tag{9.25}$$

For sets of size three, Eq. (9.8) gives
$$\ell_{hbb} = 1 + \frac{1}{3}\left(\frac{k-1}{k}\ell_{bbb'} + \frac{1}{k}\ell_{bb} + 2\ell_{hb}\right)$$
$$\ell_{hbb'} = 1 + \frac{1}{3}\left(\frac{k-2}{k}\ell_{bb'b''} + \frac{1}{k}\ell_{bbb'} + \frac{1}{k}\ell_{bb'} + \ell_{hbb'} + \ell_{hb}\right)$$
$$\ell_{bbb'} = 1 + \frac{1}{3}\left(\ell_{bb'} + \ell_{hbb'} + \frac{1}{2}\ell_{bbb'} + \frac{1}{2}\ell_{hbb}\right)$$
$$\ell_{bb'b''} = 1 + \frac{1}{2}\ell_{bb'b''} + \frac{1}{2}\ell_{hbb'},$$

and the solution is

$$\ell_{hbb} = \frac{21k-6}{2k+3}, \qquad \ell_{hbb'} = \frac{26k-6}{2k+3}, \qquad \ell_{bbb'} = \frac{25k}{2k+3},$$
$$\ell_{bb'b''} = \frac{30k}{2k+3}. \qquad (9.26)$$

***9.6.2. Conditions for collective help or harm.*** We now compute the cost-benefit thresholds $\kappa_{S,g}$, where set $S$ consists of the two vertices on a single blade. For help or harm to the hub, evaluating Eq. (9.10b) gives:

$$\kappa_{S,h} = \frac{d_h \left(\frac{1}{2}\ell_{hbb} + \frac{1}{2k}\ell_{bb} + \frac{k-1}{2k}\ell_{bbb'} - \ell_{hbb}\right)}{d_b \left(\frac{1}{4}\ell_{hb} + \frac{1}{4k}\ell_{bb} + \frac{k-1}{2k}\ell_{bb'}\right)} = \frac{k(4k-9)}{2(14k-9)}. \qquad (9.27)$$

As $k \to \infty$, $\kappa_{S,h} = \frac{k}{7} + \mathcal{O}(1)$.

To a vertex within $S$:

$$\kappa_{S,b} = \frac{d_b \left(\frac{1}{4}\ell_{bb} + \frac{1}{4}\ell_{hbb} + \frac{1}{2k}\ell_{bb} + \frac{k-1}{2k}\ell_{bbb'} - \ell_{bb}\right)}{d_b \left(\frac{1}{4}\ell_{hb} + \frac{1}{4k}\ell_{bb} + \frac{k-1}{2k}\ell_{bb'}\right)} = \frac{41k-36}{4(14k-9)}. \qquad (9.28)$$

As $k \to \infty$, $\kappa_{S,b}$ converges to $\frac{41}{56}$.

To a vertex in a different blade:

$$\kappa_{S,b'} = \frac{d_b \left(\frac{1}{4}\ell_{bbb'} + \frac{1}{4}\ell_{hbb} + \frac{1}{2k}\ell_{bb} + \frac{k-1}{2k}\ell_{bbb'} - \ell_{bbb'}\right)}{d_b \left(\frac{1}{4}\ell_{hb} + \frac{1}{4k}\ell_{bb} + \frac{k-1}{2k}\ell_{bb'}\right)} = -\frac{k+9}{14k-9}. \qquad (9.29)$$

As $k \to \infty$, $\kappa_{S,b'}$ converges to $-1/14$. These results are summarized in Figure S1e.

**9.7. Spider.** The Spider graph (Fig. 1d of the main text) contains one hub vertex (labeled $h$), $k$ inner vertices (labeled $i$), and $k$ outer vertices (labeled $o$). The hub vertex connects to all inner vertices, and each inner vertex connects to a single outer vertex. The weighted degrees are $d_h = k$, $d_i = 2$, and $d_o = 1$.

***9.7.1. Coalescence lengths.*** We solve for coalescence lengths using Eq. (9.8). We again use primes ($'$) and double primes ($''$) to indicate vertices on different "legs". For example, $\ell_{io'}$ is the coalescence length from an inner vertex and an outer vertex on different legs, while $\ell_{io'o''}$ refers to one inner and two outer vertices on three different legs. We again omit the superscripts $\circ$, understanding all coalescence lengths to be computed at neutrality.

For sets of size two Eq. (9.8) gives the following recurrence relations:

$$\ell_{io} = 1 + \frac{1}{4}\ell_{ho}$$
$$\ell_{ho} = 1 + \frac{1}{2k}\ell_{io} + \frac{k-1}{2k}\ell_{io'} + \frac{1}{2}\ell_{hi}$$
$$\ell_{hi} = 1 + \frac{1}{4}\ell_{ho} + \frac{k-1}{2k}\ell_{ii'}$$
$$\ell_{ii'} = 1 + \frac{1}{2}\ell_{io'} + \frac{1}{2}\ell_{hi}$$
$$\ell_{io'} = 1 + \frac{1}{4}\ell_{oo'} + \frac{1}{4}\ell_{ho} + \frac{1}{2}\ell_{ii'}$$
$$\ell_{oo'} = 1 + \ell_{io'}.$$

For sets of size three, Eq. (9.8) gives

$$\ell_{hio} = 1 + \frac{k-1}{3k}\ell_{ioi'} + \frac{1}{3k}\ell_{io} + \frac{1}{3}\ell_{ho} + \frac{1}{3}\ell_{hi}$$

$$\ell_{hii'} = 1 + \frac{k-2}{3k}\ell_{ii'i''} + \frac{2}{3k}\ell_{ii'} + \frac{1}{3}\ell_{hio'} + \frac{1}{3}\ell_{hi}$$

$$\ell_{hio'} = 1 + \frac{1}{3k}\ell_{io'} + \frac{1}{3k}\ell_{ioi'} + \frac{k-2}{3k}\ell_{ii'o''} + \frac{1}{6}\ell_{ho} + \frac{1}{6}\ell_{hoo'} + \frac{1}{3}\ell_{hii'}$$

$$\ell_{hoo'} = 1 + \frac{2}{3k}\ell_{ioo'} + \frac{k-2}{3k}\ell_{io'o''} + \frac{2}{3}\ell_{hio'}$$

$$\ell_{ioi'} = 1 + \frac{1}{6}\ell_{io'} + \frac{1}{6}\ell_{hio'} + \frac{1}{3}\ell_{ii'} + \frac{1}{6}\ell_{hio} + \frac{1}{6}\ell_{ioo'}$$

$$\ell_{ioo'} = 1 + \frac{1}{6}\ell_{hoo'} + \frac{1}{6}\ell_{oo'} + \frac{1}{3}\ell_{io'} + \frac{1}{3}\ell_{ioi'}$$

$$\ell_{ii'i''} = 1 + \frac{1}{2}\ell_{hii'} + \frac{1}{2}\ell_{ii'o''}$$

$$\ell_{ii'o''} = 1 + \frac{1}{3}\ell_{io'o''} + \frac{1}{3}\ell_{hio'} + \frac{1}{3}\ell_{ii'i''}$$

$$\ell_{io'o''} = 1 + \frac{1}{6}\ell_{oo'o''} + \frac{1}{6}\ell_{hoo'} + \frac{2}{3}\ell_{ii'o''}$$

$$\ell_{oo'o''} = 1 + \ell_{io'o''}.$$

For brevity, we omit the explicit solutions for $\ell_S$; they all have the form $(ak^2 + bk + c)/(12k^2 + 35k + 1)$ for some integer coefficients $a$, $b$, and $c$.

**9.7.2. Conditions for collective help or harm.** We now compute the critical benefit-cost thresholds $\kappa_{S,g}$, where $S$ is a set of two vertices on a single leg. For collective help or harm to $S$'s own outer vertex, evaluating Eq. (9.10b) gives:

$$\kappa_{S,o} = \frac{d_o\left(\frac{1}{2}\ell_{io} + \frac{1}{2}\ell_{hio} - \ell_{io}\right)}{\frac{1}{2}\left(d_o\left(\frac{1}{2}\ell_{ho}\right) + d_i\left(\frac{k-1}{2k}\ell_{ii'}\right)\right)} = \frac{42k^2 - 25k - 1}{84k^2 - 67k - 1}. \tag{9.30}$$

As $k \to \infty$, $\kappa_{S,o}$ converges to $1/2$.

To $S$'s own inner vertex:

$$\kappa_{S,i} = \frac{d_i\left(\frac{k+1}{2k}\ell_{io} + \frac{k-1}{2k}\ell_{ioi'} - \ell_{io}\right)}{\frac{1}{2}\left(d_o\left(\frac{1}{2}\ell_{ho}\right) + d_i\left(\frac{k-1}{2k}\ell_{ii'}\right)\right)} = \frac{4(k-1)(25k+1)}{84k^2 - 67k - 1}. \tag{9.31}$$

As $k \to \infty$, $\kappa_{S,i}$ converges to $25/21$.

To the hub:

$$\kappa_{S,h} = \frac{d_h\left(\frac{1}{2k}\ell_{io} + \frac{k-1}{2k}\ell_{ioo'} + \frac{1}{2}\ell_{hio} - \ell_{hio}\right)}{\frac{1}{2}\left(d_o\left(\frac{1}{2}\ell_{ho}\right) + d_i\left(\frac{k-1}{2k}\ell_{ii'}\right)\right)} = \frac{k(12k^2 - 19k - 9)}{84k^2 - 67k - 1}. \tag{9.32}$$

As $k \to \infty$, $\kappa_{S,h} = \frac{k}{7} + \mathcal{O}(1)$.

To an inner vertex in a different leg:

$$\kappa_{S,i'} = \frac{d_i\left(\frac{2k-1}{2k}\ell_{ioi'} + \frac{1}{2k}\ell_{io} - \ell_{ioi'}\right)}{\frac{1}{2}\left(d_o\left(\frac{1}{2}\ell_{ho}\right) + d_i\left(\frac{k-1}{2k}\ell_{ii'}\right)\right)} = -\frac{4(25k+1)}{84k^2 - 67k - 1}. \tag{9.33}$$

As $k \to \infty$, $\kappa_{S,i'} = -\frac{1}{7k} + \mathcal{O}\left(\frac{1}{k^2}\right)$.

**Allen et al.** 

To an outer vertex in a different leg:

$$\kappa_{S,o'} = \frac{d_o \left(\frac{1}{2}\ell_{ioo'} + \frac{1}{2}\ell_{hio} - \ell_{ioo'}\right)}{\frac{1}{2}\left(d_o \left(\frac{1}{2}\ell_{ho}\right) + d_i \left(\frac{k-1}{2k}\ell_{ii'}\right)\right)} = -\frac{12k^2 + 35k + 1}{84k^2 - 67k - 1}. \tag{9.34}$$

As $k \to \infty$, $\kappa_{S,o'}$ converges to $-1/7$. These results are summarized in Figure S1f.

**9.8. Numerical computation for spatial networks.** To numerically compute critical benefit-cost thresholds on an arbitrary weighted graph $G$, we first compute the relevant coalescence lengths $\ell_S^\circ$ using Eq. (4.8), and then apply Eq. (9.10b). We make use of the fact that Eq. (4.8) is recursive in set sizes: for each set $S$ of size $k \geq 2$, Eq. (4.8) expresses $\ell_S^\circ$ in terms of coalescence lengths of sets of size $k$ or $k-1$. Because of this, Eq. (4.8) can be solved first for sets of size two, then size three, and so on up to the largest set size that is needed. To obtain the cost-benefit thresholds $\kappa_{S,g}$ for a particular set $S$ requires computing coalescence lengths for sets up to size $|S| + 1$.

Obtaining the coalescence lengths $\ell_S^\circ$ for sets of size $|S| = k$, given those for size $|S| = k-1$, requires solving a system of $\binom{N}{k}$ linear equations. For standard algorithms based on Gaussian elimination this takes $\mathcal{O}\left(\binom{N}{k}^3\right)$ time, although more efficient scaling is possible in theory [85, 86].

All computations and algorithms for Delaunay and weaver graphs (see below) were performed using MATLAB (version R2022a). Code is available at https://github.com/Emmanuel-Math-Bio-Research-Group/Collective-Action.

***9.8.1. Delaunay triangulation.*** Delaunay triangulations [87] are a model of random spatial networks, used by Archetti et al. [88] to model populations of cancer cells in solid tumors. We generated 50 Delaunay networks of 16 nodes each. Each was generated by placing random points on a $[0,1] \times [0,1]$ square and using MATLAB's built-in `delaunayTriangulation` function. We weighted each edge inversely proportional to Euclidean distance, so that edge weight corresponds to physical proximity.

To identify subcommunities (which we took as collectives for the Collective Action Dilemma), we used a spatial variant of the Girvan-Newman algorithm [89]. We removed edges from the graph one by one, from longest (least weight) to shortest (most weight), until a new connected component was created. Each time a new connected component was created, the components were considered to be a possible subdivision into communities. We computed the modularity of this subdivision using the formula for weighted networks in Eq. (9) of Newman [90]. The process was then continued until the graph was totally disconnected. We retained the subdivision with the greatest modularity, and used the corresponding subcommunities as collectives in our analysis.

***9.8.2. Sociable weavers.*** Sociable weavers (*Philetairus socius*, Fig. S2) live in large communal nests that persist over generations. They engage in a variety of collective behaviors, including cooperative offspring raising [92] and nest maintenance [91, 93, 94]. The majority of these cooperative behaviors are performed by males [91, 92, 94]. Most individuals—especially males—remain at the nest rather than dispersing [95]. Males disperse locally within a nest [91], in close analogy with the death-Birth process.

To infer the spatial structure of nest chambers within sociable weaver nests, we used pairwise distance matrices provided in the dataset [96] associated to van Dijk et al. [91]. These were based on photographs of 23 communal nests at Benfontein Game Farm, Kimberley, South Africa, taken in September–December 2010 and 2011, and January–February 2013. Each matrix represents one communal nest; each row a single nest chamber.

To clean the data, we first eliminated empty rows and columns. Matrices that had partially empty rows and columns were eliminated from the dataset, leaving 13 matrices remaining. To avoid having multiple rows that refer to the same nest chamber, we identified entries that have distance zero to each other (meaning they refer to the same nest chamber). For each such case, we retained only one corresponding row and column, the entries of which are the average of the recorded distances to each other nest chamber.

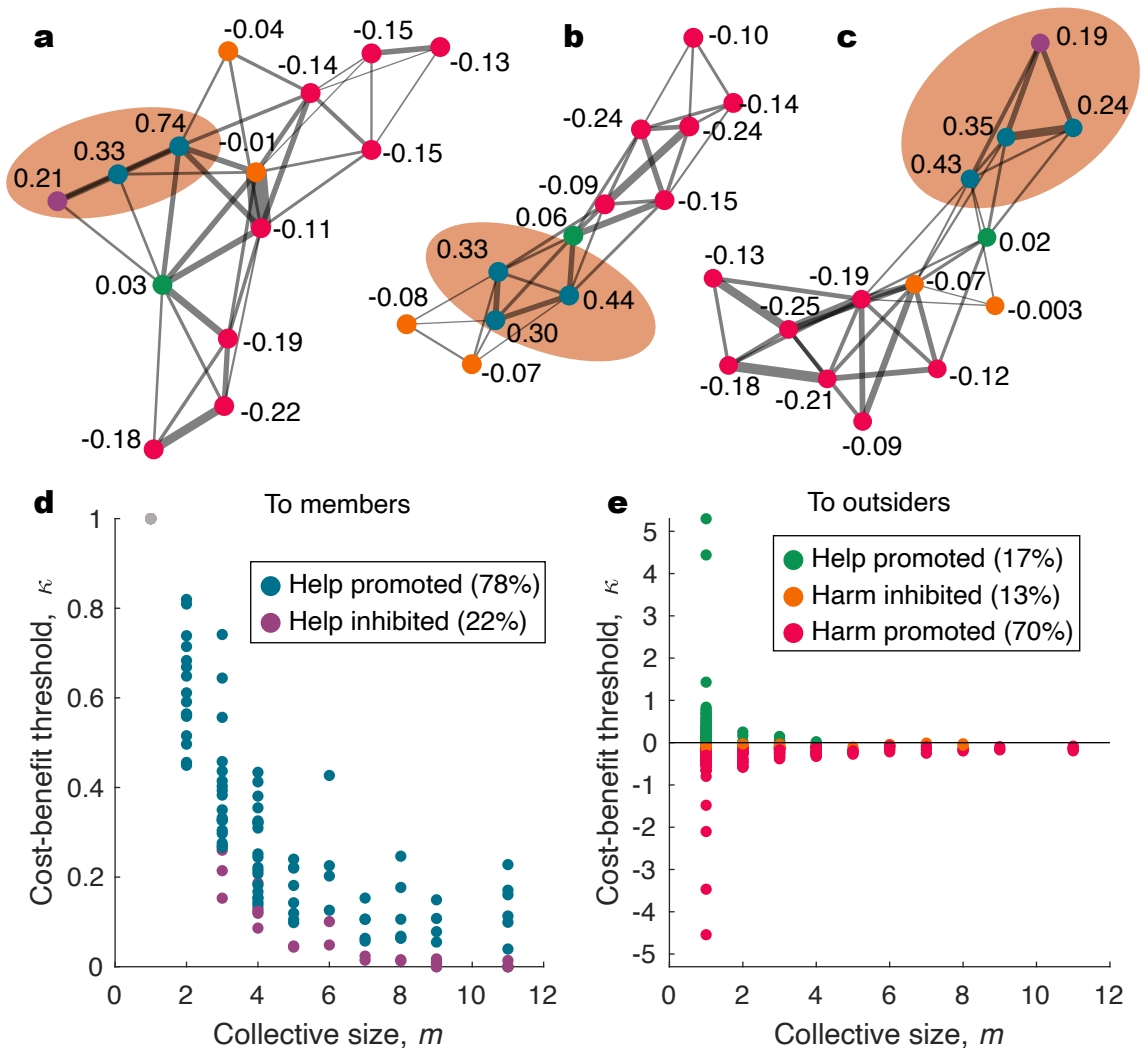

**Fig. S2. Collective action in communal sociable weaver nests. a-c** Sociable weavers live in large communal nests that persist over generations, where they collectively raise offspring and maintain the nest. Males, who do most of the helping, disperse locally within the nest [91]. Using pairwise distance measurements by van Dijk et al. [91], we inferred networks representing the spatial structure of nest chambers in 13 communal nests (see Methods). We then used a spatial variant of the Girvan-Newman algorithm [89] to partition these networks into a total of 43 subcommunities. We computed the cost-benefit threshold $\kappa_{S,g}$ from each subcommunity to each node. Three examples are shown here. **c** Similarly to Delaunay networks, spatial structure promotes help ($\kappa_{S,g} > (n-m)/(m(n-1))$) to most targets within collectives, but inhibits help to a significant minority, especially in larger collectives. **d** Spatial structure promotes help ($\kappa_{S,g} > 0$) to 17% of external targets, of which 91% were neighbors of the collective and the rest were two-step neighbors. Harm is promoted ($\kappa_{S,g} < -1/(n-1)$) to the majority of external targets. Values of $\kappa_{S,g}$ far outside the range of -1 to 1 (which were not observed for Delaunay networks) arise here from a single "collective" of size 1, corresponding to a partiuclar nest chamber that is spatially removed from all others in its nest. This chamber has low reproductive value, effectively reducing the cost of help and harm to others.

Since entrances to nest chambers are located on the underside of a communal nest, they have a two-dimensional spatial structure. To capture this structure, we represented each nest as a planar graph, with nodes representing nest chambers. These graphs were constructed by the following algorithm: Starting from a completely disconnected graph, we identified the shortest pairwise distance between nest chambers and drew an edge between the two corresponding nodes. We repeated this process for all pairwise distances, from shortest to the longest, skipping over any pairs for which an adding an edge would make the graph non-planar. This process terminated when there were no more edges to add without violating planarity.

As in the Delaunay networks, we weighted each edge inversely proportional to distance. Subcommunities were detected using the algorithm described in Section 9.8.1.

## 10. Maximization of inclusive fitness for a single collective

Does selection lead collectives to act as if maximizing inclusive fitness? We find that it does under particular, idealized assumptions: (i) each mutant allele affects the actions of only a single class of collectives, and (ii) effects due to multiple collective actors in this class combine additively.

### 10.1. Formal result.
To define classes of collectives, we use the definition of symmetry from Section 1.8. Two collectives $S, T \subseteq G$ are equivalent if there is some symmetry $\sigma$ such that $\sigma(S) = T$. This equivalence relation partitions the set of nonempty subsets of $G$ into equivalence classes. We denote equivalence class of a given set $S \subseteq G$ by $[S]$; that is

$$[S] = \{T \subseteq G \mid T = \sigma(S) \text{ for some symmetry } \sigma \in \mathrm{Sym}(G, p)\}. \tag{10.1}$$

The assumption required for our maximization result is that there is a particular class $[S]$, such that all fitness increments can be written in the form

$$w_g(\mathbf{x}) = \sum_{T \in [S]} \left( c^A_{T,g}\, \imath^A_T(\mathbf{x}) + c^a_{T,g}\, \imath^a_T(\mathbf{x}) \right). \tag{10.2}$$

This representation has the form of Eq. (5.7), but with $C^A_{T,g} = C^a_{T,g} = 0$ for all collectives $T$ not in class $[S]$. In this sense, we assume that only collectives in class $[S]$ affect fitness, and their effects combine additively.

**Theorem 10.1.** *Assuming that fitness increments are given by* Eq. (10.2) *for a particular $S \subseteq G$, with collective fitness effects $c^A_{T,g}$ and $c^a_{T,g}$ satisfying* Eq. (5.8), *allele A is favored if*

$$\sum_{g \in G} c^A_{S,g}\, r^A_{S,g} > \sum_{g \in G} c^a_{S,g}\, r^a_{S,g}. \tag{10.3}$$

*Proof.* Since every $T \in [S]$ can be written as $T = \sigma(S)$ for some symmetry $\sigma \in \mathrm{Sym}(G, p)$, we can rewrite Eq. (10.2) as

$$w_g(\mathbf{x}) = \sum_{\sigma \in \mathrm{Sym}(G,p)} \left( c^A_{\sigma(S),g}\, \imath^A_{\sigma(S)}(\mathbf{x}) + c^a_{\sigma(S),g}\, \imath^a_{\sigma(S)}(\mathbf{x}) \right). \tag{10.4}$$

Now applying Theorem 5.1, allele $A$ is favored if

$$\sum_{g \in G} \sum_{\sigma \in \mathrm{Sym}(G,p)} c^A_{\sigma(S),g}\, r^A_{\sigma(S),g} > \sum_{g \in G} \sum_{\sigma \in \mathrm{Sym}(G,p)} c^a_{\sigma(S),g}\, r^a_{\sigma(S),g}. \tag{10.5}$$

For every symmetry $\sigma \in \mathrm{Sym}(G, p)$, the inverse permutation $\sigma^{-1}$ is also a symmetry [6]. Furthermore, we have $c^A_{\sigma(S),g} = c^A_{S,\sigma^{-1}(g)}$ and $c^a_{\sigma(S),g} = c^a_{S,\sigma^{-1}(g)}$ by Eq. (5.18), and $r^A_{\sigma(S),g} = r^A_{S,\sigma^{-1}(g)}$ and $r^a_{\sigma(S),g} = r^a_{S,\sigma^{-1}(g)}$ by Eq. (4.17). Substituting, we obtain

$$\sum_{\sigma \in \mathrm{Sym}(G,p)} \sum_{g \in G} c^A_{S,\sigma^{-1}(g)}\, r^A_{S,\sigma^{-1}(g)} > \sum_{\sigma \in \mathrm{Sym}(G,p)} \sum_{g \in G} c^a_{S,\sigma^{-1}(g)}\, r^a_{S,\sigma^{-1}(g)}. \tag{10.6}$$

Since $\sigma^{-1}$ permutes the elements of $G$, the effect of $\sigma^{-1}$ in the inner summations of Eq. (10.6) is to permute the terms of $\sum_{g \in G} c_{S,g}^A r_{S,g}^A$ and $\sum_{g \in G} c_{S,g}^a r_{S,g}^a$ without affecting the total. Thus, for each $\sigma \in \text{Sym}(G, p)$, we have

$$\sum_{g \in G} c_{S,\sigma^{-1}(g)}^A r_{S,\sigma^{-1}(g)}^A = \sum_{g \in G} c_{S,g}^A r_{S,g}^A$$
$$\sum_{g \in G} c_{S,\sigma^{-1}(g)}^a r_{S,\sigma^{-1}(g)}^a = \sum_{g \in G} c_{S,g}^a r_{S,g}^a. \tag{10.7}$$

Substituting these for the inner sums in Eq. (10.6), and dividing by the total number of symmetries, we obtain Condition (10.3). □

**10.2. Interpretation.** To relate Theorem 10.1 to the idea of maximization, let us suppose there is a resident allele $a$ and a number of alternative alleles. Now assume that for each alternative allele $A$, there is a corresponding class of collectives $[S_A]$ such that fitness increments for $A$ relative to $a$ can be written in the form $w_g(\mathbf{x}) = \sum_{T \in [S_A]} c_{T,g}^A \iota_T^A(\mathbf{x})$. This assumption means that each alternative allele $A$ affects the actions of only a single class of collectives $[S_A]$, and the fitness effects of these actions combine additively. Then, by Theorem 10.1, the resident allele $a$ is selected over every alternative if and only if

$$\sum_{g \in G} c_{S_A,g}^A r_{S_A,g}^A < 0 \qquad \text{for each alternative allele } A. \tag{10.8}$$

In words, the resident allele resists replacement by any alternative allele affecting the actions of a single class of collectives, if and only if every such alternative has a negative inclusive fitness effect on the affected class.

In light of the above remarks, Theorem 10.1 can be seen as a maximization result, with some important caveats. First, for a given set of possible alleles, there does not necessarily exist any allele that is favored over all others. Nontransitive (e.g. "rock-paper-scissors") competition is possible whenever there are more than two alleles. Second, there is not necessarily an "objective function" that any collective is selected to act as if maximizing (in the sense, for example, of Grafen [97]). This is because the collective fitness effects $c_{S,g}^A$ and $c_{S,g}^a$ characterize the relative effects of two competing alleles with respect to each other, rather than absolute quantities. Indeed, the existence of such an objective function precludes nontransitive competition, violating the first caveat. Third, for nonweak selection, the collective relatednesses $r_{S_A,g}^A$ also depend the alternative allele $A$, since selection may affect genetic assortment. For weak selection, however, the neutral relatednesses $r_{S_A,g}^\circ$ may be used. Fourth, this maximization result depends crucially on the assumption that each alternative allele $A$ affects only the actions of a single class $[S_A]$ of collectives, with the fitness effects of multiple such collectives combining additively, as formalized in Eq. (10.2). In general, a mutant allele may affect multiple classes of collectives, perhaps with a positive inclusive fitness effect for some and a negative effect for others. Selection then aggregates these effects over all collectives—as formalized in Theorem 5.1—and does not necessarily lead to maximizing behavior by any actor.

**10.3. Additive fitness effects.** An important special case arises for additive fitness effects. Suppose each fitness increment $w_g(\mathbf{x})$ is a linear function of the allelic variables $x_g$. Then fitness increments can be written in the form

$$w_g(\mathbf{x}) = \sum_{h \in G} c_{hg}^A x_h, \tag{10.9}$$

where $c_{hg}^A$ quantifies the effect of an $A$ allele in site $h$ on $g$'s fitness. In this case, fitness effects are additive in the sense that the combined effect of two sites $h$ and $k$ having allele $A$ equals the sum, $c_{hg}^A + c_{kg}^A$, of the two separate effects. This linearity may be posited *a priori*, or it may be derived from supposing that mutations have small phenotypic effects (see Section 5.6). For diploid populations, Eq. (10.9) applies if the fitness effects of individual actions combine additively and if there is no genetic dominance ($h = \frac{1}{2}$, in the notation of Section 1.7).

If we further suppose that the population is homogeneous—as defined in Section 1.8—then there is only one class of sites, and Eq. (10.9) consequently has the form of Eq. (10.2). Then, by Theorem 10.1, a resident allele $a$ is favored over an alternative allele $A$ if and only if

$$\sum_{h \in G} c_{gh}^A \, r_{gh}^A < 0, \tag{10.10}$$

with $r_{gh}^A$ shorthand for $r_{\{g\},h}^A$. Thus, in a homogeneous population with additive fitness effects, the resident allele is favored over all others if and only if each alternative allele has a negative effect on site-level inclusive fitness. This result can be extended to individual-level inclusive fitness in homogeneous diploid populations if there is no genetic dominance. Condition (10.10) provides the clearest point of connection to the classical concept of inclusive fitness maximization [70, 97].

If the population is heterogeneous—meaning there are multiple classes of sites—then we must assume (as in Section 10.1) that each alternative allele affects the actions of only a single class. That is, for each alternative allele $A$, we assume there is some corresponding class $[h_A]$ of sites, such that $c_{kg} = 0$ for all $k \notin [h_A]$ and all $g \in G$. Then, for states $\mathbf{x}$ containing alleles $a$ and $A$, the fitness of each site $g$ can be written

$$w_g(\mathbf{x}) = \sum_{k \in [h_A]} c_{kg}^A \, x_k. \tag{10.11}$$

Then, as in Condition (10.8), the resident allele $a$ is favored over each alternative $A$ if and only if

$$\sum_{k \in G} c_{h_A k}^A \, r_{h_A k}^A < 0 \qquad \text{for each alternative allele } A. \tag{10.12}$$

This can again be understood as a limited maximization result, applicable under the assumptions that (i) mutations alter the fitness effects of individual sites in a single class only, and (ii) these fitness effects combine additively.

**10.4. Comparison to Lehmann and Rousset (2020).** Lehmann and Rousset [29] also obtain a general result on inclusive fitness maximization, within the framework developed by Lehmann et al. [27] (see Section 1.9.6). Specifically, they identify conditions for a resident allele to be uninvadable, and show that that these conditions imply that individuals act as if maximizing an inclusive fitness quantity, in the sense of exhibiting Nash equilibrium behavior with respect to this quantity. Their maximization result—like ours—assumes that each mutation from the resident allele affects the actions of only a single class of actors. However, unlike our result, their maximization results apply to individuals (rather than collectives) even when the fitness effects of multiple individuals combine non-additively.

The key difference between Lehmann and Rousset's [29] maximization results and ours lies in the way fitness effects are attributed to specfic actors. Lehmann and Rousset use a linear regression of the form

$$w_g(\mathbf{x}) = c_{\varnothing,g} + \sum_{h \in G} c_{hg}^A x_h + \epsilon_{g,\mathbf{x}} \tag{10.13}$$

to attribute fitness effects (quantified by the regression coefficients $c_{hg}^A$) to individual actors, even in the presence of synergy. In contrast, our framework attributes synergistic fitness effects to collective actors, by representing fitness in the form of Eq. (5.7). While both approaches are mathematically valid, ours allows for more direct analysis of collective action.

## 11. Empirical estimation of collective fitness effects

We now consider the question of how collective fitness effects can be estimated from empirical data. To this end, we develop a polynomial regression model, in which these collective fitness effects appear as regression coefficients.

**11.1. Observable variables.** We first consider what variables are to be used in the statistical model. One approach would be formulate a statistical analogue of Eq. (5.1), in which the fitness increments $w_g(\mathbf{x})$ are modeled as a polynomial function of the state $\mathbf{x}$. However, these fitness increments, being expectations over all possible transition from a given state, are not directly observable in empirical data. For empirical applications, it is preferable to have variables corresponding to directly observable quantities.

For this reason we define the *realized fitness increment $W_g$* as the difference between site $g$'s reproductive value and the total reproductive value of $g$'s copies (including self) in the next generation:

$$W_g = \sum_{h \in \alpha^{-1}(g)} v_h - v_g. \tag{11.1}$$

$W_g$ is a random variable, whose distribution depends on the state $\mathbf{x}$ via the parentage map $\alpha$—which is itself random with probability distribution $\{p_\mathbf{x}(\alpha)\}$. Since $W_g$ depends on realized (rather than expected) survival and reproduction, it is observable in a given population state, insofar as survival and reproduction can be recorded and reproductive values can be empirically quantified. The realized and expected fitness increments are related by

$$\mathbb{E}_\mathbf{x}[W_g] = w_g(\mathbf{x}), \tag{11.2}$$

as can be seen by comparing Eq. (3.5) and Eq. (11.1).

We suppose that the observable data consists of pairs $(\mathbf{x}, \mathbf{W})$, where $\mathbf{x} \in \{0,1\}^G$ is the population state and $\mathbf{W} = (W_g)_{g \in G} \in \mathbb{R}^G$ is a vector of realized fitness increments. We regard $\mathbf{x}$ as an independent variable and $\mathbf{W}$ as a dependent variable.

**11.2. Statistical model.** To formulate a statistical model, the modeler chooses, for each site $g \in G$, a set $\mathcal{S}_g$ of subsets of $G$ to be used as predictors for $g$'s fitness. The realized increment of each site $g \in G$ is then statistically modeled as

$$W_g = \sum_{S \in \mathcal{S}_g} c_{S,g}\, \underline{x}_S + \epsilon_{g,\mathbf{x}} \qquad \text{for all } g \in G. \tag{11.3}$$

This is a polynomial regression model, with monomials $\underline{x}_S = \iota_S^A(\mathbf{x}) = \prod_{h \in S} x_h$, and associated regression coefficients $c_{S,g}$, for each $S \in \mathcal{S}_g$. The residual, $\epsilon_{g,\mathbf{x}}$, captures all deviations of $W_g$—including inherent randomness—from the predicted value $\sum_{S \in \mathcal{S}_g} c_{S,g} \iota_S^A(\mathbf{x})$. Eq. (11.3) is a statistical analogue of the representation of fitness increments in Eq. (5.1).

The choice of which set of subsets $\mathcal{S}_g$ to use in modeling $W_g$ amounts to choosing an appropriate set of predictors. A set $S$ should be included in $\mathcal{S}_g$ only if (a) the sites in $S$ plausibly affect $g$'s fitness beyond effects due to proper subsets of $S$, and (b) upon applying least-squares polynomial regression (see below), the coefficient $c_{S,g}$ meets a threshold level of statistical significance.

**11.3. Least-squares polynomial regression.** Now suppose we have a dataset of $m$ observations, each consisting of a pair $(\mathbf{x}, \mathbf{W})$. We characterize this dataset by its empirical frequency distribution $\hat{\pi}$, so that $\hat{\pi}(\mathbf{x}, \mathbf{W})$ equals the fraction of observations in which the pair $(\mathbf{x}, \mathbf{W})$ is observed.

For each $g \in G$, we estimate the coefficients $c_{S,g}$ by least-squares regression. The estimators $\hat{c}_{S,g}$ minimize the sum of residuals squared,

$$\mathbb{E}_{\hat{\pi}}\left[\hat{\epsilon}_{g,\mathbf{x}}^2\right] = \mathbb{E}_{\hat{\pi}}\left[\left(W_g - \sum_{S \in \mathcal{S}_g} \hat{c}_{S,g}\underline{x}_S\right)^2\right]. \tag{11.4}$$

Setting the partial derivatives of Eq. (11.4) with respect to $\hat{c}_{S,g}$ equal to zero, we obtain

$$\mathbb{E}_{\hat{\pi}}\left[\underline{x}_S\left(W_g - \sum_{T \in \mathcal{S}_g} \hat{c}_{T,g}\underline{x}_T\right)\right] = \mathbb{E}_{\hat{\pi}}\left[\underline{x}_S \hat{\epsilon}_{g,\mathbf{x}}\right] = 0. \tag{11.5}$$

As $S$ ranges over $\mathcal{S}_g$, Eq. (11.5) forms a system of equations that uniquely determines the estimators $\hat{c}_{S,g}$ for a given $g \in G$.

**11.4. Recovering conditions for selection.** It remains to show that the estimators $\hat{c}_{S,g}$ converge to the true values of $c_{S,g}$ if the data consists of a large sample from the underlying stochastic process. For this, we must first identify the relevant theoretical distribution on $(\mathbf{x}, \mathbf{W})$. This distribution, which we denote $\tilde{\pi}$, is characterized as follows:

(i) The marginal distribution on $\mathbf{x} \in \{0,1\}^G$ is given by

$$\tilde{\pi}(\mathbf{x}) = \begin{cases} 0 & \mathbf{x} = \mathbf{a} \text{ or } \mathbf{x} = \mathbf{A} \\ \lim_{u \to 0} \frac{\pi(\mathbf{x})}{1 - \pi(\mathbf{A}) - \pi(\mathbf{a})} & \text{otherwise,} \end{cases} \quad (11.6)$$

and

(ii) For a given state $\mathbf{x}$, the conditional distribution $\tilde{\pi}(\mathbf{W}|\mathbf{x})$ is obtained by sampling $\alpha$ from $p_{\mathbf{x}}$ and then computing $W_g$ for each $g \in G$ according to Eq. (11.1).

The marginal distribution on $\mathbf{x}$ in Eq. (11.6) was termed the *rare-mutation conditional distribution* by Allen and McAvoy [3]. This distribution has the property that $\tilde{\pi}(\mathbf{x})$ is proportional to the expected time spent in state $\mathbf{x}$ by $\mathcal{M}_0$ with initial state sampled from $\mu$; see Corollary 2 of McAvoy and Allen [4].

We now prove that the estimators $\hat{c}_{S,g}$ converge to the theoretical values $c_{S,g}$:

**Theorem 11.1.** *Suppose the statistical model, Eq. (11.3), is constructed so that, for each $g \in G$, $\mathcal{S}_g$ contains all sets $S$ for which $c_{S,g} \neq 0$. Let $\hat{\pi}_m$, for $m \geq 1$, be the frequency distribution on $\mathbf{x}$ and $\mathbf{W}$ obtained by sampling $m$ times independently from $\tilde{\pi}$. Then, as $m \to \infty$, the estimators $\hat{c}_{S,g}$ obtained from $\hat{\pi}_m$ converge almost surely to $c_{S,g}$ for each $S \subseteq G$ and $g \in G$.*

*Proof.* By the strong law of large numbers, the empirical frequency distribution $\hat{\pi}_m$ converges almost surely, as $m \to \infty$, to $\tilde{\pi}$. It therefore suffices to show that the estimators $\hat{c}_{S,g}$, obtained using the theoretical distribution $\tilde{\pi}$, match the $c_{S,g}$ as they are defined in Eq. (5.1). Since the solution to Eq. (11.5) is unique, we need only show that the $c_{S,g}$ satisfy Eq. (11.5) with distribution $\tilde{\pi}$; i.e., that

$$\mathbb{E}_{\tilde{\pi}}\left[\underline{x}_S \left(W_g - \sum_{T \in \mathcal{S}_g} c_{T,g} \underline{x}_T\right)\right] = 0, \quad (11.7)$$

for all $g \in G$. Since $\mathcal{S}_g$ contains all sets $T \subseteq G$ for which at least one $c_{T,g}$ is nonzero, we have $\sum_{T \in \mathcal{S}_g} c_{T,g} \underline{x}_T = w_g(\mathbf{x})$ for all $g \in G$ and $\mathbf{x} \in \{0,1\}^G$. So the left-hand side of Eq. (11.7) becomes

$$\mathbb{E}_{\tilde{\pi}}[\underline{x}_S (W_g - w_g(\mathbf{x}))] = \sum_{\mathbf{x} \in \{0,1\}^G} \tilde{\pi}(\mathbf{x}) (\underline{x}_S (\mathbb{E}_{\mathbf{x}}[W_g] - w_g(\mathbf{x}))) = 0, \quad (11.8)$$

by Eq. (11.2), completing the proof. $\square$

**11.5. Symmetry-reduced statistical model.** Eq. (11.3) statistically models the fitness of each site $g \in G$ separately. This raises challenges for empirical application, since it is usually impracticable to obtain a statistically sufficient number of fitness observations for every individual organism. Rather, empirical studies usually aggregate over individuals, perhaps separating them into classes based on sex, age, or other factors. This reflects a viewpoint that individuals in the same class can be considered interchangeable.

To implement this idea, a group $P$ of permutations of $G$ is specified as part of the statistical model. This permutation group, $P$, represents the modeler's guess of the "true" symmetry group, $\text{Sym}(G, p)$ (see Section 1.8). For example, if the population is hypothesized to be well-mixed (i.e., unstructured), then $P$ consists of all permutations of $G$. If there are distinct classes, but individuals within a given class are considered interchangeable and unstructured, then $P$ consists of all permutations that preserve the division into classes. Likewise, for a spatially-structured population, $P$ consists of permutations that preserve the spatial structure (e.g., translations of a lattice).

Any such permutation group $P$ partitions $G$ into non-overlapping classes (orbits, in the group-theoretic sense [11]). The statistical model is then formed by choosing a representative $g$ from each class, modeling $g$'s realized fitness as

$$W_g = \frac{1}{|P|} \sum_{\sigma \in P} \sum_{S \in \mathcal{S}_g} c_{\sigma(S),\sigma(g)} \, \underline{x}_{\sigma(S)} + \epsilon_{g,\mathbf{x}}. \tag{11.9}$$

Above $|P|$ is the size of $P$; i.e., the number of symmetries in the model. Since Eq. (11.9) involves only one representative $g$ of each class, it may be greatly reduced compared to the original statistical model Eq. (11.3). In particular, if $P$ acts transitively on $G$ (meaning the population is hypothesized to be homogeneous; see Section 1.8), the statistical model Eq. (11.9) consists of only one equation.

The conclusion of Theorem 11.1—that the estimators $\hat{c}_{S,g}$ converge to the true collective fitness effects $c_{S,g}$ for large samples from the underlying stochastic process—holds for the symmetry-reduced model Eq. (11.9) as well, under the additional stipulation that the permutation group $P$ in the statistical model matches the true symmetry group $\mathrm{Sym}(G,p)$. We state this as a corollary:

**Corollary 11.2.** *Suppose that the statistical model* Eq. (11.9) *has* $P = \mathrm{Sym}(G,p)$ *and, for each $g$ used as a representative of a class of sites, $\mathcal{S}_g$ contains all sets $S$ for which $c_{S,g} \neq 0$. Let $\hat{\pi}_m$, for $m \geq 1$, be the frequency distribution on $\mathbf{x}$ and $\mathbf{W}$ obtained by sampling $m$ times independently from $\tilde{\pi}$. Then, as $m \to \infty$, the estimators $\hat{c}_{S,g}$ for $\hat{\pi}_m$ converge almost surely to $c_{S,g}$ for each $S \subseteq G$ and $g \in G$.*

The proof is essentially the same as that of Theorem 11.1, making use of the additional fact that $c_{S,g} = c_{\sigma(S),\sigma(g)}$ for all symmetries $\sigma \in \mathrm{Sym}(G,p)$.